\documentclass[times,3p,sort&compress]{elsarticle}
\usepackage[utf8x]{inputenc}
\usepackage{graphicx,epsfig,color,booktabs}
\usepackage{amssymb,amsmath,caption}
\usepackage[dvipdfm,colorlinks,breaklinks,linkcolor=blue,citecolor=blue,anchorcolor=blue,urlcolor=blue]{hyperref}

\journal{Physics Reports}

\begin{document}

\begin{frontmatter}

\title{Percolation on complex networks: Theory and application}

\author[ustc]{Ming Li}
\author[hnu]{Run-Ran Liu}
\author[uestc,hnu,bcsrc]{Linyuan L\"u \corref{cor1}}
\ead{linyuan.lv@uestc.edu.cn}
\cortext[cor1]{Corresponding author}
\author[ustc]{Mao-Bin Hu}
\author[uestc]{Shuqi Xu}
\author[uf]{Yi-Cheng Zhang}

\address[ustc]{Department of Thermal Science and Energy Engineering, University of Science and Technology of China, Hefei, 230026, P. R. China}
\address[hnu]{Alibaba Research Center for Complexity Sciences, Hangzhou Normal University, Hangzhou, 310036, P. R. China}
\address[uestc]{Institute of Fundamental and Frontier Sciences, University of Electronic Science and Technology of China, Chengdu, 610054, P. R. China}
\address[bcsrc]{Beijing Computational Science Research Center, Beijing, 100193, P. R. China}
\address[uf]{Department of Physics, University of Fribourg, Fribourg CH-1700, Switzerland}

\begin{abstract}
In the last two decades, network science has blossomed and influenced various fields, such as statistical physics, computer science, biology and sociology, from the perspective of the heterogeneous interaction patterns of components composing the complex systems. As a paradigm for random and semi-random connectivity, percolation model plays a key role in the development of network science and its applications. On the one hand, the concepts and analytical methods, such as the emergence of the giant cluster, the finite-size scaling, and the mean-field method, which are intimately related to the percolation theory, are employed to quantify and solve some core problems of networks. On the other hand, the insights into the percolation theory also facilitate the understanding of networked systems, such as robustness, epidemic spreading, vital node identification, and community detection. Meanwhile, network science also brings some new issues to the percolation theory itself, such as percolation of strong heterogeneous systems, topological transition of networks beyond pairwise interactions, and emergence of a giant cluster with mutual connections. So far, the percolation theory has already percolated into the researches of structure analysis and dynamic modeling in network science. Understanding the percolation theory should help the study of many fields in network science, including the still opening questions in the frontiers of networks, such as networks beyond pairwise interactions, temporal networks, and network of networks. The intention of this paper is to offer an overview of these applications, as well as the basic theory of percolation transition on network systems.
%In this paper we will first introduce the methods for dealing with the classical percolation on complex networks, and the corresponding findings will also be reviewed. Second, we will review some typical percolation models and their critical phenomena in the study of networks. Third, we introduce the rapidly growing research on applications of percolation theory on networked systems. At last, a summary of this paper and an outlook on frontiers will be given.
\end{abstract}

\begin{keyword}
Percolation  \sep Complex Network \sep Network Structure \sep Network Dynamics \sep Phase Transition and Critical Phenomena
\end{keyword}

\end{frontmatter}

\newpage
\tableofcontents
\pagenumbering{arabic}
\newpage

%**************************************************************************
% Section I
% Introduction
%**************************************************************************

\section{Introduction}

In contrast to many other modern research fields, the network problem is often easy to define by abstracting from everyday life \cite{Newman2010,Barabasi2016}. For examples, how many people an epidemic can infect in a social contact network, whether a communication network can maintain its function after an intentional attack, which node has the largest impact in a social network, and so on \cite{Albert2002,Dorogovtsev2002a,Boccaletti2006,Dorogovtsev2008}. The key point of these problems with network involved can be summarized as a cluster forming process within a chosen fraction of nodes, those might be infected people, preserved nodes after an intentional attack, or individuals with the same opinion. In principle, these processes are easy to define, however, not so easy to solve.

Fortunately, in statistical physics, a profound theoretical system, called percolation theory, just touches this problem, \textit{i.e.}, the behaviors of a networked system when some of nodes or links are not available \cite{Stauffer1991}. Indeed, when the network science was just a new rising topic, the percolation theory has already been widely used for explaining empirical results, and solving models \cite{Barabasi2016}. Now, after more than twenty years' development of network science, the percolation theory, including conceptions, analytical methods, and algorithms, can be found almost in all the research fields of network science.

It is known that the classical percolation in statistical physics only considers regular lattices, therefore, with these applications to complex networks, the percolation theory itself also has been enriched and developed. In recent hot areas of network science, such as higher-order networks \cite{Boccaletti2014,Bianconi2018} and networks beyond pairwise interactions \cite{Battiston2020}, models and methods of percolation have still been widely touched. This is obviously because the connection property must always be a key point to understand network structure and dynamics.

Due to importance of the percolation theory in the study of complex networks, almost all the review articles about networks have the relevant sections to introduce distinguishable developments and applications of percolation theory on complex networks, however, a systematic comparison and summary specifically from the perspective of percolation is still absent. This review article aims to fill this gap, and comb the scattered discussions on the network percolation and its applications, which can facilitate a wide area of sciences, ranging from physics and computer science to biology and sociology, as well as various branches of probability theory in mathematics.

\subsection{Classical percolation}

Percolation now usually refers to a class of models that describe geometric features of random media. In statistical physics, percolation theory is often accompanied by scaling law, fractal, self-organization criticality, and renormalization, which are all of immense importance theoretically in many diverse fields of physics \cite{Stauffer1991}. Therefore, percolation has long served as a basic ideal model for demonstrating phase transition and critical phenomena. However, quite apart from the role percolation theory now plays, it originates from an honest applied problem in the study of gelation in the 1940s \cite{Flory1941,Flory1941a,Flory1941b}. To be a mathematical subject, it first starts from Broadbent and Hammersley's paper in 1957 \cite{Broadbent1957}, in which its name, and the geometrical and probabilistic concepts were introduced.

The study of percolation then becomes popularized in the physics community, and many of the open problems have been proposed and solved \cite{Stauffer1979,Essam1980,Isichenko1992,Sahimi1993,Nakayama1994,Araujo2014,Saberi2015}. Now, the percolation theory has also been found to be of a broad range of applications to diverse problems. The applications in network science that this article focuses on are one of the typical representatives. This is also benefited from the development of computational technology, as the simulation plays a crucial role in the study of percolation \cite{Hammersley1983}.

In the following, we will firstly give a brief introduction to the percolation model, as well as some physical concepts and quantities involved.

\subsubsection{Percolation process in random media}

Imagine a large porous stone immersed in water. Does the water come into the core of the stone? If so, there must be some paths formed by the pores running through the stone. However, as a whole, the connection of any two adjacent pores is probabilistic, supposing the probability is $p$. The problem thus reduces to whether there exist such paths for a given probability $p$. This actually is a typical example of percolation problems. The connection of two adjacent pores, in the terminology of percolation, is called occupying the corresponding bond between the two pores, and hence $p$ is the occupied probability. Obviously, if $p$ is large enough, the core of the porous stone can be wetted. In that case, the connected pores are able to form a cluster that penetrates the stone. Accordingly, percolation theory is mainly concerned with the existence of such a cluster and its structure with respect to the occupied probability $p$.

The above process is just a special case of the percolation process. The same question can also be proposed for many other systems constituted by random medium. Another typical example is the forest fire model. Suppose a burning tree can only ignite the trees in the adjacent sites, the destructiveness of the forest fire depends on the density of the trees, \textit{i.e.}, the probability $p$ of finding a tree on a site. This is also obviously a percolation process -- when the adjacent trees form a cluster that can penetrate the entire forest, the fire could raze almost the entire forest; otherwise, the fire is constrained in a small area. Furthermore, similar regulations can also be applied to model the spreading of epidemics among individuals, where infected individuals can infect their neighbors probabilistically.

Although the soaking of porous stone, the forest fire and the spreading of epidemics seemingly belong to separated fields, the three are now converging in an intriguing manner. That is, is there a cluster of connected sites through the system? As is quite typical, it is actually easier to examine such an infinite cluster in an infinite system than just large ones. With the increasing of the probability $p$, there must be a critical $p_c$, called percolation threshold or critical point, below which such a cluster cannot be found. For a more detailed understanding of this criticality, the percolation model needs to be defined explicitly. That is what we are going to talk about.

\subsubsection{Percolation model on lattices}

\begin{figure}
\centering
\includegraphics[width=0.9\columnwidth]{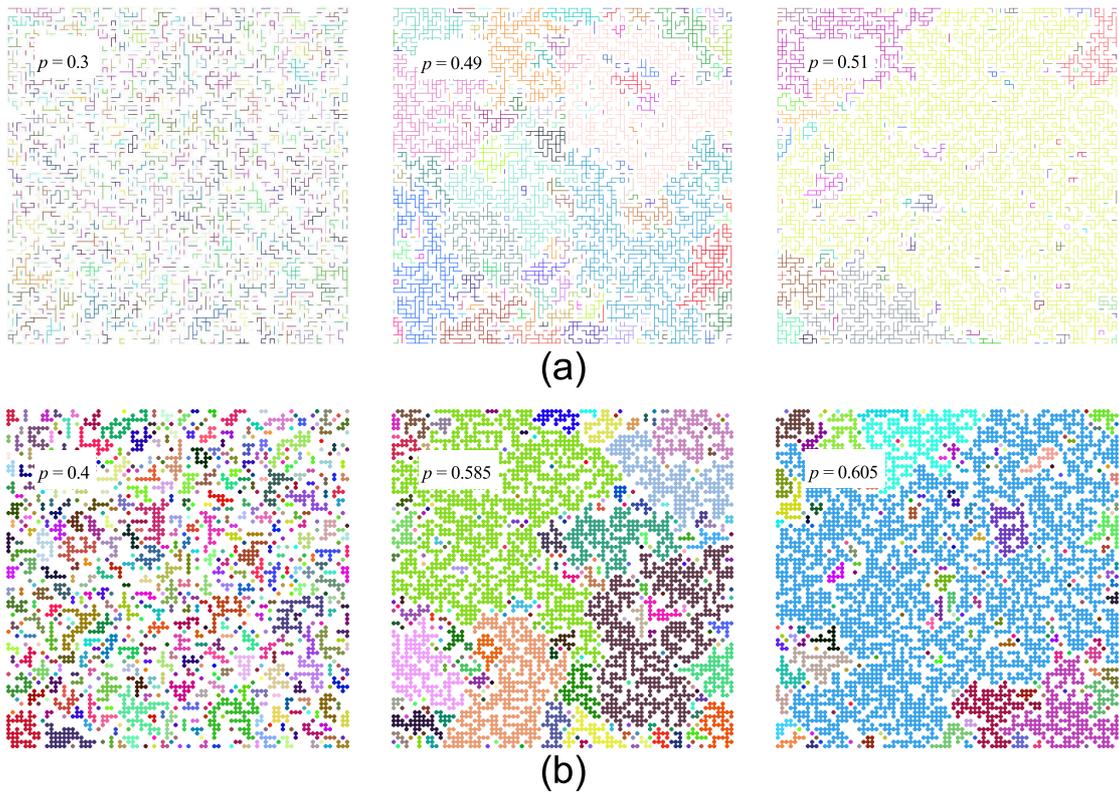}\\
\caption{(Color online) Schematic diagrams of the classical percolation on square lattice, where different colors denote different clusters. The size of the system used here is $N=L\times L=80\times 80$. The values of $p$ labeled in the figures are the corresponding site/bond occupied probabilities. (a) Bond percolation. For ease of identification, the sites and the unoccupied bonds are not shown here. For $p=0.51$, a giant cluster exists indicated by yellow. (b) Site percolation. For ease of identification, the bonds and the unoccupied sites are not shown here. For $p=0.605$, a giant cluster exists indicated by blue.} \label{fig1-pclc}
\end{figure}

To facilitate presentation, we consider a two-dimensional lattice here as shown in Fig.\ref{fig1-pclc} (a). Mathematically, each bond is occupied with probability $p$ or unoccupied with probability $1-p$. Then, the occupied bonds connect the sites into clusters. This model is called bond percolation, which can be used to model the process of the soaking of porous stone and the spreading of epidemics. For the forest fire model, one often uses a slightly different percolation model, the so-called site percolation model. For this model, we occupy each site with probability $p$ rather than bonds as shown in Fig.\ref{fig1-pclc} (b). In general, the bond percolation is considered less general than the site percolation due to the fact that the bond percolation can be reformulated as a site percolation on a different lattice, but not vice versa.

The percolation theory mainly focuses on the emergence of the infinite cluster with the increasing of probability $p$. To characterize this phenomenon, one often adopts the size of the giant cluster, which is defined as
\begin{equation}
S=\lim_{N\to\infty}\frac{s_1}{N}.
\end{equation}
Here, $N$ is the size of the system (site number), and $s_1$ is the number of sites in the largest cluster. As shown in Fig.\ref{fig1-pclc}, with the increasing of $p$, there must be a critical point $p_c$, above which a non-zero $S$ can be found. This figures out the percolation transition of the system with respect to the control parameter $p$, and $S$ is the corresponding order parameter.

In addition, there is another commonly used order parameter called wrapping probability, which is defined as the probability that a cluster wraps around the periodic boundary conditions on a regular lattice. For large systems, this probability is equal to the probability that the system percolates. This parameter is usually used to estimate the position of the percolation threshold, since it is a size-independent parameter at the critical point. Specifically, cluster wrapping can be defined in a number of different ways, such as wrapping around one direction, wrapping around one direction but not the other, and wrapping around both directions. This probability, however, cannot be well defined in network systems without spatial constraints, so that the size of the giant cluster is the preferred order parameter in network science.

To describe the features of finite clusters, the distribution of cluster sizes $p_s=m_s/\sum_sm_s$ is also used, where $m_s$ is the number of the clusters with size $s$. Sometimes, the normalized cluster number $n_s=m_s/N$ is also used to feature the cluster size distribution. It is obvious that $p_s=n_sN/\sum_sm_s=n_s\langle s\rangle$, where $\langle s\rangle= N/\sum_s m_s=\sum_s sp_s$ is the average cluster size.

It must be pointed out that the average size $\langle s\rangle$ is not the mean cluster size $\chi$ commonly used in percolation theory, which is defined as $\chi=\sum_s s^2n_s$. It is not hard to know that $sn_s$ is the probability that a randomly chosen site belongs to a cluster with size $s$. Thus, the mean cluster size $\chi$ actually is the average size of the cluster that a randomly chosen site belongs to. Together with the characteristic length $\xi$ and the characteristic size $s_\xi$, above which the clusters are exponentially scarce, these statistics are often used to describe and characterize the percolation transition \cite{Stauffer1991}. The critical behaviors of these parameters will be briefly introduced in the next section.

Besides, there exist several other variants of percolation. For example, one could drop the assumption of independence of occupation, so that the occupation of a site or a bond could depend on the state of other sites or bonds. This type of percolation is often called dependent percolation, which is widely used in network science to model the spreading dynamics and the cascading process \cite{Albert2002,Boccaletti2006,Dorogovtsev2008,Boccaletti2014,Pastor-Satorras2015}. Such models will also be discussed in subsequent parts of this paper.

\subsubsection{Percolation transition}

The prerequisite for the study of critical phenomena is determining the percolation threshold $p_c$. Figure \ref{fig1-pclc} indicates that the site percolation and the bond percolation have different critical points. In fact, the threshold of bond percolation on square lattice can be solved exactly by the duality transformation or real-space renormalization \cite{Wu1982,Stauffer1991,Christensen2005,Ziff2006,Ohzeki2013}, that is $p_c=1/2$. However, there is no exact solution for site percolation on square lattice. Through Monte Carlo simulation, a good estimate of the critical point can be obtained, such as $p_c=0.59274621(13)$ in Ref.\cite{Newman2000}, and $p_c=0.5927465(4)$ in Ref.\cite{Deng2005}. In Tab.\ref{tb1-cp}, we list the percolation thresholds for some typical systems.

\begin{table}[ht]
\centering \footnotesize
\caption{Selected percolation thresholds for various networks. Here, $z$ is the coordination number of the Bethe lattice, $m$ is the minimum degree of SF network, $\zeta(s,a)$ is the Hurwitz zeta function, $G_1(x)$ is the generating function of the excess-degree distribution, and $\langle k^n\rangle$ is the $n$th moment of the degree distribution.} \label{tb1-cp}
\begin{tabular}{lp{0.35\columnwidth}p{0.35\columnwidth}}
\toprule
Network & Site & Bond \\
\hline
square lattice & $0.59274621(13)$ \cite{Newman2000}, $0.5927465(4)$ \cite{Deng2005} & $1/2$ \\
triangular lattice & $1/2$ & $2\sin(\pi/18)$ \cite{Sykes1964} \\
honeycomb lattice & $0.697040230(5)$ \cite{Jacobsen2014}, $0.6970402(1)$ \cite{Feng2008}, $0.6970413(10)$ \cite{Ziff2009} & $1-2\sin(\pi/18)$ \cite{Sykes1964}  \\
simple cubic & $0.3116077(4)$ \cite{Deng2005}, $0.3116077(2)$ \cite{Wang2013}, $0.3116060(48)$ \cite{Koza2016}, $0.3116004(35)$ \cite{Skvor2009}, $0.31160768(15)$ \cite{Xu2013} & $0.24881182(10)$ \cite{Wang2013}, $0.2488125(25)$ \cite{Dammer2004}, $0.2488126(5)$ \cite{Lorenz1998} \\
hypercubic $d=4$ & $0.1968861(14)$ \cite{Grassberger2003}, $0.196889(3)$ \cite{Paul2001}, $0.196901(5)$ \cite{Ballesteros1997}, $0.19680(23)$ \cite{Kotwica2019}, $0.1968904(65)$ \cite{Koza2016}, $0.19688561(3)$ \cite{Mertens2018} & $0.1601314(13)$ \cite{Grassberger2003}, $0.160130(3)$ \cite{Paul2001}, $0.1601310(10)$ \cite{Dammer2004}, $0.1601312(2)$ \cite{Xun2020}, $0.16013122(6)$ \cite{Mertens2018} \\
hypercubic $d=5$ & $0.1407966(15)$ \cite{Grassberger2003}, $0.1407966(26)$ \cite{Koza2016}, $0.14079633(4)$ \cite{Mertens2018}  & $0.118172(1)$ \cite{Grassberger2003}, $0.1181718(3)$ \cite{Dammer2004}, $0.11817145(3)$ \cite{Mertens2018} \\
hypercubic $d=6$ & $0.109017(2)$ \cite{Grassberger2003}, $0.1090117(30)$ \cite{Koza2016}, $0.109016661(8)$ \cite{Mertens2018}  & $0.0942(1)$ \cite{Adler1990a}, $0.0942019(6)$ \cite{Grassberger2003}, $0.09420165(2)$ \cite{Mertens2018} \\
hypercubic $d=7$ & $ 0.0889511(9)$ \cite{Grassberger2003}, $0.0889511(90)$ \cite{Koza2016}, $0.088951121(1)$ \cite{Mertens2018}  & $0.078685(30)$ \cite{Adler1990a}, $0.0786752(3)$ \cite{Grassberger2003}, $0.078675230(2)$ \cite{Mertens2018} \\
Bethe lattice & $1/z$ & $1/z$  \\
ER network & $1/\langle k\rangle$ \cite{Bollobas2001,Newman2001} & $1/\langle k\rangle$ \cite{Bollobas2001,Newman2001}  \\
SF network & $\zeta(\lambda-1,m)/[\zeta(\lambda-2,m)-\zeta(\lambda-1,m)]$ \cite{Newman2001} & $\zeta(\lambda-1,m)/[\zeta(\lambda-2,m)-\zeta(\lambda-1,m)]$ \cite{Newman2001}  \\
tree-like network & $1/G_1'(1)=\langle k\rangle/[\langle k^2\rangle-\langle k\rangle]$ \cite{Newman2001} & $1/G_1'(1)=\langle k\rangle/[\langle k^2\rangle-\langle k\rangle]$ \cite{Newman2001}  \\
\bottomrule
\end{tabular}
\end{table}

In the subcritical regime ($p<p_c$), all clusters are finite and the size distribution has a tail which decays exponentially; in the supercritical regime ($p>p_c$), an infinite cluster can be found in the system, and the size distribution of other finite clusters has a tail which decays slower than exponential. Near the critical point, some asymptotic behaviors can be found, referred to as critical phenomena. In the percolation theory, these behaviors are characterized by the critical exponents \cite{Stauffer1991},
\begin{align}
S     &\varpropto (p-p_c)^{\beta}, \\
\chi  &\varpropto |p-p_c|^{-\gamma}, \\
\xi   &\varpropto |p-p_c|^{-\nu},    \\
s_\xi &\varpropto |p-p_c|^{-1/\sigma},   \\
p_s   &\varpropto s^{-\tau}.
\end{align}
with relations
\begin{align}
\beta  &= \frac{\tau-2}{\sigma},  \\
\gamma &= \frac{3-\tau}{\sigma}.
\end{align}
Here, $\beta$, $\gamma$, $\nu$, $\sigma$, and $\tau$ are the so-called critical exponents, which determine the universal class of the percolation transition. Besides, at the critical point the characteristic length $\xi$ and size $s_\xi$ also have a relation
\begin{equation}
s_\xi \varpropto \xi^{d_f}.
\end{equation}
The exponent $d_f$ is often called fractal dimension \cite{Stauffer1991,Bunde2012}, characterizing the structure of the infinite cluster at the critical point. Assuming the dimension of the system is $d$, there is another relation between critical exponents, called hyperscaling,
\begin{equation}
d_f=d-\frac{\beta}{\nu}.
\end{equation}
Note that this relation is also universal, and independent of the topological structure of the system. The phase transition theory points out that there is an upper critical dimension $d_c$ (for percolation $d_c=6$), above which the critical exponents become the same as that in mean-field theory \cite{Stanley1971,Ma2018}. So this hyperscaling relation holds only for $d\leq d_c$. It is also worth mentioning that the geometric structure of high-dimensional percolation clusters cannot be fully accounted for by the counterparts of random networks, though both of them have the mean-field nature \cite{Huang2018}.

In addition, the critical behavior below $d_c$ is different from the mean-field approximation which is valid away from the phase transition. For these systems, the renormalization group theory has made remarkable predictions about the behavior of the percolation near and at the threshold \cite{Pelissetto2002,Zinn-Justin2007}. In the renormalization group theory, there is also a lower critical dimension (for percolation $d_l=2$), below which there is no phase transition.

Furthermore, the universality principle states that the value of $p_c$ is determined by the local structure of the system, whereas the behavior of clusters that is observed near $p_c$ is independent of the local structure (lattice type and percolation type). In this sense, percolation is believed to be a substrate-dependent but model-independent process, and therefore, the critical exponents of the transition are only determined by the geometry of the system, and identical for the bond and site percolation. The critical exponents are thus more natural to be considered than the threshold $p_c$, and there is no need to deal with the site and bond percolation, individually.

To end this subsection, it must be pointed out that the critical exponent $\beta$ is distinguishable for the site and bond percolation on scaling-free (SF) networks \cite{Radicchi2015}. This is a special case derived from the vanished percolation threshold, which has no effect on the other properties discussed above.

\subsection{Networks}

This paper mainly focuses on the percolation theory on networks and its applications. So, in this section we will provide an overview of some simple and general properties and models of networks.

\subsubsection{Network representation}

In network science, the elements of a system and the connections between them are no longer known as site and bond. Instead, they are often called node and link, or vertex and edge, respectively. The number of links a node has is called degree, labeled $k$.

To exactly represent the connection pattern of a network, one often uses a $N\times N$ matrix $\mathbf{A}$, called adjacent matrix, where $N$ is the number of nodes in the network. In this way, the element $a_{ij}$ of $\mathbf{A}$ is one when there is a link from node $i$ to node $j$, and zero when there is no link. For undirected networks, it must have $a_{ij}=a_{ji}$, thus $\mathbf{A}$ is a symmetric matrix. For weighted networks, the element $a_{ij}$ can further be any non-zero value to represent the corresponding link weight.

\subsubsection{Statistical properties}

Instead of studying a special network with a given adjacent matrix, the network science is more on discovering the common nature of a class of networks, \textit{i.e.}, a network ensemble. For that, networks are often featured by the degree distribution $p_k$, which provides the probability that a randomly selected node in the network has degree $k$.

A typical network ensemble is the one with Poisson degree distribution
\begin{equation}
p_k=\frac{e^{-\langle k\rangle}\langle k\rangle^k}{k!},
\end{equation}
where $\langle \rangle$ means the average over all the nodes. This just is the ensemble of the known Erd\H{o}s--R\'{e}nyi (ER) random networks/graphs \cite{Bollobas2001}, and can be simply realized by randomly connecting a given number of links among a set of nodes, or connecting each pairs of nodes with a given probability. Quite obviously, the generation of ER networks is actually a percolation process. Only when the system percolates ($\langle k\rangle>1$), the giant cluster can be found. Specifically, when $\langle k\rangle<1$ almost surely each cluster has size $\log N$, at $\langle k\rangle=1$, the largest cluster has a size of order $N^{2/3}$, and at $\langle k\rangle>1$, there exists a single giant cluster of size of order $N$ and all other clusters have a size of order $\log N$. Since such systems have no spatial constraint, the critical phenomena in this ensemble just recover the mean-field solution.

Besides, distinguishing from the degree distribution, there is another kind of commonly used networks, which represents a deeper organizing principle of real networks called the SF property \cite{Barabasi2016}. In mathematical terms, the SF property translates into a power-law function of the form
\begin{equation}
p_k \propto k^{-\lambda}, ~~~~ k\geq m.  \label{eq1-pk}
\end{equation}
For a real network, there must also be an upper bound of degree, called degree cutoff $K$.

The main difference between an ER and a SF network comes in the tail of the degree distribution. For ER networks, most nodes have comparable degrees and hence the degree cutoff is in the order of the average degree. In contrast, nodes with very large degrees are expected in SF networks, called hubs of the networks. Indeed, the degrees of hubs grow with the network size, and thus can grow quite large. This indicates that SF networks have strong heterogeneity, and hence the percolation transition bears a strong dependence upon the degree distribution. Beyond this, there are many other measurements to further subdivide the network ensembles, such as clustering, correlation, and community. In recent years, temporal networks, multiplex networks, and high-order networks have also received a lot of attention. Because of space limitation and the focus of this paper, the details will not be included here and brief discussions will be given in this article when necessary. Further details on these can also be found in Refs.\cite{Albert2002,Boccaletti2006,Dorogovtsev2008,Newman2010,Barabasi2016}.

\subsubsection{Network models}

After a review of the fundamental properties of complex networks, we explore the mathematical modeling of network ensembles, and mainly focus on the configuration model and the hidden parameter model \cite{Newman2010,Barabasi2016}. In principle, the two models can generate any network ensembles with a meaningful degree distribution. In fact, these two models are usually only used to realize SF property, since there exist easy-implemented models for Poisson degree distribution (ER networks) and small-world (SW) networks \cite{Newman2010,Barabasi2016}. Note that the SF network also has an easy-implemented model, Barab\'{a}si--Albert (BA) model \cite{Barabasi1999}, nevertheless, the generated SF network has its own structural features and limitations rather than a network with programmable and tunable SF properties.

From random graph theory \cite{Bollobas2001}, there are two ways to generate a network with Poisson degree distribution. One is randomly connecting nodes until the excepted number of links is met, and the other is connecting each pair of nodes with a given probability. For a given average degree, networks implemented by the former method must be a subset of those by the later one. In spite of this, the two methods will become equivalent in the thermodynamic limitation ($N\to\infty$).

The SW network grows out of a regular network, such as the triangle lattice and the square lattice, by randomly rewiring a small fraction of links \cite{Watts1998,Newman1999,Newman2000a,Newman2000b}. Since there is no spatial constraint for the rewired link, the average distance of nodes becomes much smaller than that of the underlying regular network. This is why it is called small-world. In turn, the regular backbone of the SW network usually leads to a high clustering, namely, the neighbors of a node are also connected. This is another characteristic of the SW network. For this type of networks, the degree distribution is often not the concern, and one often uses the backbone network and the rewiring probability to define a SW network ensemble.

The configuration model can help us build a network with a pre-defined degree distribution \cite{Bender1978,Bollobas1980,Newman2010,Barabasi2016}. The algorithm consists of two steps. First, assigning degrees to each node drawn randomly from the pre-defined degree distribution, represented as stubs. Then, two stubs are selected randomly and connected. Repeating this procedure until all stubs are paired up. If there is nothing in this procedure to forbid self- and multi-connections, the obtained network is probably not a simple graph as usually studied in network science. While rejecting such connections could add much more overhead to the program's execution time.

\begin{figure}
\centering
\includegraphics[width=0.7\columnwidth]{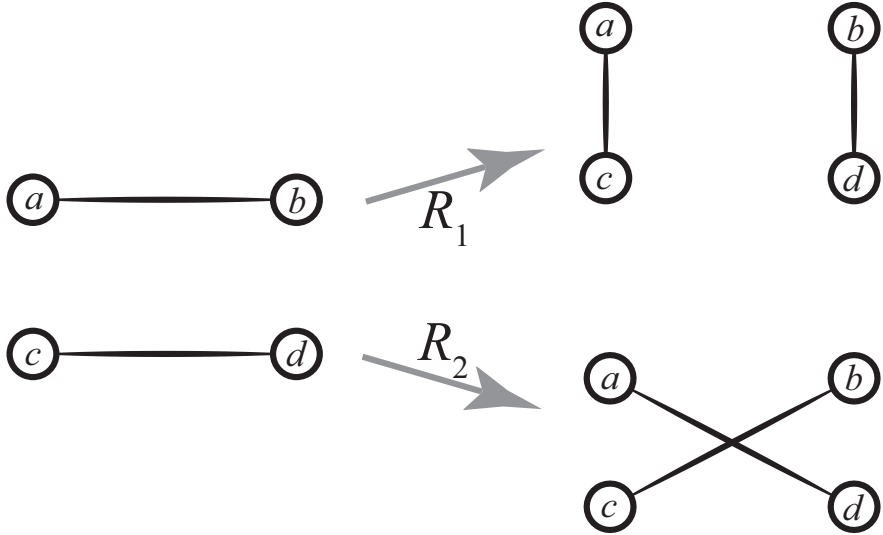}
\caption{Schematic diagram of the rewiring procedure. Here, the rewiring means that swapping the ends of two links. There are two ways to rewire, labeled $R_1$ and $R_2$ respectively in the figure. Note that, to ensure the ergodicity and the detailed balance, the two ways must be chosen randomly.} \label{fig1-rp}
\end{figure}

Indeed, one of the effective methods to generate random networks by the configuration model without self- and multi-connections can be as follows. First, connecting the stubs in any fast and easy-to-implement ways (the details are dependent on the specific degree distribution), and recording the self- and multi-connections. Then, rewiring a fraction of links of the obtained network, and the self- and multi-connections must be included. Note that the rewiring leading to self- or multi-connections is forbidden.

A schematic diagram of the rewiring procedure is shown in Fig.\ref{fig1-rp}. For the degree distribution with the maximum degree $K$ larger than $\sqrt{N}$, this method could be much more effective than the original one, which may not even complete the network. This is because there must be some degree correlation in the network with $K>\sqrt{N}$ \cite{Burda2003,Boguna2004}, randomly connecting stubs with self- and multi-connection forbidden is a very inefficient procedure. Of course, the efficiency of the algorithm also depends on the implementing way and the detailed property of the network, thus this just is a general discussion and not always the case.

By the way, the rewiring operation as shown in Fig.\ref{fig1-rp} is often called degree preserving randomization, which can randomize the connections of a network without changing the degree distribution. With this operation, one can eliminate other properties from the network, for example, clustering, and assure that a certain network feature is predicted by its degree distribution alone. If the network is large enough and all the degrees are much smaller than the network size $N$, the degree preserving randomization could turn the network to be locally tree-like.

Note that the correlation derived from the degree distribution cannot be undone by this randomization procedure. A typical example is a network with $K>\sqrt{N}$. Consequently, when we study networks with hubs, \textit{i.e.}, SF networks, the configuration model and the degree preserving randomization must be carefully dealt with. In turn, with some constraints, this rewiring operation can also ``randomize" the network into the one with a given high-order property \cite{Volz2004,AngelesSerrano2005}.

There is another model, hidden parameter model, to generate networks with a pre-defined degree distribution \cite{Goh2001,Caldarelli2002,Soederberg2002,Chung2002,Boguna2003}. In this model, each node $i$ is assigned a hidden parameter $\eta_i$, chosen from a pre-defined distribution $p_\eta$. Then, we connect each node pair with probability $p(\eta_i,\eta_j)=\eta_i\eta_j/N\langle \eta\rangle$. The expected number of links is thus $L=\sum_{i,j}\eta_i\eta_j/2N\langle \eta\rangle$. For SF networks, we can set $\eta_i\propto i^{-\alpha}$, and the obtained network has the degree distribution $p_k\propto k^{-(1+1/\alpha)}$.

The SF networks generated by the configuration model and the hidden parameter model also have their own properties, thus are not exactly equivalent, at least for small systems. First of all, the boundaries of degrees must be given for the configuration model to normalize the degree distribution. However, the boundaries of degrees are not needed for the hidden parameter model, in which a degree cutoff occurs naturally at $N^{1/(\lambda-1)}$, thus known as the natural cutoff of the SF network \cite{Cohen2001,Hong2007,Waclaw2008,Castellano2008,Baek2012}. This cutoff can be also found in the configuration model, if the upper boundary $M$ used to normalize the degree distribution is larger than $N^{1/(\lambda-1)}$. Then, the hidden parameter model generates no self- and multi-connections, since only a single connection would occur between each node pair. Another is that the degree distribution of the SF network generated by the hidden parameter model has negative deviations for small degrees, while the one generated by the configuration model almost perfectly matches the pre-defined degree distribution. In addition, setting the total number of links is not allowed in the configuration model, since it is fixed by the pre-defined degree distribution. Rather, the hidden parameter model allows controlling the link number, accurately. Finally, what need to be pointed out is that both the networks generated by the configuration model and by the hidden parameter model are only a possible realization of the pre-defined degree distribution. To check the property induced by a certain degree distribution, the ensemble average must be done.

In recent years many works have also been devoted to the so-called multiplex networks (or multilayer networks, interdependent networks) \cite{Boccaletti2014,Bianconi2018}. In this network ensemble, links are divided into different layers to represent different types of connections between nodes. To generate such a network ensemble, one can first generate several networks using the methods mentioned above, and then bundle nodes from different networks together to form the layered connections. It can also be interpreted as inserting some inter-connections between nodes in different networks (layers), which represent the dependence relations between them. The bundling of nodes (or inter-connection) could be one-to-one, one-to-many, or with some other rules, that depending on the inter-correlation between layers. Of course, there are many other network models with different properties, one can find them in special books and reviews about complex networks.

\subsection{Outline of the report}

This paper contains six sections. In the next section, we start by reviewing the basic properties of the percolation process on networks, including the model definition, the analytical methods, and the corresponding discussions. Section \ref{sct3} reviews some typical percolation models on networks, mainly focusing on the phase transition and the critical phenomena. The percolation model is by nature featuring the cluster forming process in random media. Therefore, the percolation theory including conceptions, theoretical methods, and algorithms, is widely used in network structure analysis. This is the main content of Sec.\ref{sct4}. Moreover, the percolation process is often used to model and analyze network dynamics, which will be reviewed in Sec.\ref{sct5}. Finally, we summarize and prospect to this paper in Sec.\ref{sct6}.

In addition, the main notations used in this paper are listed in Tab.\ref{tb1-nl} for readability.

\begin{table}[ht]
\centering \footnotesize
\caption{The main notations used in this paper.} \label{tb1-nl}
\begin{tabular}{cl}
\toprule
Notation & Meaning \\
\hline
$d$ & system dimension \\
$d_f$ & fractal dimension \\
$G_0(x)$ & generating function of degree distribution \\
$G_1(x)$ & generating function of excess-degree distribution \\
$\mathcal{H}$ & Hamiltonian of Potts model \\
$H_0(x)$ & generating function of $\pi_s$ \\
$H_1(x)$ & generating function of $\rho_s$ \\
$i_k(t)$ & fraction of infected individuals in all degree-$k$ nodes at time step $t$ \\
$k$ & node degree \\
$K$ & degree cutoff of a network \\
$\langle k\rangle$ & average degree \\
$\langle k^n\rangle$ & $n$-th moment of degree distribution \\
$\mathbf{M}$ &  Hashimoto or non-backtracking matrix of graph \\
$N$ & node number of a network \\
$p$ & occupied probability \\
$p_c$ & percolation threshold \\
$p_k$ & degree distribution \\
$p_s$ & distribution of the cluster sizes \\
$q_k$ & excess-degree distribution \\
$r_{i\to j}$ & probability that link $i\to j$ belongs to the giant cluster \\
$r_k(t)$ & fraction of removed individuals in all degree-$k$ nodes at time step $t$ \\
$R$ & probability that a link belongs to the giant cluster \\
$S$ & probability that a node belongs to the giant cluster \\
$s$ & cluster size \\
$s_i$ & probability that node $i$ belongs to the giant cluster \\
$s_k(t)$ & fraction of susceptible individuals in all degree-$k$ nodes at time step $t$ \\
$s_\xi$ & characteristic size \\
$Z$ & partition function of Potts model \\
$\mathcal{Z}$ & rescaled partition function of Potts model \\
$z_i$ & partition function of the branching from node $i$ of Potts model \\
$\beta$ & critical exponent for the giant cluster; infection rate \\
$\gamma$ & critical exponent for the mean cluster size\\
$\delta_{x,y}$ & Kronecker delta function \\
$\zeta(x)$ & Riemann zeta function \\
$\zeta(x,a)$ & Hurwitz zeta function \\
$\Theta_k(t)$ & probability that a neighbor of a degree-$k$ node is infected \\
$\kappa$ & ratio of $\langle k^2\rangle$ and $\langle k\rangle$ \\
$\lambda$ & exponent of SF degree distribution, $p_k\propto k^{-\lambda}$ \\
$\mu$ & recovery rate \\
$\nu$ & critical exponent for the characteristic length \\
$\xi$ & characteristic length \\
$\pi_s$ & size distribution of the cluster that a randomly chosen node belongs to \\
$\rho_s$ & size distribution of the cluster at the end of a link \\
$\sigma$ & critical exponent for the characteristic size \\
$\tau$ & Fisher exponent \\
$\Phi(x,s,a)$ & Lerch’s transcendent \\
$\chi$ & mean cluster size \\
\bottomrule
\end{tabular}
\end{table}

%*****************************************************************************
% Section II
% Classical percolation on networks
%*****************************************************************************

\section{Classical percolation on networks}

In statistical physics, the percolation model is often displayed on regular lattices. However, the regular topological structure is not always the case in reality. In this section, we will review the basic properties of the classical percolation on heterogeneous network systems, mainly focusing on analytic methods, critical phenomena, and Monte Carlo algorithms of the classical percolation. It should be pointed out that there are a lot of theoretical studies of percolation transition on networks from the perspective of random graphs, one can refer to Refs.\cite{Molloy1995,Aiello2001,Bollobas2001,Chung2002a,Janson2009,Bollobas2012,Hatami2012} and the relevant references to learn more. Here we mainly focus on the network percolation from the perspective of statistical physics and network science.

\subsection{Problem description}  \label{sct2-drnp}

As a lattice model, the percolation model can be easily extended to networked systems, namely, nodes or links are randomly designated either occupied (with probability $p$) or unoccupied  (with probability $1-p$). For convenience, in network science one usually removes each node or link with probability $1-p$ to realize the percolation model. The parameters and the corresponding problems can be defined on the preserved network as that on regular lattices. As a result, there is no need to repeat the arguments. Here we will concentrate on the relationships between the classical percolation model and some typical issues of network science, such as network robustness, and epidemics spreading on network.

For site percolation, the network science often translates the unoccupied nodes as failed/removed ones. In this way, the percolation process is just the performance of the network under random nodal failures. If the giant cluster remains after the percolation process, the resulted network can be thought of as a whole, and still functional. Thus the percolation threshold $p_c$ can be used to evaluate the robustness of the network \cite{Callaway2000,Cohen2000,Albert2000,Cohen2001,Cohen2010}. A large threshold $p_c$ indicates that only a small amount of failed nodes can disconnect the network, so its robustness is poor, and vice versa. Furthermore, if the unoccupied nodes are chosen with preference, this model can further explain the robustness of networks under intentional attacks. A remarkable knowledge of this model is the finding of the strong robustness of SF networks under random failures, and extremely fragile for intentional attacks \cite{Albert2000}. A specific discussion on this problem can be found in Sec.\ref{sct4}.

In network science, bond percolation often refers to the spreading process \cite{Pastor-Satorras2015}. The connection between the spreading of epidemic and percolation is in fact one of the original motivations for the percolation model itself \cite{Essam1980,Stauffer1991,Isichenko1992,Christensen2005}. For this process, the occupation of a link means a successful infection between the two connected nodes. That is, the occupied probability reflects the infectiousness of the epidemic. However, different from the mapping between site percolation and models for network robustness, the link occupied probability is not simply equivalent to the infectious probability, but an integrated probability of transmission of the disease between two individuals \cite{Newman2002}. In this way, the emergence of the giant cluster corresponds to the outbreak of an epidemic, so that the percolation threshold can also be used to evaluate the infectiousness of the epidemic. An epidemic with smaller percolation threshold $p_c$ has a stronger infectious ability. Specific discussion on this problem can be found in Sec.\ref{sct5}.

The basic mechanisms of the network percolation and their extensions can be further applied in various fields of network science from structure analysis to dynamics modeling, which this article explores next.

\subsection{Analytic method based on branching process}  \label{sct2-ampb}

Although the percolation model is easy to define, the exact solutions are absent for most systems. If the network topology is restricted to be tree-like, like that of Bethe lattice, there is a mean-field method based on the branching process for solving the percolation model. Here we introduce the framework of this method, including the solutions for the size of the giant cluster and the distribution of finite clusters \cite{Newman2001,Newman2002,Newman2010}.

\subsubsection{Behaviors of the giant cluster}

The percolation process is tantamount to dilute links and thus changes the degree distribution $p_k$. So, for convenience and generality, the occupation of nodes or links will not be considered in the following discussion. After obtaining the result, we can revise it by considering the diluting effect of the percolation on the degree distribution $p_k$.

When a network percolates, there must be an infinite cluster, in which the branching process is endless. Specifically, when arriving at a node by following a link, the node must have some other links (at least one) to ensure the continuity of branching. In terms of this, assuming a link belongs (connects) to the giant cluster with probability $R$, we have a self-consistent equation
\begin{align}
R &= \sum_{k=1}q_k\left[ 1-(1-R)^{k-1}\right]  \nonumber \\
  &= 1-G_1(1-R), \label{eq2-r}
\end{align}
where $q_k=p_kk/\langle k\rangle$ is the probability that the node reached by following a link has degree $k$, known as the excess-degree distribution, and $G_1(x)=\sum_kq_kx^{k-1}$ is the corresponding generating function. The term $1-(1-R)^{k-1}$ in the square brackets just means that at least one excess link is needed to keep branching.

After obtaining $R$ from Eq.(\ref{eq2-r}), the order parameter $S$, \textit{i.e.}, the probability that a randomly chosen node belongs to the giant cluster (the infinite cluster), can be expressed as
\begin{align}
S &= \sum_{k=0}p_k\left[ 1-(1-R)^k\right]  \nonumber \\
  &= 1-G_0(1-R),   \label{eq2-s}
\end{align}
where $G_0(x)=\sum_kp_kx^k$ is the generating function of the degree distribution. The right hand side of this equation means that at least one of the neighbors must belong to the giant cluster. From the perspective of branching, this equation can also be interpreted as the start point of the branching process, while Eq.(\ref{eq2-r}) is the intermediate process. In the infinite cluster any nodes can be the start point, so $S$ is an average probability, as well as the parameter $R$.

Furthermore, for the percolation with occupied probability $p$, we only need to replace the generating functions in Eqs.(\ref{eq2-r}) and (\ref{eq2-s}) with those of the diluted network. Next, let us calculate the degree distribution of the diluted network. Regardless of the type of the percolation (site percolation or bond percolation), the diluting ratio of links is $p$, \textit{i.e.}, each link is preserved with probability $p$. So the generating functions $g_0(x)$ and $g_1(x)$ of the diluted networks can be written as
\begin{align}
g_0(x) &= \sum_{k^\prime=0}^\infty p_{k^\prime} x^{k^\prime}  \nonumber \\
&= \sum_{k^\prime=0}^\infty \sum_{k=k^\prime}^\infty p_k \binom{k}{k^\prime} p^{k^\prime} (1-p)^{k-k^\prime} x^{k^\prime} \nonumber \\
&= G_0(1-p+px),   \label{eq2-g0p} \\
g_1(x) &= \sum_{k^\prime=1}^\infty \frac{p_{k^\prime}k^\prime}{p\langle k\rangle} x^{k^\prime-1}  \nonumber \\
&= \sum_{k^\prime=1}^\infty \sum_{k=k^\prime}^\infty p_k \binom{k}{k^\prime} p^{k^\prime} (1-p)^{k-k^\prime} \frac{k^\prime}{p\langle k\rangle} x^{k^\prime-1} \nonumber \\
&= G_1(1-p+px). \label{eq2-g1p}
\end{align}
Then, inserting them into Eqs.(\ref{eq2-r}) and (\ref{eq2-s}), we have \cite{Newman2001,Newman2002}
\begin{align}
R &= 1-g_1(1-R) \nonumber \\
  &=1-G_1(1-pR),  \label{eq2-rp}    \\
S &= 1-g_0(1-R)  \nonumber \\
  &=1-G_0(1-pR).  \label{eq2-sp}
\end{align}
Before continuing to consider the details of these two equations, we must point out that the order parameter $S$ here is the size respected to the diluted network. For bond percolation, the dilution is only for links, so the diluted network has the same size as the original network and $S$ is also the size respecting to the original network. However, for site percolation the diluted network consists of only occupied nodes (fraction $p$ respected to the original network). As a result, Eq.(\ref{eq2-sp}) should be revised for site percolation, that is
\begin{equation}
S = p\left[1-G_0(1-pR)\right]. \label{eq2-sp1}
\end{equation}

\begin{figure}
\centering
\includegraphics[width=0.8\columnwidth]{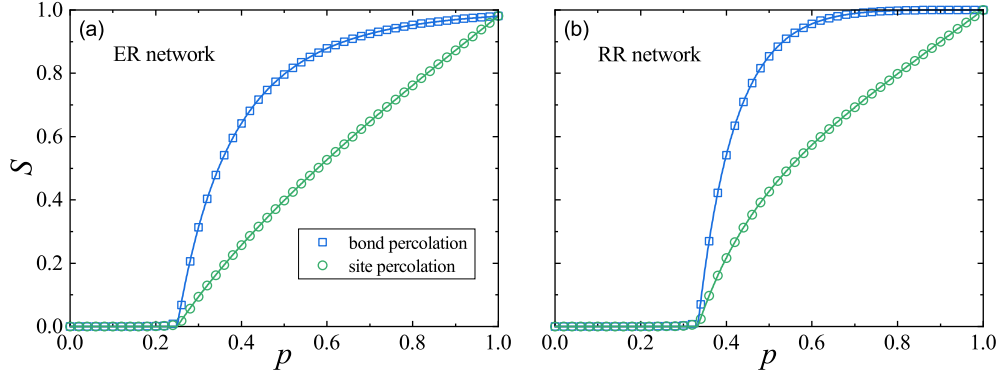}
\caption{Percolation on ER networks and random regular (RR) networks. The scatters are the simulation results on networks with size $N=10^5$, the corresponding lines are the theoretical prediction of Eqs.(\ref{eq2-rp})-(\ref{eq2-sp1}). (a) ER networks with average degree $\langle k\rangle=4$. For this case the generating functions can be simplified as $G_0(x)=G_1(x)=e^{-\langle k\rangle (1-x)}$. (b) RR networks with all the nodes having identical degree $k_0=4$. For this case the generating functions can be simplified as $G_0(x)=x^{k_0}$ and $G_1(x)=x^{k_0-1}$.}  \label{fig2-perrr}
\end{figure}

In general, one can solve Eq.(\ref{eq2-rp}) first, and then insert $R$ into Eq.(\ref{eq2-sp}) or (\ref{eq2-sp1}) to get the order parameter $S$ for the bond percolation and the site percolation, respectively. From Eqs.(\ref{eq2-rp})-(\ref{eq2-sp1}), we can also find that the two types of percolation models give the same probability $R$, but different giant clusters. For mathematical tractability, sometimes one just uses $R$ to represent the meaning of $pR$ for site percolation in Eqs.(\ref{eq2-rp}) and (\ref{eq2-sp1}), that is
\begin{align}
R &= p[1-G_1(1-R)],    \label{eq2-sr} \\
S &= p[1-G_0(1-R)].    \label{eq2-ss}
\end{align}
These two equations are fully equivalent to Eqs.(\ref{eq2-rp}) and (\ref{eq2-sp1}).

Although the mean-field approximation is used in the above discussion, \textit{i.e.}, all the nodes have the same probability $S$ and all the links have the same probability $R$, Eqs.(\ref{eq2-rp})-(\ref{eq2-ss}) can give an exact solution for the percolation on tree-like networks, see Fig.\ref{fig2-perrr} as an example. A comparison between the mean-field prediction and the numerical simulation on some real-world networks can also be found in Ref.\cite{Gleeson2012}.

Following the idea of branching, the microstructure of the giant cluster in random networks \cite{Tishby2018}, such as degree distribution, and degree correlations, as well as that of temporal networks \cite{Starnini2014}, can also be obtained. For networks with loops, different links could lead to the same nodes in the branching process. In other words, the branching processes starting from different links are no longer independent of each other, thus all the equations (\ref{eq2-r})-(\ref{eq2-ss}) are not valid. That is why the above method is only applicable to the tree-like structure.

\subsubsection{Behaviors of finite clusters}

Besides the giant cluster, the percolation theory also concerns the behaviors of finite clusters, such as mean cluster size, and cluster size distribution. Next, along with the idea of the branching process used above, we will further show how to obtain the information of finite clusters in percolation model.

As discussed in Ref.\cite{Newman2001}, it is more convenient to study the distribution $\pi_s=n_ss=p_ss/\langle s\rangle$ rather than $n_s$ or $p_s$, which means the probability distribution of the size of the cluster that a randomly chosen node belongs to. One can further define $\rho_s$ as the size distribution of the cluster at the end of a link. Accordingly, the generating functions have the forms $H_0(x)=\sum_{s=0}\pi_sx^s$ and $H_1(x)=\sum_{s=0}\rho_sx^s$. For convenience, the occupation of links and nodes is not taken into account in the following. When necessary, one can use the revised generating functions Eqs.(\ref{eq2-g0p}) and (\ref{eq2-g1p}) to get the corresponding results.

Note that a zero size has no physical senses, so $\pi_0=0$, while $\rho_0$ can be non-zero corresponding a node with degree one. Since $H_0(1)$ and $H_1(x)$ only contain the finite clusters, and the giant cluster is excluded (if there is one), we have $H_0(1)=\sum_{s}\pi_s=1-S$ and $H_1(1)=\sum_{s}\rho_s=1-R$, where $S$ and $R$ are defined as that for Eqs.(\ref{eq2-r}) and (\ref{eq2-s}). Below the critical point, $R=0$ and $S=0$, so $H_0(1)=1$ and $H_1(1)=1$.

\begin{figure}
\centering
\includegraphics[width=0.7\columnwidth]{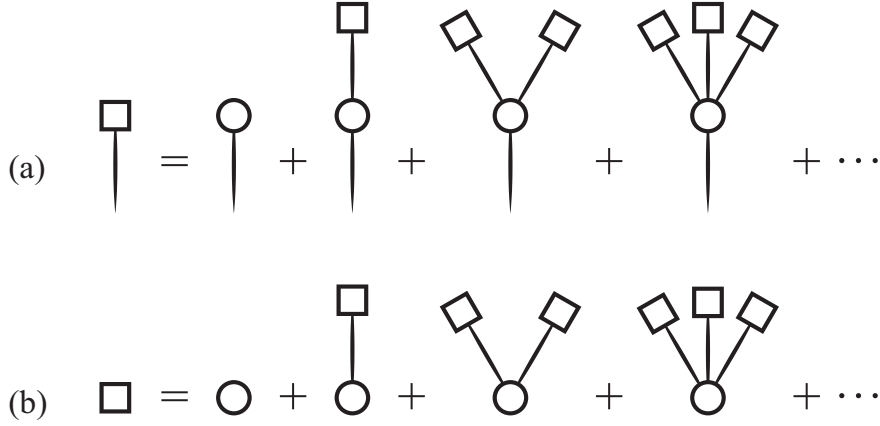}
\caption{Schematic diagrams of the self-consistent equations for the cluster size distribution. The line, circle and square represent link, node and cluster, respectively. (a) The possible cases of the branchings at the end of a link, corresponding to Eq.(\ref{eq2-h1}). (b) The possible cases of the branchings starting from a node, corresponding to Eq.(\ref{eq2-h0}). Such schematic diagrams are originally shown in Ref.\cite{Newman2001}.}  \label{fig2-csd}
\end{figure}

As the schematic in Fig.\ref{fig2-csd}, the cluster size distributions $\pi_s$ and $\rho_s$ are dependent on both the degree/excess-degree distribution and the branching size at the end of a link. Thus, Fig.\ref{fig2-csd} can be translated as the equations
\begin{align}
H_0(x) &=x\sum_{k=0}p_k[H_1(x)]^k  \nonumber   \\
       &=xG_0[H_1(x)],    \label{eq2-h0} \\
H_1(x) &=x\sum_{k=1}q_k[H_1(x)]^{k-1} \nonumber \\
       &=xG_1[H_1(x)].    \label{eq2-h1}
\end{align}
Note that there is one more factor $x$ on the right hand sides of the two equations for the contribution of the root node. Using the relations $H_0(1)=1-S$ and $H_1(1)=1-R$, these two equations just recover Eqs.(\ref{eq2-r}) and (\ref{eq2-s}) when $x=1$.

In principle, we can solve Eqs.(\ref{eq2-h0}) and (\ref{eq2-h1}) together to find $H_0(x)$, and then use
\begin{equation}
\pi_s=\left.\frac{d^{s-1}}{dx^{s-1}}\frac{H_0(x)}{x}\right|_{x=0}            \label{eq2-pis}
\end{equation}
to find the cluster size distribution. However, usually no closed form can be found for $H_0(x)$. Instead, we can substitute Eqs.(\ref{eq2-h0}) and (\ref{eq2-h1}) into Eq.(\ref{eq2-pis}), then do a change of variables via Cauchy formula, and finally get
\begin{equation}
\pi_s=\left.\frac{\langle k\rangle}{(s-1)!}\frac{d^{s-2}}{dx^{s-2}}[G_1(x)]^s\right|_{x=0}.
\end{equation}
With this, the cluster size can be obtained directly from $G_1(x)$, therefore avoiding solve Eqs.(\ref{eq2-h0}) and (\ref{eq2-h1}) \cite{Newman2007,Kryven2017,Kryven2017a}. As $\pi_s\propto sn_s\propto sp_s$, the result must have the form $\pi_s\propto s^{1-\tau}$, where $\tau$ is the Fisher exponent.

Furthermore, the mean cluster size $\chi$ can also be obtained from Eqs.(\ref{eq2-h0}) and (\ref{eq2-h1}). From the definition, the mean cluster size $\chi$ can be represented as
\begin{align}
\chi &=\frac{\sum_s s^2n_s}{\sum_s sn_s} \nonumber \\
     &=\frac{\sum_s s\pi_s}{\sum_s \pi_s} \nonumber \\
     &=\frac{H_0'(1)}{H_0(1)}  \nonumber \\
     &=\frac{1+G_0'(1)H_1'(1)}{H_0(1)}.
\end{align}
In the last equation the differential of Eq.(\ref{eq2-h0}) is used. We can further employ Eq.(\ref{eq2-h1}) to replace the term $H_1'(1)$, and yield
\begin{equation}
\chi = \frac{1}{H_0(1)} \left[ 1+\frac{G_0'(1)}{1-G_1'(1)}\right].  \label{eq2-chi}
\end{equation}
It is obvious that $\chi$ is divergent when
\begin{equation}
G_1'(1)=1. \label{eq2-mrc}
\end{equation}
This is the Molloy--Reed criterion $\langle k(k-2)\rangle=0$ found in random graph theory \cite{Molloy1995}, above which ($G_1'(1)>1$) the giant cluster exists.

In addition, the discussion of finite clusters based on generating functions $H_0(x)$ and $H_1(x)$ can also work without the help of generating functions $G_0(x)$ and $G_1(x)$, so it is not limited to uncorrelated tree-like networks, see Sec.\ref{sct3-ptng} for an example of the percolation on growing networks.

\subsection{Potts model formulation}  \label{sct2-pmf}

The Potts model is a generalization of the Ising model with more than two components \cite{Potts1952}, and related to a number of other outstanding problems in lattice systems \cite{Wu1982}, including percolation model. Although the Potts model is also unsolved for an arbitrary network, the connection between the Potts model and the percolation model has made it possible to explore the network percolation from the known information on the Potts model. This section will provide the mapping between Potts model and percolation model, as well as some discussions.

\subsubsection{Fortuin-Kasteleyn cluster representation}

It is known that the $q\to1$ limit of the $q$-state Potts model without external field corresponds to the percolation problem, which is thus applied to the study of percolation on networks \cite{Stephen1976,Wu1978,Essam1980,Wu1982}. To consider the percolation on a network $\mathcal{G}$, we introduce a $q$-state Potts model with the Hamiltonian
\begin{equation}
\mathcal{H}=-J\sum_{\langle i,j\rangle}\delta_{\alpha_i,\alpha_j} -H\sum_{i}\delta_{\alpha_i,\alpha},   \label{eq2-hm}
\end{equation}
where $\alpha_i=1,2,3,\ldots,q$ is the spin state of node $i$, and the sums $\sum_{\langle i,j\rangle}$ and $\sum_{i}$ are over all the links and nodes of the network, respectively. $\delta_{x,y}$ is the Kronecker delta function, \textit{i.e.}, $\delta_{x,y}=0,1$ if $x\neq y$ and $x=y$, respectively. Here, the ferromagnetic interaction is used $J>0$, and the magnetic field $H>0$ is applied to the spin $\alpha_H$. Then, the partition function can be written as
\begin{align}
Z &= \sum_{\{\alpha\}} e^{-\beta\mathcal{H}}  \nonumber \\
  &= \sum_{\{\alpha\}} e^{K\sum_{\langle i,j\rangle}\delta_{\alpha_i,\alpha_j}} e^{L\sum_{i}\delta_{\alpha_i,\alpha_H}}  \nonumber \\
  &= \sum_{\{\alpha\}} \prod_{\langle i,j\rangle}e^{K\delta_{\alpha_i,\alpha_j}} \prod_{i}e^{L\delta_{\alpha_i,\alpha_H}}    \nonumber \\
  &= \sum_{\{\alpha\}} \prod_{\langle i,j\rangle}\left[1+(e^K-1)\delta_{\alpha_i,\alpha_j}\right]  \prod_{i}\left[1+(e^L-1)\delta_{\alpha_i,\alpha_H}\right], \label{eq2-z}
\end{align}
where $K=\beta J$, $L=\beta H$, and the sum $\sum_{\{\alpha\}}$ is over all the spin configurations of the system. We can further expand the products and use the subnetworks of $\mathcal{G}$ to represent the terms in the expansion, which is often called Fortuin--Kasteleyn cluster representation \cite{Kasteleyn1969}.

Next, we give a brief derivation for the Fortuin--Kasteleyn cluster representation. For arguments $a_i$, $i\in I$, the product $\prod_{i\in I} (1+a_i)$ is equivalent to the sum of all the possible polynomials formed by $a_i$, $i\in I$, so that $\prod_{i\in I} (1+a_i)=\sum_{I'\subseteq I}\prod_{i\in I'}a_i$, where $I'$ is a subset of $I$, and the sum $\sum_{I'}$ is over all the possible cases. By this formula, the product $\prod_{\langle i,j\rangle}$ in Eq.(\ref{eq2-z}) can be represented by the sum of the subnetworks $\mathcal{G}'$ of $\mathcal{G}$, that is
\begin{align}
Z &= \sum_{\{\alpha\}} \sum_{\mathcal{G}'\subseteq\mathcal{G}} \prod_{ \langle i,j\rangle \in \mathcal{G}'}(e^K-1)\delta_{\alpha_i,\alpha_j} \prod_i\left[1+(e^L-1)\delta_{\alpha_i,\alpha_H}\right] \nonumber \\
  &= \sum_{\mathcal{G}'\subseteq\mathcal{G}} (e^K-1)^{l(\mathcal{G}')} \sum_{\{\alpha\}}  \prod_{ \langle i,j\rangle \in \mathcal{G}'}\delta_{\alpha_i,\alpha_j} \prod_{ i}\left[1+(e^L-1)\delta_{\alpha_i,\alpha_H}\right],
\end{align}
where $l(\mathcal{G}')$ is the number of links in subnetwork $\mathcal{G}'$. Note that the term $\prod_{ (i,j) \in \mathcal{G}'}\delta_{\alpha_i,\alpha_j}$ gives nonzero value only when the nodes in the same clusters are in the same states (any of the $q$ states $\alpha=1,2,\ldots,q$). This can further simplify the partition function to be
\begin{align}
Z &= \sum_{\mathcal{G}'\subseteq\mathcal{G}} (e^K-1)^{l(\mathcal{G}')} \sum_{\{\alpha_c\}}  \prod_{c} \left[1+(e^L-1)\delta_{\alpha_c,\alpha_H}\right]^{s_c}  \nonumber \\
  &= \sum_{\mathcal{G}'\subseteq\mathcal{G}} (e^K-1)^{l(\mathcal{G}')} \prod_{c} \left(e^{Ls_c}+q-1\right).    \label{eq2-fkz}
\end{align}
Here, the product $\prod_{c}$ is over all the clusters formed by the links in subnetwork $\mathcal{G}'$, and $s_c$ is the number of nodes in cluster $c$. The sum over spin $\{\alpha\}$ in Eq.(\ref{eq2-z}) has now been replaced by a sum over subnetwork $\{\mathcal{G}'\}$, this is just the Fortuin--Kasteleyn cluster representation of the partition function of Potts model.

Let $e^K-1=p/(1-p)$, the partition function Eq.(\ref{eq2-fkz}) can be rescaled as
\begin{align}
\mathcal{Z} &\equiv (1-p)^{l(\mathcal{G})}Z        \nonumber \\
            &= \sum_{\mathcal{G}'\subseteq\mathcal{G}} p^{l(\mathcal{G}')} (1-p)^{l(\mathcal{G})-l(\mathcal{G}')} \prod_{c} \left(e^{Ls_c}+q-1\right)          \nonumber \\
            &= \sum_{\mathcal{G}'\subseteq\mathcal{G}} P_{\mathcal{G}'} \prod_{c} \left(e^{Ls_c}+q-1\right),           \label{eq2-zrs}
\end{align}
where $l(\mathcal{G})$ is the total number of links in $\mathcal{G}$. If $p$ is interpreted as the occupied probability of links, the term $P_{\mathcal{G}'}=p^{l(\mathcal{G}')}(1-p)^{l(\mathcal{G})-l(\mathcal{G}')}$ is just the probability of a percolation configuration with $l(\mathcal{G}')$ occupied links. Thus, Eq.(\ref{eq2-zrs}) bridges the Potts model and the percolation model.

From Eq.(\ref{eq2-zrs}), we can write the free energy per node as
\begin{equation}
f(K,L,q)=\lim_{N\to\infty}\frac{\ln \mathcal{Z}}{N}.
\end{equation}
For convenience, we further define
\begin{align}
h(K,L) &\equiv \left.\frac{\partial}{\partial q} f(K,L,q)\right|_{q=1}  \nonumber \\
       &= \frac{\sum_{\mathcal{G}'\subseteq\mathcal{G}} P_{\mathcal{G}'}\sum_ce^{-Ls_c}}{N\sum_{\mathcal{G}'\subseteq\mathcal{G}} P_{\mathcal{G}'}}    \nonumber \\
       &= \frac{\left\langle \sum_ce^{-Ls_c}\right\rangle}{N}.
\end{align}
For $L>0$ the sum in this equation is over all the finite clusters, then the size of the percolating cluster is
\begin{align}
S &= 1- \frac{\left\langle \sum_c s_c \right\rangle}{N}    \nonumber \\
  &= 1+ \left.\frac{\partial}{\partial L} h(K,L)\right|_{L=0},     \label{eq2-slnz}
\end{align}
and the mean cluster size is
\begin{equation}
\chi = \left.\frac{\partial^2}{\partial L^2} h(K,L)\right|_{L=0}.
\end{equation}
By the above two equations, the percolation property can then be extracted from the Potts model.

The product $\prod_{c}$ in Eq.(\ref{eq2-zrs}) is over all the clusters. It thus can be rewritten in another form of
\begin{equation}
\mathcal{Z} = \sum_{\mathcal{G}'\subseteq\mathcal{G}} P_{\mathcal{G}'} \prod_{s} \left(e^{Ls}+q-1\right)^{m_s},    \label{eq2-zns}
\end{equation}
where the product $\prod_{s}$ is over the size of the clusters, and $m_s$ is the number of clusters with size $s$. By this formula, function $h(K,L)$ can also be expressed as
\begin{equation}
h(K,L)=\frac{\left\langle \sum_s m_s e^{-Ls}\right\rangle}{N}=\sum_s n_s e^{-Ls}.
\end{equation}
Here, $n_s=\langle m_s/N \rangle$ is the cluster size distribution. This is just the generating function $h(K,x)=\sum_s n_s x^s$ with $x=e^{-L}$, which can be used to study the cluster size distribution.

Specifically, once the Hamiltonian (Eq.(\ref{eq2-hm})) of a network is known, all the percolation parameters can be obtained by the above equations. However, for most network models, it is difficult to simplify the sum over links in Eq.(\ref{eq2-hm}). A solvable case is the hidden parameter model, for which the pre-defined connection probability is given for each pair of nodes. Thus, the sum over links can be translated as a weighed (connection probability) sum over all the node pairs. With this technique, the Potts model formulation has been used in solving the percolation on SF networks \cite{Lee2004,Lee2005}, on correlated hypergraphs \cite{Bradde2009}, and in generalized canonical random network ensembles \cite{Bradde2009a}.

\subsubsection{Relation with the mean-field equations}

Through the so-called recurrence relation, the mean-field equations for tree-like networks can be recovered by the Potts model formulation \cite{Dorogovtsev2004}. For this purpose, the partition function can be represented as an integration of a root, labeled node $0$, and the branchings from it, that is
\begin{align}
Z &= \sum_{\alpha_0=1}^q e^{L\delta_{\alpha_0,\alpha_H}} \prod_{i=1} z_i(\alpha_0)  \nonumber \\
&= e^L\prod_{i=1} z_i(\alpha_H) +(q-1) \prod_{i=1} z_i(\alpha).   \label{eq2-ztn}
\end{align}
Here, $z_i(\alpha_0)$ is the partition function of the branching from node $i$, and the product $\prod_{i}$ runs over all the root's nearest neighbors (nodes in the first shell of node $0$). Due to the symmetry, $\alpha$ in the second term of Eq.(\ref{eq2-ztn}) can be any state excluding $a_H$. Obviously, this formulation is only for the tree-like networks, if not, the product $\prod_{i}$ must include some double counting.

Different from Eq.(\ref{eq2-z}), the partition function equation (\ref{eq2-ztn}) represents the effective state of the root node. In this way, the free energy per nodes is $\ln Z$, then we have
\begin{align}
h(K,L) &= \left.\frac{\partial}{\partial q} \ln Z \right|_{q=1}  \nonumber \\
 &= e^{-L} \frac{\prod_{i=1} z_i(\alpha)}{\prod_{i=1} z_i(\alpha_H)}.
\end{align}
To study the percolation model, we assume here that $z_i(\alpha)$ describes the state for $q=1$ and $L=0$, thus it is irrelevant to the differentials $\partial/\partial q$ and $\partial/\partial L$. For convenience, we further rescale $z_i(\alpha)$ by $z_i(\alpha_H)$, \textit{i.e.}, $x_i\equiv z_i(\alpha)/z_i(\alpha_H)$. Then, by Eq.(\ref{eq2-slnz}), we can find the order parameter of the percolation transition
\begin{align}
S &= 1 +\left.\frac{\partial}{\partial L} h(K,L) \right|_{L=0}  \nonumber \\
&= 1 -\prod_{i=1} x_i. \label{eq2-sxi}
\end{align}
From this equation, the physical meaning of $x_i$ becomes clear, that is the probability that the branching from node $i$ is not the giant cluster.

To obtain the order parameter, we need to further find $x_i(\alpha)$. Comparing Eqs.(\ref{eq2-ztn}) and (\ref{eq2-z}), we have
\begin{equation}
z_i(\alpha_0)=\sum_{\{\alpha\}} e^{K\sum_{\langle j,k\rangle} \delta_{\alpha_j,\alpha_k} +L\sum_{j}\delta_{\alpha_H,\alpha_j} +K\delta_{\alpha_0,\alpha_i}},
\end{equation}
where the sums $\sum_{j}$ and $\sum_{\langle j,k\rangle}$ run over the nodes and the links in the branching from node $i$, respectively. Note that the interaction between node $i$ and the root $K\delta_{\alpha_0,\alpha_i}$ is also included in this partition function. Furthermore, it is not hard to find that $z_i(\alpha_0)$ satisfies the following recurrence relation
\begin{equation}
z_i(\alpha_0) = \sum_{\alpha_i=1}^q e^{K\delta_{\alpha_0,\alpha_i}+L\delta_{\alpha_H,\alpha_i}} \prod_{m} z_m(\alpha_i).   \label{eq2-zn0}
\end{equation}
Here, the product $\prod_{m}$ is over all the neighbors of node $i$ except the root (nodes in the second shell of node $0$), and $z_m(\alpha_i)$ is the partition function for the subnetwork branching from node $m$.

For infinite system, the recursive relation Eq.(\ref{eq2-zn0}) holds for any two adjacent shells. If $\alpha_0\neq\alpha_H$, it can be rewritten as
\begin{equation}
z_i(\alpha_0) = e^{K}\prod_{m} z_m(\alpha_0) +e^{L}\prod_{m} z_m(\alpha_H)  +(q-2)\prod_{m} z_m(\alpha).
\end{equation}
While for $\alpha_0=\alpha_H$, Eq.(\ref{eq2-zn0}) reduces to
\begin{equation}
z_i(\alpha_H) = e^{K+L}\prod_{m} z_m(\alpha_H) +(q-1)\prod_{m} z_m(\alpha).
\end{equation}
Then, we can find the recursive equation for $x_i$,
\begin{align}
x_i(\alpha_0) &= \frac{z_i(\alpha_0)}{z_i(\alpha_H)}   \nonumber \\
&= \frac{e^{K}\prod_{m} z_m(\alpha_0) +e^{L}\prod_{m} z_m(\alpha_H)  +(q-2)\prod_{m} z_m(\alpha)}  {e^{K+L}\prod_{m} z_m(\alpha_H) +(q-1)\prod_{m} z_m(\alpha)}  \nonumber \\
&= \frac{e^{K}\prod_{m} x_m(\alpha_0) +e^{L} +(q-2)\prod_{m} x_m(\alpha)} {e^{K+L} +(q-1)\prod_{m} x_m(\alpha)}.
\end{align}
For $q=1$ and $L=0$, it reduces to
\begin{align}
x_i &= e^{-K} +(1-e^{-K}) \prod_{m} x_m  \nonumber \\
&= 1-p +p\prod_{m} x_m,  \label{eq2-xixm}
\end{align}
where $p=1-e^{-K}$.

From the view of mean-field theory, the probabilities $x_i$ for different nodes can all be replaced by an average probability $x$. Then, Eqs.(\ref{eq2-sxi}) and (\ref{eq2-xixm}) become
\begin{align}
S &=  1 -\sum_kp_k x^k =1-G_0(x),    \\
x &=  1-p +p\sum_k \frac{p_kk}{\langle k\rangle} x^{k-1}=1-p+pG_1(x).
\end{align}
These are just the mean-field equations (\ref{eq2-rp}) and (\ref{eq2-sp}) found in Sec.\ref{sct2-ampb} with $x=1-pR$. By this equation, we bridge the Potts model formulation of the percolation model with that of the mean-field method based on the branching process \cite{Dorogovtsev2004}. Note that the tree-like structure is also required here.

\subsection{Message passing method}   \label{sct-mpm}

In recent years, there has been a growing interest in the analysis of the percolation problem on real networks. These systems all have finite sizes and are featured by the adjacency matrix rather than the degree distribution. As a consequence, the above mean-field method does not work in this case. Instead, the most common method to estimate the percolation threshold on these networks is the so-called message passing method \cite{Karrer2014,Newman2014,Hamilton2014,Radicchi2015a,Radicchi2015b,Allard2015,Radicchi2016,Kuehn2016,Timar2017,Bianconi2016,Bianconi2017}. Note that the percolation threshold (transition) is theoretically defined in the thermodynamic limit ($N\to\infty$), here we assume that the system is large enough to observe a percolation transition. Next, we will go to the details of this method.

Different from the above mean-field method, the message passing method allows each node $i$ has its own probability $s_i$ to express whether it is a part of the giant cluster. From this perspective, the size of the giant cluster can be written as
\begin{equation}
S=\frac{\sum_is_i}{N}.
\end{equation}
Similar to Eqs.(\ref{eq2-sr}) and (\ref{eq2-ss}), we can also express $s_i$ as
\begin{align}
s_i        &= p\left[1-\prod_{j\in \mathcal{N}_i}(1-r_{i\to j})\right],   \\
r_{i\to j} &= p\left[1-\prod_{k\in \mathcal{N}_j, k\neq i}(1-r_{j\to k})\right],
\end{align}
where $r_{i\to j}$ is the probability that the link $i\to j$ leads to the giant cluster, and $\mathcal{N}_i$ is the set of neighbors of node $i$. Furthermore, by evaluating the logarithm of the above two equations, we can turn the products into sums,
\begin{align}
\ln\left(1-\frac{s_i}{p}\right) &= \sum_{j\in \mathcal{N}_i}\ln(1-r_{i\to j}),      \label{eq2-mps}   \\
\ln \left(1-\frac{r_{i\to j}}{p}\right) &= \sum_{k\in \mathcal{N}_j,k\neq i}\ln \left(1-r_{j\to k}\right)  \nonumber \\
                                        &= \sum_{k} A_{kj} \ln \left(1-r_{j\to k}\right) - A_{ji} \ln \left(1-r_{j\to i}\right).     \label{eq2-mpr}
\end{align}
Here $\mathbf{A}$ is the adjacency matrix of the network, \textit{i.e.}, $A_{ij}=1$ if there is a link $i\to j$, otherwise $A_{ij}=0$.

Let $w_{i\to j}=\ln(1-r_{i\to j}/p)$ and $v_{i\to j}=\ln(1-r_{i\to j})$, Eq.(\ref{eq2-mpr}) becomes equivalent to the vectorial equation
\begin{equation}
\mathbf{w}=\mathbf{M}\mathbf{v},  \label{eq2-wmv}
\end{equation}
where $\mathbf{M}$ is a $2L\times2L$ matrix ($L$ is the number of links). From Eq.(\ref{eq2-mpr}), we can find that only when $j=k$ and $i\neq l$, the entry $M_{i\to j,k\to l}$ is non-zero. In other words, the two links must be head to tail, and excluding the case of backtracking. This matrix is known as the Hashimoto or non-backtracking matrix of graphs \cite{Hashimoto1989,Krzakala2013}. Mathematically, $M_{i\to j,k\to l}=\delta_{j,k}(1-\delta_{i,l})$, where $\delta_{x,y}$ is the Kronecker delta function $\delta_{x,y}=1$ if $x=y$, and $\delta_{x,y}=0$, otherwise.

When $p\to p_c$, all $r_i$ trend to zero. Then, expanding Eq.(\ref{eq2-mpr}) around the critical point, we obtain an eigenvalue equation of matrix $\mathbf{M}$,
\begin{equation}
\frac{1}{p_c}\mathbf{r} = \mathbf{M}\mathbf{r}.  \label{eq2-mpc}
\end{equation}
According to the Perron-Frobenius theorem, only the largest eigenvalue $\lambda_{max}$ of matrix $\mathbf{M}$ can give a meaningful eigenvector $\mathbf{r}$ (with all elements non-negative). Therefore, it gives a lower bound for the site percolation threshold on an infinite graph $p_c=1/\lambda_{max}$ \cite{Karrer2014}. Considering the effects of triangles, this framework can also be used to establish a tighter lower bound of the bond percolation threshold on clustered networks \cite{Zhang2017}.

The above discussion is for site percolation, it can be easily extended to bond percolation, that is
\begin{align}
s_i &= 1-\prod_{j\in \mathcal{N}_i}(1-pr_{i\to j}),     \\
r_{i\to j} &= 1-\prod_{k\in \mathcal{N}_j,k\neq i}(1-pr_{j\to k}).
\end{align}
Further discussion of the two equations can be done like that of the site percolation.

In addition, as was done by the mean-field method, the message passing method can also be adopted to study the cluster size distribution. One can refer to Refs.\cite{Karrer2014,Newman2014,Hamilton2014,Radicchi2015a,Radicchi2015b,Allard2015,Radicchi2016,Kuehn2016,Timar2017,Bianconi2016,Bianconi2017} for details.

\subsection{Phase transition and critical phenomena}

Due to the heterogeneous structure, the classical percolation on networks demonstrates many interesting phenomena. In this subsection we will review the findings on tree-like network ensembles, for which the percolation problem can be solved exactly by the methods reviewed in the above subsections.

\subsubsection{Percolation threshold}  \label{sct2-pt}

\begin{figure}
\centering
\includegraphics[width=0.6\columnwidth]{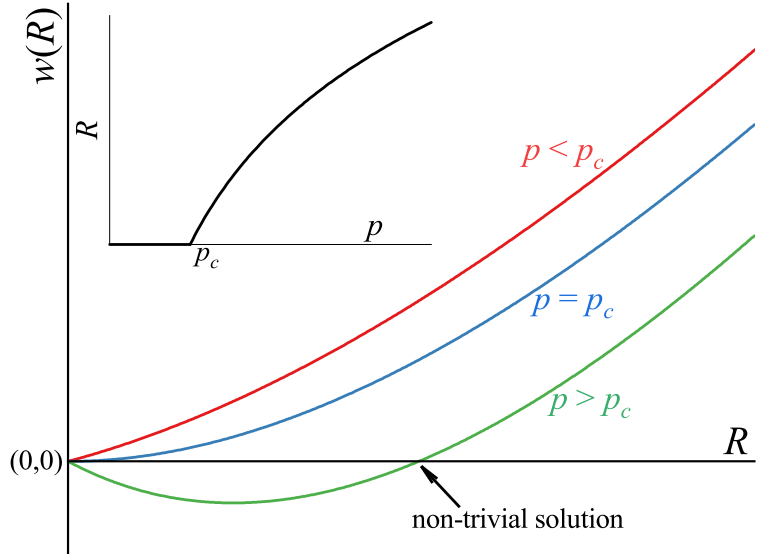}
\caption{(Color online) Schematic of the solution of Eq.(\ref{eq2-rp}). Only when $p$ exceeds $p_c$, function $w(R)$ has a non-trivial crossing point with $R$-axis. The critical point corresponds to the tangency of the function $w(R)$ and $R$-axis, which gives a vanished $R$. The inset shows the non-trivial solution $R$ vs. $p$.}  \label{fig2-sltc}
\end{figure}

To discuss the critical phenomena, we first need to determine the critical point. For tree-like networks, it is easy to know from Eqs.(\ref{eq2-rp})-(\ref{eq2-ss}) that only a non-zero $R$ can lead to a non-zero $S$. Thus the percolation threshold $p_c$ can be found by analyzing the non-trivial solution of Eq.(\ref{eq2-rp}), and the site and bond percolations have the same critical point. For this purpose, we can construct a function
\begin{equation}
w(R)=R-1+G_1(1-pR).
\end{equation}
The solution of Eq.(\ref{eq2-rp}) corresponds to the crossing point of $w(R)$ and $R$-axis. Since $w(R)$ is continuous with $w(0)=0$ and $w(1)>0$, we can draw a qualitative curve of $w(R)$ as shown in Fig.\ref{fig2-sltc}. It is easy to find that the critical point $p_c$ corresponds to the tangency of $w(R)$ and $R$-axis with $R_c=0$, so we have
\begin{equation}
\frac{dw(R)}{dR}=1-p_cG_1'(1)=0.
\end{equation}
Thus, the percolation threshold can be written as
\begin{equation}
p_c=\frac{1}{G_1'(1)}.       \label{eq2-pc}
\end{equation}
This is a general form for any tree-like networks. Using the expression of the generating function $G_1(x)$, it can be rewritten as
\begin{equation}
p_c=\frac{\langle k\rangle}{\langle k^2\rangle-\langle k\rangle}.  \label{eq2-pc1}
\end{equation}
For $p=1$, this equation reduces to Eq.(\ref{eq2-mrc}), \textit{i.e.}, the Molloy--Reed criterion $\langle k(k-2)\rangle=0$ \cite{Molloy1995}. Note that this is an approximate expression for tree-like infinite networks. A general case for finite systems is the one found by message passing method, see the last subsection.

For ER networks, node degrees obey the Poisson distribution $p_k=e^{-\langle k\rangle}\langle k\rangle^k/k!$, so that
\begin{align}
G_0(x) &=G_1(x)= e^{\langle k\rangle (x-1)},   \\
\langle k^2\rangle &= \langle k\rangle^2+\langle k\rangle.
\end{align}
This allows the percolation threshold has a simple form $p_c=1/\langle k\rangle$. Besides, due to the Poisson degree distribution, ER networks can have a closed solution of percolation threshold for many network percolation models. For comparison, we list the percolation thresholds of ER networks for different models in Tab.\ref{tb2-cpp}.

\begin{table}[ht]
\centering \footnotesize
\caption{The percolation thresholds of ER networks for some solvable models. The order parameter of the first part of the table is the occupied probability $p$, \textit{i.e.}, removing each node with probability $1-p$. In the second part of the table, the order parameter is the link inserting probability $p$.} \label{tb2-cpp}
\begin{tabular}{ll}
\toprule
Model & Threshold \\
\hline
Classical percolation & $1/\langle k\rangle$ \cite{Newman2001}  \\
Biconnected cluster & $1/\langle k\rangle$ \cite{Newman2008}  \\
Core percolation & $2.7182818/\langle k\rangle$ \cite{Liu2012a} \\
Greedy articulation point removal  & $3.39807/\langle k\rangle$ \cite{Tian2017} \\
Two interdependent networks  & $2.4554/\langle k\rangle$ \cite{Buldyrev2010,Parshani2010} \\
Single networks with dependence links  & $\sqrt{2.4554/\langle k\rangle}$ \cite{Parshani2011} \\
\hline
Classical percolation & $1/N$ \cite{Bollobas2001} \\
Clique percolation & $\{(k−l)!/[\binom{k}{l}−1]\}^{2/(k−l)(k+l−1)} N^{−2/(k+l−1)}$  \cite{Bollobas2009,Li2015} \\
Growing network & $1/8$  \cite{Callaway2001} \\
\bottomrule
\end{tabular}
\end{table}

A more interesting case is the SF network with $2<\lambda<3$, for which $\langle k^2\rangle$ is divergent, indicating a vanished percolation threshold. Employing the Hurwitz zeta function $\zeta(x,a)=\sum_{n=a}n^{-x}$, the percolation threshold of SF networks can be represented as
\begin{align}
p_c&=\frac{\sum_{k=m}^\infty k^{1-\lambda}}{\sum_{k=m}^\infty k^{2-\lambda} -\sum_{k=m}^\infty k^{1-\lambda} }  \nonumber \\
&=\frac{\zeta(\lambda-1,m)}{\zeta(\lambda-2,m)-\zeta(\lambda-1,m)}.
\end{align}
One can find that for $m=1$, this formula gives a percolation threshold $p_c$ larger than $1$ when $\lambda>3.47875\ldots$, indicating there is no percolation transition. This is due to the absence of the spanning cluster in such SF networks \cite{Aiello2001}. The typical value $\lambda\approx3.47875$ can be obtained by $2\zeta(\lambda-1)=\zeta(\lambda-2)$, where $\zeta(x)$ is the Riemann zeta function. This problem can be overcome by simply setting $m\geq2$. Furthermore, the Hurwitz zeta function $\zeta(x,a)$ is divergent for $x\leq1$. Thus a vanished percolation threshold can be found for SF networks with $\lambda\leq3$.

The zero percolation threshold is not unique to SF networks, and a general class of self-similar and hierarchical networks can also have zero percolation threshold \cite{Dorogovtsev2002,Serrano2011,Auto2008}. This property can be derived from the hierarchy of nested subgraphs in self-similar networks and real-space renormalization group technique, and does not require the assumption that networks are tree-like. In addition, by embedding SF networks into physical dimensions, a non-zero percolation threshold can also be reconstructed \cite{Warren2002}. This is mainly because the spatial constraint induces a high clustering and dilutes the global connections.

\subsubsection{Scaling behaviors}  \label{sct2-sb}

Varied percolation thresholds in different networks are excepted, since it is not universal, whereas things get interesting in that the heterogeneous structures could produce a different universal property of percolation transition, \textit{i.e.}, the critical exponent. Indeed, we can expand the analytical results (see Sec.\ref{sct2-ampb}) around the critical point to find the critical exponents. It is also worth pointing out that the critical exponents derived from the results obtained in Sec.\ref{sct2-ampb} are still the sense of mean-field solutions, although they might be different from the regular mean-field nature.

For a few cases that have a closed form of the generating functions $G_0(x)$ and $G_1(x)$, such as ER networks and RR networks, they give the critical exponents above the critical dimension (the mean-field critical exponents), since these systems have no spatial constraint. Mathematically, for these networks, the degree distribution only affects the coefficients of the expansion series around the percolation threshold, then yields the same leading order. For the calculation of these cases, one can refer to Refs.\cite{Newman2001,Newman2002,Newman2007,Kryven2017} for details.

A special case is SF networks, which have strong heterogeneity. Although there is also no spatial constraint, it shows a $\lambda$-dependent critical behavior ($\lambda$ is the exponent of the degree distribution $p_k=ck^{-\lambda}$). Even for a vanished percolation threshold ($2<\lambda<3$), the critical behavior is also well defined. The critical exponents can be also derived from the results obtained in Sec.\ref{sct2-ampb}, but bear a strong heterogeneity of the degree distribution.

Although the generating functions $G_0(x)$ and $G_1(x)$ have no closed form for SF networks, they can be represented by the Lerch's transcendent $\Phi(x,s,a)=\sum_{n=0}^\infty x^n(n+a)^{-s}$,
\begin{align}
G_0(x) &= \sum_{k=m} ck^{-\lambda}x^k   \nonumber \\
&= \frac{x^m\Phi(x,\lambda,m)}{\zeta(\lambda,m)}  \nonumber \\
&=cx^m\Phi(x,\lambda,m),  \\
G_1(x) &= \frac{\sum_{k=m}k^{1-\lambda}x^{k-1}}{\sum_{k=m}k^{1-\lambda}}   \nonumber \\
&= \frac{x^{m-1}\Phi(x,\lambda-1,m)}{\zeta(\lambda-1,m)}  \nonumber \\
&= \frac{1}{\zeta(\lambda-1,m)}\frac{\partial}{\partial x}\left[x^m\Phi(x,\lambda,m)\right]    \nonumber \\
&=\frac{c}{\langle k\rangle}\frac{\partial}{\partial x}\left[x^m\Phi(x,\lambda,m)\right],
\end{align}
where $\langle k^n\rangle=c\sum_{k=m}k^{n-\lambda}=c\zeta(\lambda-n,m)$. To find the critical behaviors of the percolation on SF networks, we need further to know the series expansion of $G_0(x)$ and $G_1(x)$ at the critical point $p_c$, which corresponds to $x\to1$, see Sec.\ref{sct2-pt}. However, $\Phi(x,s,a)$ has a singularity at $x=1$, which is dependent on $s$. This is the mathematical origin of the specific critical behaviors of SF networks.

From Ref.\cite{Bateman1953}, for $|\ln x|<2\pi$ and $a\neq 0,-1,-2,\cdots$, the Lerch's transcendent $\Phi(x,s,a)$ can be expanded as
\begin{equation}
\Phi(x,s,a) = x^{-a}\left[\Gamma(1-s)(-\ln z)^{s-1} +\sum_{n=0}\frac{\zeta(s-n,a)}{n!}(\ln x)^n\right], ~s\neq1,2,3,\cdots.
\end{equation}
Let $x\equiv1-\epsilon\to1^-$, we also have
\begin{equation}
-\ln(1-\epsilon) = \epsilon+\frac{1}{2}\epsilon^2+\frac{1}{3}\epsilon^3+\cdots.
\end{equation}
Then, we can further write $\Phi(x,s,a)$ as a series of $\epsilon$,
\begin{equation}
\Phi(x,s,a) = x^{-a}\left[\Gamma(1-s)\left(\sum_{n=1}\frac{\epsilon^n}{n!}\right)^{s-1} +\sum_{n=0} \frac{(-\epsilon)^n}{n!} \sum_{l=0}^nS(n,l)\zeta(s-l,a) \right], ~s\neq1,2,3,\cdots,  \label{eq2-phi}
\end{equation}
where $S(n,l)$ is the signed Stirling numbers of the first kind. With this equation, we can find the leading terms of the generating functions $G_0(x)$ and $G_1(x)$ for $x\to1$, \textit{i.e.}, the asymptotic series to the percolation threshold. Then, substituting into the corresponding equations in Sec.\ref{sct2-ampb}, the $\lambda$-dependent critical exponents can be obtained, which are summarized in Tab.\ref{tb2-ce}. We can find that the case of $3<\lambda<4$ is just a mirror copy of the case of $2<\lambda<3$.

\begin{table}[ht]
\centering  \footnotesize
\caption{The mean-field results of the critical exponents for the classical percolation transition on SF networks \cite{Cohen2002}.} \label{tb2-ce}
\begin{tabular}{cccccc}
\toprule
$\lambda$ & $\beta$ & $\gamma$ & $\nu$ & $\sigma$ & $\tau$ \\
\hline
$(2,3)$ & $1/(3-\lambda)$ & $-1$ & $(\lambda-1)/(3-\lambda)$ & $(3-\lambda)/(\lambda-2)$ & $(2\lambda-3)/(\lambda-2)$ \\
$(3,4)$ & $1/(\lambda-3)$ & $1$ & $(\lambda-1)/(\lambda-3)$ & $(\lambda-3)/(\lambda-2)$  & $(2\lambda-3)/(\lambda-2)$ \\
$(4,\infty)$ & 1 & 1 & $3$ & $1/2$ & $5/2$ \\
\bottomrule
\end{tabular}
\end{table}

In essence, a strong heterogeneity just means a broad degree distribution. Therefore, when the emergence of hubs in SF networks is suppressed (large $\lambda$), the regular mean-field exponents as that found in ER networks are excepted. Cohen \emph{et al}. pointed out that the regular mean-field results can be found when $\lambda>\lambda_c=4$ \cite{Cohen2002}.

Moreover, Radicchi and Castellano pointed out that the site percolation on SF networks gives a different critical exponent $\beta$ for $2<\lambda<3$ \cite{Radicchi2015}. This comes from the singularity of Eq.(\ref{eq2-sp1}), which has the leading term $S\propto p(1-R)$. For $\lambda>3$, the system has a non-trivial $p_c$, and the leading term reduces to $S\propto 1-R$, which gives the same critical exponent $\beta$ as that of the bond percolation. However, for $2<\lambda<3$, it can be rewritten as $S\propto (p-p_c)(1-R)$, where $p_c=0$. This leads to $\beta = 1/(3-\lambda)+1=(4-\lambda)/(3-\lambda)$.

It is also worth noting that with the Potts model formulation, Lee \emph{et al}. shows $\tau=\lambda$ for $2<\lambda<3$ \cite{Lee2004,Lee2005}, which is contradict with that listed in Tab.\ref{tb2-ce} \cite{Cohen2002}. By this, the validity of the treatments used in the two works requires further study. Besides, Cohen \emph{et al}. pointed out that due to the vanished threshold for $2<\lambda<3$, the exponent $\tau=(2\lambda-3)/(\lambda-2)$ is calculated at a small but fixed occupied probability \cite{Cohen2002}.

With the framework shown above, many other features about the percolation transition on SF networks were also studied, such as the fractal dimensions of percolating networks \cite{Cohen2004}, the giant cluster in the large-network limit \cite{Janssen2016}, branching trees \cite{Lee2007}, statistical ensemble \cite{Burda2001,Goltsev2003,Burda2004}, width of percolation transition \cite{Kalisky2006}, the upper critical dimension \cite{Wu2007}, and the cluster forming in SF networks with exponent less than two \cite{Seyed-allaei2006} or one \cite{Timar2016}.

Although the above discussions are concentrated on the degree distributions of networks, we must point out that the spatial constraint still plays a key role in the nature of percolation transition. By embedding networks in a physical dimension, the network percolation can also exhibit a range of phase behaviors, as well as reconstructing the universality class of physical dimensions, depending on the dominant structure properties, such as spatial constraint, small-world, fractal, and hierarchy \cite{Boettcher2009,Rozenfeld2010,Hasegawa2010,Li2011,Hasegawa2013a}. In Refs.\cite{Newman1999,Newman2000b,Moore2000,Newman2002a}, Newman and the coauthors also developed a method to study the percolation transition on SW networks, suggesting that the SW network resembles more a random network in infinite dimension.

%In a recent work, we also point out that $\lambda_c=4$ is just a result for a special degree cutoff $M=N^{1/(\lambda-1)}$. For any cutoff $N^{1/\alpha}$, the critical value is $\lambda_c=3+\alpha/3$. This reinforces that the hubs are the key to determine the dynamics, and have a significant impact on the critical behaviors.

\subsection{On clustered and correlated networks} \label{sct2-ccn}

Real world networks are commonly clustered and correlated rather than a tree-like random structure \cite{Barabasi2016}. Some works also contributed to the percolation of clustered and correlated networks. Note that the clustering and the correlation discussed here are local, \textit{i.e.}, they are the properties of adjacent nodes.

\subsubsection{Networks with low clustering}

\begin{figure}
\centering
\includegraphics[width=0.8\columnwidth]{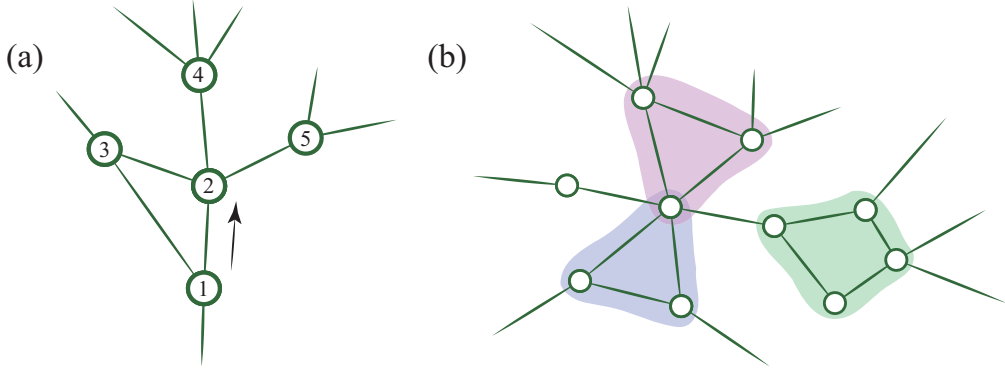}
\caption{(Color online) Schematic of the branching process in clustered networks. (a) The branching process starts from node $1$ takes the breadth-first strategy, \textit{i.e.}, searching all the neighbors of node $1$ firstly. At the end of link $l_{12}$ indicated by the arrow, node $2$ has three excess links, one of which ($l_{23}$) leads to a branching (node $3$) that has already been reached as a neighbor of node $1$. Therefore, only links $l_{24}$ and $l_{25}$ can lead to new branchings, \textit{i.e.}, there are only two effective excess degrees. In general, a local backward link, such as $l_{23}$, corresponds to the formation of a clustering, thus occurs with probability $C$ on average. (b) Considering the non-tree-like motifs as special links/nodes indicated by different colors, the network can be seen as a tree-like one.}  \label{fig2-cn}
\end{figure}

As a starting point, we first review the method proposed by Berchenko \emph{et al.} to find the approximate solution of the percolation model for networks with low clustering \cite{Berchenko2009}. Network science often employs the clustering coefficient $C$ to feature how well connected the neighborhood of a node is \cite{Wasserman1994,Watts1998,Newman2010,Barabasi2016}. If the neighbors are fully connected, the clustering coefficient is $1$, while a value close to $0$ means that there are hardly any connections in the neighborhood. Indeed, as the definition of the clustering coefficient, it also represents the connection probability of a node's two neighbors. In the branching process one can exclude the backward links from the excess degrees, see Fig.\ref{fig2-cn} (a). Therefore, the effective excess-degree $k'$ of a node with degree $k$ can be expressed as
\begin{equation}
k'=\binom{k-1}{k'}C^{k-1-k'}(1-C)^{k'}.     \label{eq2-kc}
\end{equation}
For $C=0$, this reduces to the tree-like structure $k'=k-1$. This equation has no physical meaning for $C=1$, since it is only an approximate treatment for low clustering. With Eq.(\ref{eq2-kc}), the generating function of the excess degrees can be revised as
\begin{eqnarray}
\mathcal{G}_1^c(x) &=& \sum_{k=1}\frac{p_kk}{\langle k\rangle}\sum_{k'=0}\binom{k-1}{k'}C^{k-1-k'}(1-C)^{k'}x^{k'}       \nonumber \\
         &=& G_1(C+x-Cx),         \label{eq2-g1c}
\end{eqnarray}
where $G_1(x)$ is that with no thought of clusterings. With this revised generating function, we can solve the percolation model on networks with low clustering, which gives the critical point \cite{Berchenko2009}
\begin{equation}
p_c=\frac{1}{(1-C)G_1'(1)}.   \label{eq2-pcc}
\end{equation}
This indicates that for a given degree distribution, the clustering leads to a larger percolation threshold. This mainly because for a fixed number of links, the clustering structure reinforces the core of the network with the price of diluting the global connections \cite{Kiss2008,Newman2009,Miller2009,Gleeson2010}. However, the clustering cannot restore a finite percolation threshold for SF networks with $2<\lambda<3$ \cite{Serrano2006a}, since the divergence of $G_1'(1)$ only depends on the degree distribution in this approximation.

It is important to notice that Eq.(\ref{eq2-pcc}) works only for networks with low clusterings. Strong clustering could induce the core-periphery structure \cite{Rombach2014,Rombach2017}, in which the core and periphery might percolate at different critical points \cite{Colomer-de-Simon2014,Allard2017,Hebert-Dufresne2019}, and the above approximate treatment is not applicable. In this perspective, the clustered networks are often characterized as a robust system \cite{Newman2003,Serrano2006,Gleeson2009}.

\subsubsection{Networks with high clustering}

For high clustering, the triangles formed by adjacent nodes could share links, which cannot be directly reflected by the clustering coefficient, Eq.(\ref{eq2-kc}) thus overcounts the excess links. Of course, one can further use the polynomials of the clustering coefficient to estimate these high-order clustering structures \cite{Volz2004}. However, to completely eliminate the overcounting, the history of the branching process has to be considered, which is beyond the compass of the mean-field method. In addition, this problem in a sense relates to the minimum spanning tree of networks \cite{Graham1985,Cheriton1976,Dobrin2001}, which is also a problem without a unified solution.

If the triangles or other non-tree-like motifs do not share links with each other, there is another framework to solve the corresponding percolation problem \cite{Newman2009,Miller2009,Karrer2010,Mann2020}. Beyond the degree distribution, this approach requires to know the distribution of all other motifs that a node has, see Fig.\ref{fig2-cn} (b). Taking the triangle structure as an example, when the number of triangles that a node has is known, one can treat the triangle as a special link in the branching process. Then, the network can percolate by means of not only single links but also triangles. Under this premise, the clustered networks can also be regarded as tree-like, and the processing method provided above can thus be used. In principle, this method can be applied to networks with any motifs. However, it is a tedious work to classify diverse clustered structures and get the corresponding distribution. In addition, this approach can be well summarized as mapping the clustered network into a tree-like hyper-network \cite{Ghoshal2009}.

The general thought in dealing with this problem, namely, recognizing a certain connection structure as an individual (link or node), is also used in the so-called clique percolation \cite{Derenyi2005}. In this percolation problem, the $k$-clique, a complete subnetwork with $k$ nodes, is treated as a node, and two $k$-cliques sharing $l$ nodes are considered to be connected. This percolation model is usually used as an algorithm for the detection of communities with overlap \cite{Palla2005}. For percolation transition, the main finding is the seemingly discontinuous transition, which essentially is a continuous one \cite{Li2015}. The details will be discussed in Sec.\ref{sct4-cd}.

\begin{figure}
\centering
\includegraphics[width=0.8\columnwidth]{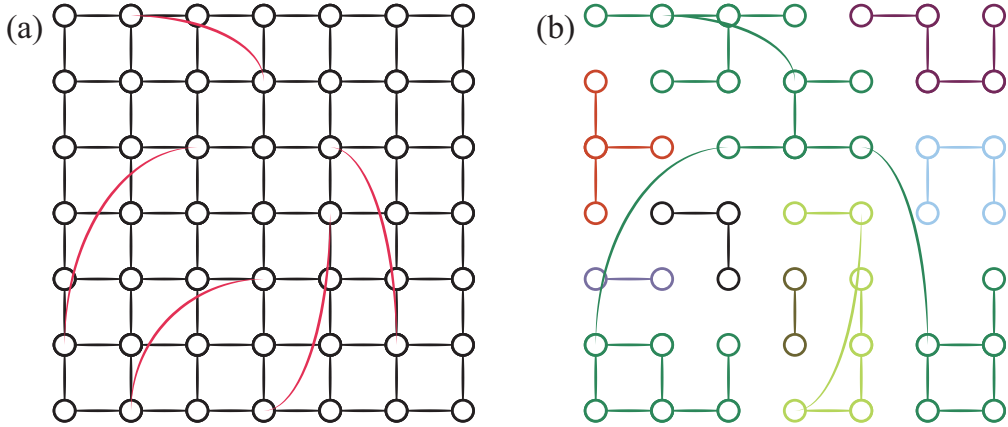}
\caption{(Color online) Schematic of the percolation configuration of a SW network. (a) A SW network consisted of an underlying two-dimensional square lattice and several shortcuts indicated by red. (b) A possible configuration after removing some links from the SW network shown in (a). Since the shortcuts are sparse, it is nearly impossible to form loop structures with shortcuts near or below the percolation threshold. The loop structures only exist in the clusters extracted from the underlying lattice.}  \label{fig2-swp}
\end{figure}

Another application of this framework is the solution of the percolation model on SW networks \cite{Moore2000,Newman2002a}, which typically show a highly clustering effect. In Refs.\cite{Moore2000,Newman2002a} the authors treated the finite percolation clusters extracted from the underlying lattice of the SW network as effective nodes, then the system is just a set of effective nodes connected by the shortcuts, which are links added between randomly selected pairs of nodes in the underlying lattice \cite{Watts1998,Newman2000a}. As the setting of SW networks, the shortcuts are sparse, so near or below the percolation threshold the reduced network demonstrates a tree-like structure composed of effective nodes and shortcuts, see Fig.\ref{fig2-swp}. Consequently, the formulation of the self-consistent equation as Eq.(\ref{eq2-h1}) can be used to solve the model, too. Based on this treatment, they found that the shortcuts not only modify the percolation threshold but also the universality class.

Since the clusterings are naturally excluded in the non-backtracking matrix, with some techniques the message passing method can also be used to find the percolation threshold of clustered networks, see the references listed in Sec.\ref{sct-mpm} for details.

\subsubsection{Correlated networks}

Along with clusterings, real networks generally show a structure with degree correlation rather than random connections. For such networks, the branching process cannot be simply described by the excess-degree distribution. Instead, a joint degree-degree distribution is used to figure out the degree correlation of the two nodes at two ends of a link \cite{Vazquez2003,Goltsev2008}.

Theoretical results show that a finite amount of random mixing of the connections in SF networks with $2<\lambda<3$ is sufficient to give a divergent $\langle k^2\rangle$, and thus leads to a vanished percolation threshold \cite{Vazquez2003}. Moreover, the assortative correlation makes networks more robust, while the disassortative correlation makes networks fragile even with a divergent second moment of degree distribution. However, in the spatially constrained ER networks, degree correlations favor or do not favor percolation depending on the connectivity rules \cite{Schmeltzer2014a}.

The Monte Carlo simulations on the exponential random network show that the disassortative correlation has no effect on the critical phenomena so that the percolation transition on disassortative networks belongs to the same universality class as on uncorrelated networks \cite{Noh2007}. While assortative correlation is relevant, percolation transition shows continuously varying critical exponents \cite{Noh2007}. Recently, Mizutaka \textit{et al.} proposed a maximally disassortative network model, which realizes a maximally negative degree-degree correlation \cite{Mizutaka2020}. Both the analytical and simulation results suggest this maximally disassortative network can also give new critical exponents for SF networks with $2<\gamma<3$.

Goltsev \textit{et al.} further summarized three conditions \cite{Goltsev2008}: (i) The largest eigenvalue $\lambda_1$ of the branching matrix is finite if $\langle k^2\rangle$ is finite, or $\lambda_1 \to\infty$ if $\langle k^2\rangle\to\infty$; (ii) The second largest eigenvalues of the branching matrix is finite; (iii) The sequence of entries of the eigenvector associated with the largest eigenvalue converges to a nonzero value. When these conditions are fulfilled, the critical exponents are completely determined by the asymptotic behavior of the degree distribution at large degrees, thus the percolation transition on a correlated network belongs to the same universality class as the percolation on an uncorrelated network. If at least one of the three conditions is not fulfilled, the critical exponent becomes model-dependent and hence non-universal \cite{Valdez2011,Agliari2011,Schmeltzer2014}. Furthermore, by the technique that dividing nodes into different types with hyper-links, the site and bond percolations on clustered and correlated networks can also be solved \cite{Allard2012,Allard2012a,Allard2015}.

\subsection{On directed networks} \label{sct2-dn}

\begin{figure}
\centering
\includegraphics[width=0.7\columnwidth]{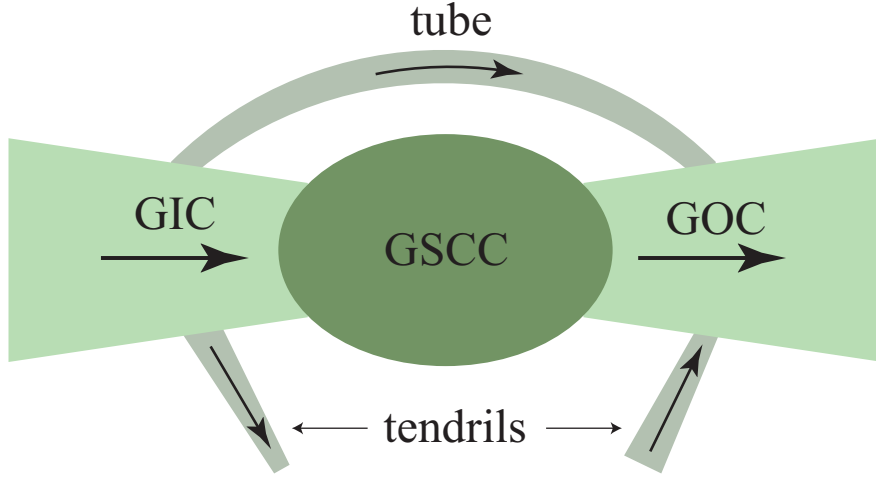}
\caption{(Color online) Schematic of the giant cluster of directed networks. GSCC is a set of nodes, in which each node is reachable from all others. The set of nodes, from which GSCC are reachable but those are not reachable from GSCC, form the GIC. In turn, GOC is the set of nodes that GSCC are not reachable but those are reachable from GSCC. Besides, GIN can also lead to some nodes which do not belong to GSCC, if these nodes can lead to GOC then called tube, otherwise called tendril. Moreover, the nodes lead to GOC but not belong to GSCC and GIC are also called tendril. These giant clusters and the attached smaller clusters, including tubes and tendrils but not just, form the GWCC, which is just the giant cluster by ignoring the direction of links. In addition to these giant clusters, there are some small clusters that are disconnected to these giant clusters, though not depicted here.}  \label{fig2-dn}
\end{figure}

The percolation transition can be also defined on directed networks. The difference is that the giant cluster of directed networks can be further specified as \cite{Newman2001}: (i) the giant strongly connected cluster (GSCC), in which each node is reachable from others; (ii) the giant in-cluster (GIC), from which GSCC are reachable but those are not reachable from GSCC; (iii) the giant out-cluster (GOC), from which GSCC are not reachable but those are reachable from GSCC; (iv) the giant weakly connected cluster (GWCC), in which each pair of nodes are reachable without regard to the direction of links. A schematic of these giant clusters is shown in Fig.\ref{fig2-dn}.

The mean-field method reviewed in Sec.\ref{sct2-ampb} can also be extended to directed networks \cite{Dorogovtsev2001,Schwartz2002,Boguna2005,AngelesSerrano2007,Gleeson2008a,Restrepo2008,Hoorn2015,Liu2017}. For this, we define $R^I$ and $R^O$ as the probabilities that a directed link leads to GSCC and comes from GSCC, respectively. Thus, similar to Eq.(\ref{eq2-r}), we have two self-consistent equations,
\begin{align}
R^{I} &= 1-\sum_{ij}\frac{p_{ij}i}{\langle k\rangle}\left(1-R^{I}\right)^j  \nonumber \\
&= 1-G_1^{I}(1,1-R^{I}), \label{eq2-ri} \\
R^{O}  &= 1-\sum_{ij}\frac{p_{ij}j}{\langle k\rangle}\left(1-R^{O}\right)^i  \nonumber \\
&= 1-G_1^{O}(1-R^{O},1), \label{eq2-ro}
\end{align}
where $p_{ij}$ is the probability that a node has in-degree $i$ and out-degree $j$, and
\begin{align}
G_1^{I}(x,y) &= \sum_{ij}\frac{p_{ij}i}{\langle k\rangle}x^{i-1}y^j,  \\
G_1^{O}(x,y) &= \sum_{ij}\frac{p_{ij}j}{\langle k\rangle}x^iy^{j-1}
\end{align}
are the generating functions of the excess-degree distribution following and against the link direction, respectively. Note that a directed network has an identical average in-degree and average out-degree $\langle k\rangle=\sum_{ij}p_{ij}i=\sum_{ij}p_{ij}j$.

When Eqs.(\ref{eq2-ri}) and (\ref{eq2-ro}) only have the trivial solution $R^{I}=R^{O}=0$, there is thus no GSCC in the system. Otherwise, the giant cluster can be expressed by $R^{I}$ and $R^{O}$ according to their definitions,
\begin{align}
S_{GIC} &= 1-G_0(1, 1-R^{I}),    \\
S_{GOC} &= 1-G_0(1-R^{O}, 1),    \\
S_{GSCC} &= 1-G_0(1-R^{O}, 1) -G_0(1, 1-R^{I}) +G_0(1-R^{O}, 1-R^{I}),    \label{eq2-sgscc}
\end{align}
where
\begin{align}
G_0(x,y) &= \sum_{ij}p_{ij}x^iy^j
\end{align}
is the generating function of the degree distribution. Similar as that for undirected networks, let $R^I (R^O)\rightarrow 0$ in Eqs.(\ref{eq2-ri}) and (\ref{eq2-ro}), we have the criterion of percolation transition,
\begin{align}
\frac{\partial G_1^{I}(1,y)}{\partial y}\bigg|_{y=1}&=1, \\
\frac{\partial G_1^{O}(x,1)}{\partial x}\bigg|_{x=1}&=1.
\end{align}
The two equations are equivalent, that is
\begin{equation}
\sum_{ij}p_{ij}(2ij-i-j)=0.  \label{eq2-gscc}
\end{equation}
This is the condition that the GSCC exists in a directed network. Considering the dilution of the initial node or link removal as that for undirected networks (see Sec.\ref{sct2-ampb}), one can also find the percolation threshold from Eq.(\ref{eq2-gscc}),
\begin{equation}
p_c=\frac{\langle k\rangle}{\langle ij\rangle}.   \label{eq2-pcd}
\end{equation}
where $\langle ij\rangle=\sum_{ij}p_{ij}ij$. If there is no correlation between in-degrees and out-degrees, \textit{i.e.}, $\langle ij\rangle=\langle k\rangle^2$, Eq.(\ref{eq2-pcd}) reduces to $p_c=1/\langle k\rangle$. This indicates that in contrast to undirected networks, a non-zero percolation threshold can exist in directed SF networks with $\lambda^I>2$ and $\lambda^O>2$ \cite{Schwartz2002}. Here, $\lambda^I$ and $\lambda^O$ are the in-degree and out-degree distribution exponents, respectively.

From the asymptotic expansion of Eqs.(\ref{eq2-ri})-(\ref{eq2-sgscc}), the critical exponents can also be obtained \cite{Schwartz2002}. In general, the GIC and GOC can give different critical exponents $\beta$, which are dependent on the in-degree and out-degree distributions. As the definition, GSCC is the intersection of GIC and GOC. Therefore, it behaves as the smaller one of GIC and GOC. For directed SF networks, the critical exponents can be also written in the form of undirected SF networks, see Tab.\ref{tb2-ce}, but with an effective $\lambda^*$, which is dependent on the existence of correlations and on the degree distribution exponents $\lambda^I$ and $\lambda^O$ \cite{Schwartz2002}.

The above discussion can also be generalized to the cases with the bidirectional links and the degree correlations \cite{Boguna2005,Meyers2006}. The result shows that the percolation threshold can be generally expressed as a function of the maximum eigenvalue of the connectivity matrices. In particular, for networks with no degree correlations, bidirectional links act as a catalyst for percolation, favoring the emergence of the GSCC, and for SF networks, only an infinitesimal fraction of bidirectional links is needed. The interface links, \textit{i.e.}, the ones joining GIC/GOC and GSCC, can also be investigated analytically in this framework \cite{AngelesSerrano2007}. In addition, the GWCC of a directed network can be obtained by mapping it into an undirected network. It should be noted that the GWCC could emerge at a smaller $p_c$ than that of GSCC, which is dependent on the degree correlations of the in-degrees and the out-degrees. This indicates that the directed network may have the GWCC and, simultaneously, may not have the GSCC.

\subsection{Algorithm for network percolation}

For most network systems, the percolation model cannot be solved exactly. In consequence, the simulation algorithm becomes very important to estimate the critical point and to study the critical phenomena. However, due to the diversity of network connections, the algorithms for classical percolation on regular lattices cannot be transplanted into network percolation directly. One needs to store the connection information of the network, while this is not necessary for that on regular lattices \cite{Hoshen1976}. Hence, the simulation of the network percolation requires more memory than of the classical percolation on physical dimensions.

In general, we can first construct the percolation configuration following the percolation rule, then use the classical algorithm of graph searching to identify the connections of the configuration, including the breadth-first searching and the depth-first searching. For a given configuration with $M$ links, these searching algorithms take time $O(M)$ to find all the clusters. This is the minimal overhead for scanning an unknown structure. If the data structure of the network does not allow to directly scan the neighbors of nodes, such as the adjacency matrix, the time complexity will be higher. This two-step algorithm, \textit{i.e.}, constructing the configuration first and then identifying the clusters, does not have a specific requirement for percolation rules and network structures, and thus is universally applicable.

To be more effective, the percolation algorithm should be a creative blend of configuration constructing and cluster identifying, not a process with two standalone steps. In other words, while constructing the percolation configuration, we must simultaneously update the cluster information. The detailed techniques are dependent on the percolation model and the network model. Here, we only introduce the algorithm proposed by Newman and Ziff for the classical percolation on networks \cite{Newman2000,Newman2001a}, which is applicable to any given networks. Other algorithms will be reviewed when necessary.

The Newman-Ziff algorithm employs an exquisite data structure \cite{Newman2000,Newman2001a}, with which the nodes/links can be occupied one by one, and the cluster size information can be updated concurrently, which improves greatly the efficiency for checking the percolation properties of a system under a sequence of occupied probability $p$. In this algorithm, each cluster is stored as a separate tree with a single root node. Each node is allocated a pointer either to the root of the cluster or to another node in the cluster, such that by following a pointer chain we can travel from any node to the root of the cluster. The root nodes can be identified by the fact that they have a null pointer. To be more efficient, we can also use the pointer of the root to store the size of the cluster.

Taking bond percolation as an example, this percolation algorithm can be summarized as follows \cite{Newman2000,Newman2001a}:
\begin{enumerate}
  \item Initially each node is its own root, and contains a record of its own size, which is $1$.
  \item Links of the network are occupied in random order. When a link is occupied, two nodes are joined together. Follow the pointer chains from the two nodes separately until we reach the root nodes of the clusters to which they belong.
  \begin{enumerate}
    \item If the two roots are the same node, we do nothing further.
    \item If the two roots are different, we examine the cluster sizes stored in them, and add a pointer from the root of the smaller cluster to the root of the larger, thereby making the smaller tree a subtree of the larger one. If the two have the same size, we may choose whichever tree we like to be the subtree of the other. We also update the size of the larger cluster by adding the size of the smaller one to it.
  \end{enumerate}
\end{enumerate}
Step $2$ is repeated until the expected number of occupied links is reached. The cluster size can be obtained from the pointer of the root, thus it allows us to evaluate the observable quantities of interest. In addition, to improve the efficiency of the algorithm, we can compress the pointer chain as much as possible. For site percolation, the algorithm is similar, see Refs.\cite{Newman2000,Newman2001a} for details.

%******************************************************************************
% Section III
% Typical models for network percolation
%******************************************************************************

\section{Network-specific percolation models}   \label{sct3}

In the previous section, we reviewed the studies of the classical percolation model on networked systems. As a framework to evaluate the system performance, the emergence of the giant cluster with respect to some control parameters is widely used in network science \cite{Albert2002,Boccaletti2006,Dorogovtsev2008,Newman2010,Cohen2010,Boccaletti2014,Barabasi2016}. However, the cluster forming rule is diversiform rather than a probabilistic occupation as that in the classical percolation. They usually contain some recursive/iterative processes or with a broad sense of occupation in the cluster forming. Thus, these derivative models are often categorized into a family called dependent percolation. Typical examples include $k$-core percolation, clique percolation, core percolation, explosive percolation, percolation on interdependent/multiplex networks, \textit{etc}. In this section we will give a brief introduction of the theoretical findings of these models.

In Tab.\ref{tb3-pm}, we list the key rules and the types of the percolation transitions for different models.

\begin{table}[ht]
\centering \scriptsize
\caption{The key rules and the types of the percolation transitions for different models. In general, there are three kinds of models. The first one is a node removal process triggered by removing a fraction $1-p$ of nodes from a network (occupying a fraction $p$ of nodes). This removal process can differ from model to model, which could contain multiple steps from one single operation to infinite iteration. It is also worth pointing out that due to a similar mechanism, although the bootstrap percolation is a recovery process for removed nodes, we also classify it into this class. The second one is a link insertion process in a set of nodes, \textit{i.e.}, at each time step a pair of nodes will be connected with probability $p$. The rule for choosing a node pair is the key of this type of models. For the last one, there could be a special definition of the connection of nodes, and the connection unit is not limited to node. Note that the probability $p$ is used as the order parameter for the first two types, however, the meaning is different. In the table these three types of models are separated by a horizontal line.} \label{tb3-pm}
\begin{tabular}{p{0.15\columnwidth}p{0.4\columnwidth}p{0.2\columnwidth}}
\toprule
Model & Rule & Type \\
\hline
$k$-core percolation \cite{Dorogovtsev2006}  & Removing all the nodes with degrees less than $k$, iteratively.  &  continuous for $k=2$, hybrid for $k\geq3$  \\
Threshold model \cite{Watts2002,Liu2012} &  Removing all the nodes with $k_i/k_{i,0}<\alpha$ iteratively, where $k_i$ and $k_{i,0}$ are the current and the initial degrees of node $i$, respectively &  continuous for small $\alpha$, discontinuous for large $\alpha$ \cite{Liu2012}  \\
Bootstrap percolation \cite{Baxter2010} &  Removed nodes will be iteratively recovered when they have at least $k$ neighbors.  &  discontinuous, hybrid \cite{Baxter2010} \\
Core percolation \cite{Liu2012a} & Removing nodes of degree $1$ along with its neighbor, iteratively.  &  continuous for undirected networks, discontinuous for directed networks  \cite{Liu2012a} \\
Greedy articulation points removal \cite{Tian2017} &  Removing all the articulation nodes iteratively, where the articulation node is the one whose removal disconnects the network.  & hybrid \cite{Liu2012a} \\
Percolation on interdependent networks \cite{Buldyrev2010,Parshani2010}  &  Removing all the failed nodes iteratively, where the failed node is the one does not belong to the giant cluster, or has a failed dependent partner in other layers. & hybrid  \cite{Buldyrev2010,Parshani2010} \\
Network observability \cite{Yang2012} & The initial observable (occupied) node makes both the node and all of its neighbors observable (occupied). & continuous \cite{Yang2012} \\
$l$-hop percolation \cite{Shang2011} & The nodes no farther than $l$ away from the initial removed nodes will also be removed.  &  continuous \cite{Shang2011} \\
History-dependent percolation \cite{Li2020} & Removing all the links between different clusters of the last generation.  &  continuous for finite generations, hybrid for infinite generation \cite{Li2020,Hu2020} \\
\hline
Explosive percolation \cite{Achlioptas2009,Friedman2009,Costa2010,DSouza2010,Moreira2010,Saberi2015}  & At each time step, more than one potential links are arbitrarily chosen. The one that suppresses the emergence of a giant cluster is inserted eventually, and other potential links are discarded.  & continuous with unusual finite size behavior \cite{Boccaletti2016}\\
Growing network \cite{Callaway2001} & At each time step, a new node is inserted into the system, then two nodes are chosen randomly from all the existing nodes and joined by a link with probability $p$.  & infinite order \cite{Callaway2001}\\
\hline
$k$-component  &  The nodes in the same cluster must connect to each other by at least $k$ independent paths. & continuous for $k=2$ \cite{Newman2008} \\
Limit path percolation \cite{Lopez2007} & After link/node removal, two nodes are considered as connected, if the new shortest path between them is shorter than $al (a\geq1)$, where $l$ is the shortest path before removal.
&  continuous \cite{Lopez2007} \\
Clique percolation \cite{Derenyi2005,Li2015} & This model considers the percolation of connected cliques in a network, and two cliques are regarded as adjacent, if they share some nodes. &  continuous \cite{Li2015,Dong2018}\\
Color-avoiding percolation \cite{Krause2017,Kadovic2018} &  Two nodes are considered as color-avoiding connected, only if they are always connected for the removal of any colored links/nodes.
 & continuous \cite{Krause2017,Kadovic2018}\\
\bottomrule
\end{tabular}
\end{table}

\subsection{k-core percolation}   \label{sct3-kcp}

$k$-core percolation is one of the typical representatives of the dependent percolation on networks\cite{Seidman1983,Bollobas1984,Pittel1996,Dorogovtsev2006}. For a given network $k$-core is the subnetwork in which nodes have at least $k$ links connecting with other nodes in this subnetwork. So, as a percolation model, it mainly studies the emergence of the giant $k$-core when a fraction $1-p$ nodes are removed, \textit{i.e.}, occupying each node with probability $p$ \cite{Dorogovtsev2006}. The case $k=1$ recovers the classical percolation. While $k=2$ the formed cluster has a structure similar to the so-called bicomponent/biconnectivity, which is a meaningful concept of network robustness \cite{Newman2008}. For $k\geq3$, $k$-core percolation transition becomes discontinuous. In this subsection we will review the theoretical researches of $k$-core percolation, as well as some related percolation models. $k$-core percolation and its variants are often used as algorithms to provide the structure information of networks \cite{Kong2019}, some of which will be included in Sec.\ref{sct4}.

\subsubsection{Model and phase transition characteristics}

We can iteratively remove nodes with degrees smaller than $k$ to obtain the $k$-core of a network. Note that even without the initial node removal, $k$-core may be absent in networks, dependent on the network topology. An extreme case is the tree-like network, for which there cannot exist any finite $k$-cores with $k\geq2$. This is not difficult to be understood from the pruning process of $k$-core percolation. In any finite and tree-like networks, there must be some nodes with degree $k=1$, located at the periphery of the network. The removal of these nodes must introduce some new periphery nodes with degree $k=1$. In this way the pruning process will destroy the whole network. However, this does nothing to the theoretical analysis of $k$-core percolation transition on tree-like networks, since an infinite spanning tree does not have periphery nodes, and all the degrees are larger than or equal to $2$. This also means that there are no finite $k$-core clusters ($k\geq2$) in tree-like networks, thus the $k$-core, if exists, is just the giant cluster, and can be used to feature the percolation transition.

As the model setting, if a node is in the $k$-core, it must have at least $k$ links connecting to other nodes in the $k$-core. By this, similar to Eqs.(\ref{eq2-rp}) and (\ref{eq2-sp1}), we can write the self-consistent equations for $k$-core percolation on tree-like networks as \cite{Dorogovtsev2006},
\begin{align}
S_k & = p \sum_{n=k}^\infty \sum_{d=n}^{\infty} p_d {d \choose n} R_k^n (1-R_k)^{d-n}   \\
    & = p \left[ 1- \sum_{n=0}^{k-1} \frac{R_k^n}{n!} \frac{d^nG_0(x)}{dx^n}\bigg|_{x=1-R_k} \right],   \label{eq3-sk}\\
R_k & = p \sum_{n=k-1}^\infty \sum_{d=n+1}^{\infty}\frac{p_dd}{\langle k\rangle}{d-1\choose n}R_k^n(1-R_k)^{d-1-n}   \label{eq3-rk1} \\
    & = p \left[  1-\sum_{n=0}^{k-2} \frac{R_k^n}{n!} \frac{d^nG_1(x)}{dx^n}\bigg|_{x=1-R_k} \right],   \label{eq3-rk}
\end{align}
where $S_k$ is the size of the $k$-core, and $R_k$ is the probability that a node reached by following a randomly chosen link belongs to the $k$-core. Here, to avoid confusion we use $p_d$ to represent the degree distribution. The sum $\sum_n$ is over the nodes with $n$ preserved neighbors, and the sum $\sum_d$ is for the degree distribution. Note that the sum $\sum_n$ in Eq.(\ref{eq3-rk1}) begins with $k-1$, since the link used to reach the node has already provided a link for the $k$-core. For both the two cases $k=1$ and $k=2$, Eq.(\ref{eq3-rk}) reduces to that of the classical percolation equation (\ref{eq2-sr}), so $1$-core percolation and $2$-core percolation have the same threshold equation (\ref{eq2-pc}). For $k>3$ the threshold of the $k$-core percolation becomes distinguishable for $k$.

\begin{figure}
\centering
\includegraphics[width=0.7\columnwidth]{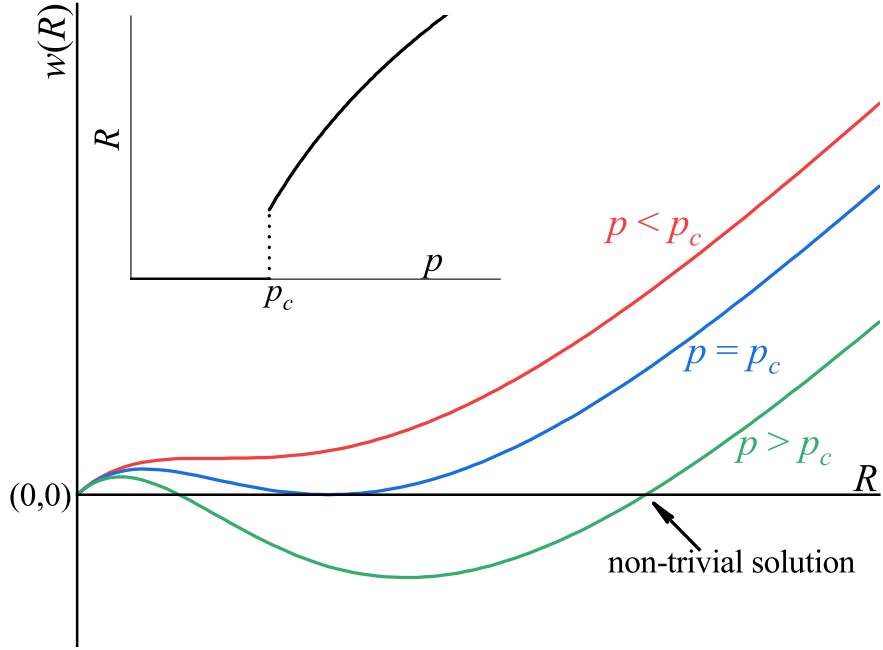}
\caption{(Color online) Schematic of the solution of Eq.(\ref{eq3-rk}). Only when $p$ exceeds $p_c$, function $w(R)$ has non-trivial crossing points with $R$-axis. The critical point corresponds to the tangency of the function $w(R)$ and $R$-axis, which gives a non-zero $R$. The inset shows the non-trivial solution $R$ for different $p$.}  \label{fig3-kcs}
\end{figure}

To get the critical point, one can also define a function $w(R_k)$ as that for classical percolation, see Sec.\ref{sct2-pt}. The critical point thus corresponds to the tangency of $w(R_k)$ and $R_k$-axis. However, since function $w(R_k)$ has more than one extreme point for $k\geq 3$, the tangency point gives a non-zero $R_{k}$, labeled $R_{k,c}$, which indicates that the percolation transition is discontinuous, see the schematic shown in Fig.\ref{fig3-kcs}.

Although there is no closed form for the solution of Eqs.(\ref{eq3-sk}) and (\ref{eq3-rk}), the asymptotic expansion shows that the scaling behaviors can also be observed near the critical point, that is
\begin{equation}
S_k-S_{k,c}\propto (p-p_c)^\beta.
\end{equation}
In consequence, the emergence of a $k$-core is a unique hybrid phase transition with a jump emergence of the $k$-core as a first-order phase transition but also with a critical singularity as the second-order phase transition. This result is first found in Bethe lattice with $\beta=1,2,1/2$ respectively for $k=1,2$ and $\geq3$ \cite{Chalupa1979}, where the model is called bootstrap percolation. Note that the bootstrap percolation now often refers to another percolation model related to $k$-core percolation, and we will give the details later. It is worth noting that this result indicates that $2$-core percolation belongs to a different universality class from that of $1$-core percolation (the classical percolation), although they have the same percolation threshold \cite{Schawe2019}. This finding is also confirmed by the Monte Carlo simulation on ER networks \cite{Zhu2017}.

For SF networks that bear a strong heterogeneous degree distribution, a ratio $\beta_{k=2}/\beta_{k=1}=\lambda-1$ for $2<\lambda<3$ and $2$ for $\lambda>3$ is found \cite{Kim2013}, where $\lambda$ is the exponent of the degree distribution of SF networks. This suggests that the $k$-core percolation could reconstitute the normal mean-field nature like that on ER networks when $\lambda>3$, while for the classical percolation the criterion is $\lambda>4$ \cite{Cohen2002,Lee2004}. Numerical results also reveal that the fluctuations of the order parameter and mean avalanche size diverge in different ways for $k\geq3$ \cite{Lee2016}. The main focus of this review is network systems, however, we need to point out that the mean-field hybrid nature of the $k$-core percolation can only survive in high dimensions \cite{Harris2005,Farrow2007,Parisi2008,Rizzo2019}.

More generally, Dorogovtsev \textit{et al.} showed that if the second moment of the degree distribution of a network is finite, $k$-core percolation has a hybrid nature, and there is no principal difference between different tree-like networks \cite{Dorogovtsev2006,Dorogovtsev2006a,Goltsev2006}. A dramatic difference takes place if the second moment of the degree distribution diverges, \textit{i.e.}, the SF networks with exponent $2<\lambda<3$, for which an infinite order transition is observed. Besides, $k$-core percolation was also analyzed for clustered networks \cite{Gleeson2009a,Bhat2017} and for biased initial node removal \cite{Yuan2016}. Moreover, $k$-core percolation is also studied as models of the jamming transition \cite{Sellitto2005,Schwarz2006}, granular gas
\cite{Alvarez-Hamelin2007}, evolution \cite{Klimek2009}, and nervous system \cite{Chatterjee2007}. Furthermore, the inducing mechanism has been found that can bridge the classical percolation and $k$-core percolation for a general $k$, which always causes a discontinuous percolation transition \cite{Zhao2013}.

In network science, the pruning process of $k$-core percolation has also attracted a lot of attention \cite{Iwata2009,Baxter2015,Wu2018,Wu2018a,Shi2018}, which is used to analyze network structures \cite{Seidman1983,Alvarez-Hamelin2005,Carmi2007,Kitsak2010,Batagelj2011,Eidsaa2013,Lue2016}. It often refers to $k$-core/$k$-shell decomposition \cite{Kong2019}, and the nodes belong to $k$-core, but not $(k+1)$-core, are named $k$-shell.

\subsubsection{Variants and related models}

Focusing on the pruning process, one can adjust the criterion to observe the emergence of other core structures of networks. Meanwhile, the corresponding percolation transitions can be also defined. Although some of them are not initially motivated by $k$-core percolation, here we review these studies roughly attributed to variants and related models of the $k$-core percolation.

\paragraph{Heterogeneous $k$-core percolation}

In the pruning process of $k$-core percolation, if different thresholds are allowed for different nodes, the model becomes a sophisticated one known as heterogeneous $k$-core percolation \cite{Branco1993,Cellai2011,Baxter2011,Cellai2013,Chae2014}. Specifically, after the initial node removal, some of nodes need $k_a$ subtrees to survive from the pruning process, some of nodes need $k_b$ subtrees, and so forth.

A representative example of the heterogeneous $k$-core is the binary mixture $\mathbf{k}=(k_a,k_b)$ that nodes have a threshold of either $k_a$ or $k_b$ distributed randomly through the network with probabilities $f$ and $1−f$, respectively. With the framework figured by Eqs.(\ref{eq3-sk}) and (\ref{eq3-rk}), one can easily express the size of the heterogeneous $k$-core as a linear combination of those for the two thresholds \cite{Branco1993,Cellai2011,Baxter2011,Cellai2013,Chae2014}, each of which takes the form as the right hand side of Eq.(\ref{eq3-sk}) or Eq.(\ref{eq3-rk}).

\begin{figure}
\centering
\includegraphics[width=0.9\columnwidth]{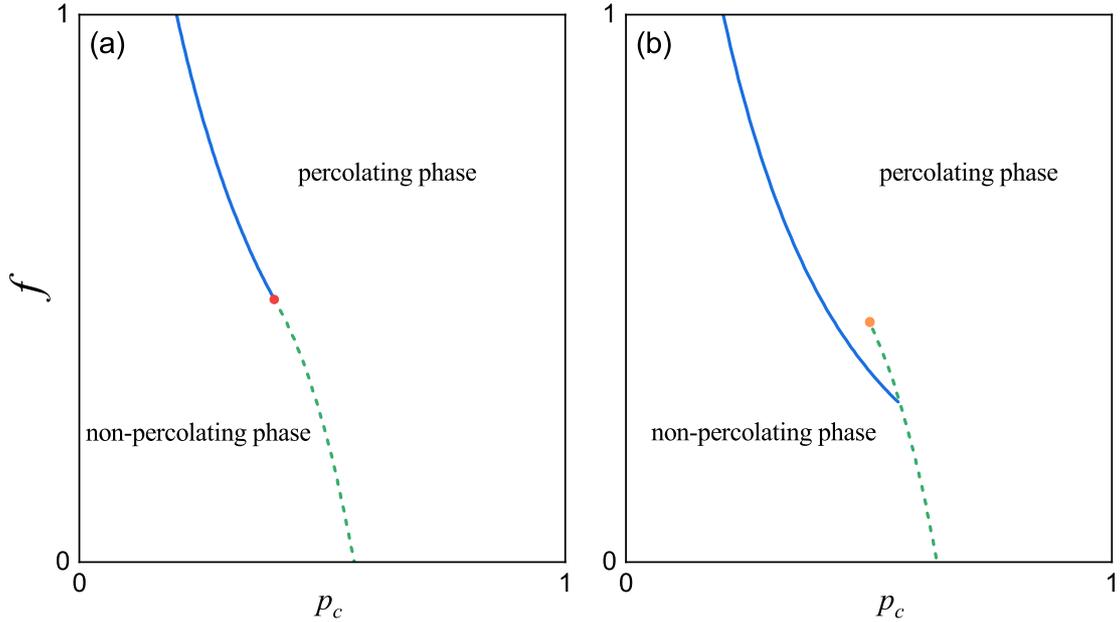}
\caption{(Color online) Schematic of two different phase diagrams of the heterogeneous $k$-core percolation. (a) The two lines indicate the continuous transition (blue solid line) and the hybrid transition (green dashed line) matching at a tricritical point (red scatter). (b) The line for the hybrid transition (green dashed line) ends at a point (orange scatter) that is out of the line for the continuous transition (blue solid line). Note that these two figures are conceptual, and do not represent the exact values.}      \label{fig3-dhkc}
\end{figure}

The transition nature of the heterogeneous $k$-core percolation obviously depends on the probabilities assigned to different thresholds, as well as the values of thresholds. Although the cases $k=1$ and $k=2$ behave similarly in $k$-core percolation, the percolation transitions of $\mathbf{k}=(1,3)$ and $\mathbf{k}=(2,3)$ are much different. For $\mathbf{k}=(2,3)$, there is a crossover of the continuous and the hybrid percolation transitions \cite{Cellai2011}, \textit{i.e.}, in the phase diagram the lines for the two types of phase transitions match at a tricritical point, see Fig.\ref{fig3-dhkc} (a). In contrast, the case $\mathbf{k}=(1,3)$ demonstrates a complex phase diagram: the line for the hybrid transition ends at a point that is out of the line of the continuous transition, see Fig.\ref{fig3-dhkc} (b). This means that a double percolation transition can be observed in the system for some appropriate probabilities $f$ \cite{Baxter2011}, namely, first showing the continuous transition, and later the discontinuous hybrid transition. In general, the tricritical point never occurs for the case $\mathbf{k}=(1,k)$, whereas in the case $\mathbf{k}=(2,k)$ it is presented only for $k=3$ \cite{Cellai2013}. It is also worth pointing out that the case $\mathbf{k}=(2,3)$ can also map onto a model of glasses on the Bethe lattice \cite{Sellitto2010}, which suggests the same universality class.

More generally, the ternary mixture \cite{Cellai2013a} and the generalized case $\mathbf{k}=(2,3,4,\ldots)$ \cite{Chae2014} were also investigated to show new types of critical phenomena. Through the analytic calculations and the numerical simulations on ER networks, Chae \textit{et al.} concluded that the heterogeneous $k$-core percolation can be featured by the series of continuous transitions with order parameter exponents $\beta=2/n,~n=1,2,3,\ldots$, discontinuous hybrid transitions with $\beta=1/2$ or $1/4$, and three kinds of multiple transitions \cite{Chae2014}.

\paragraph{Cascading failure model}

A special case of the heterogeneous $k$-core percolation is the one with threshold $\alpha k$ for nodes with degree $k$, where $\alpha\in[0,1]$ \cite{Watts2002,Gleeson2007,Centola2007,Galstyan2007,Gleeson2008,Whitney2010,Hackett2011,Liu2012,Hackett2013,Shang2015}. In other words, if a node $i$ can be preserved from the pruning process, $k_i/k_{i,0}\geq\alpha$ must be satisfied for any stages, where $k_i$ and $k_{i,0}$ are the current and the initial degree of node $i$, respectively. This node removal process is usually used to model the spreading of failures on networks, hence its name cascading failure model, sometimes also referred to as threshold model.

The simulation results suggest the existence of the crossover of the continuous percolation transition and the discontinuous percolation transition \cite{Liu2012}. However, more detailed checking of the critical phenomena as that of heterogeneous $k$-core percolation is still lacking. More often, the emergence of the giant cluster in this model is studied as a criterion for network robustness rather than a characterization for phase transition. We will review these works in the later sections.

In addition, there are two caveats to this model. First, many works on this cascading failure model focus on the clusters formed by failed nodes, not those of preserved nodes \cite{Watts2002,Gleeson2007,Gleeson2008,Whitney2010,Hackett2011,Brummitt2012,Hackett2013}. Second, there is another model in network science called cascading failure model which focuses on the overload failures, see Ref.\cite{Motter2002} and subsequent references for details.

\paragraph{Bootstrap percolation}

Along with the $k$-core percolation, there is an important and well-known model, called bootstrap percolation. In the early literature the bootstrap percolation just refers to the model of $k$-core percolation introduced above, such as Refs.\cite{Chalupa1979,Adler1991,Branco1993}. Now, the bootstrap percolation is generally referring to an activation process, which starts with a fraction $f$ of active seeds (occupied nodes) and other nodes (unoccupied) will be activated (occupied) when they have at least $k$ activated neighbors. A theoretical analysis for this model like Eqs.(\ref{eq3-sk}) and (\ref{eq3-rk}) can be found in Refs.\cite{Baxter2010,DiMuro2019,DiMuro2020}. This model and its variants are often used to describe the spreading dynamics in network science. A typical example is the cascading failure model reviewed previously.

Although the bootstrap percolation is an activation process beginning from a sparsely activated network, while the $k$-core percolation is a pruning process beginning from the whole network, a corresponding relation between them can also be established \cite{Baxter2011}. For comparison, we construct a heterogeneous $k$-core percolation with $\mathbf{k}=(1,k)$. Threshold $1$ is randomly assigned to a fraction $f$ of nodes corresponding to the seeds in bootstrap percolation, and other nodes have threshold $k$. By this mapping, the heterogeneous $k$-core percolation $\mathbf{k}=(1,k)$ can demonstrate very similar critical phenomena as the bootstrap percolation, \textit{i.e.}, a double percolation transition \cite{Baxter2010,Baxter2011}. However, this model cannot completely recover the bootstrap percolation \cite{Baxter2011}. This is because all the clusters in the bootstrap percolation must grow from the seeds, while for heterogeneous $k$-core percolation nodes can also survive from the pruning process by forming a compact cluster. Moreover, Di Muro \textit{et al.} also pointed out that the heterogeneous bootstrap percolation is the complement of the heterogeneous k-core percolation for complex networks with any degree distribution in the thermodynamic limit, as long as the thresholds of the nodes in both processes complement each other \cite{DiMuro2019}.

Due to the correlation of bootstrap percolation and $k$-core percolation, many works for $k$-core percolation also contributed to bootstrap percolation. For systems in physical dimensions and the Bethe lattice, one can also refer to Refs.\cite{Cerf1999,Holroyd2003,Holroyd2006,Balogh2006,Balogh2006a,Balogh2007,Fontes2008}. In network science, bootstrap-like processes are often used to model the cascading dynamics \cite{Albert2002,Dorogovtsev2008,Pastor-Satorras2015,Castellano2009,Newman2010,Cohen2010}, such as the spreading of information, opinion, and disease. Instead of the critical phenomena, these researches are focusing on the dynamical characteristics and their relations with the problems in real life. For this reason, there is no need to lump them into a clear category (bootstrap percolation or heterogeneous $k$-core percolation). Some of these studies will be reviewed in the later sections.

\paragraph{Core percolation} \label{sct3-cp}

The core of a network is defined as the subnetwork after a greedy leaf removal (GLR) procedure. Here, the leaf of a network refers to the node of degree $1$. The GLR procedure is just iteratively removing nodes of degree $1$ along with its neighbor. Note that the nodes with degree $1$ only have one neighbor. Apparently, the GLR procedure is more destructive than the pruning process of $2$-core percolation, and finite clusters cannot exist in tree-like networks. As the $k$-core percolation, the core percolation model is also used to identify the core structure of networks, which could dominate the dynamical properties of complex networks \cite{Liu2016}.

\begin{figure}
\centering
\includegraphics[width=0.7\columnwidth]{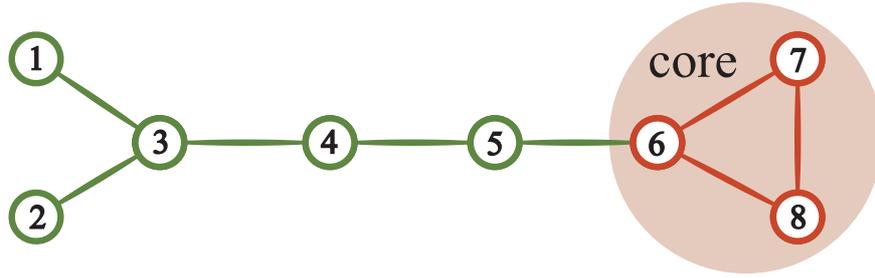}
\caption{(Color online) Removal categories of nodes in the core percolation. Red nodes are non-removable, \textit{i.e.}, they belong to the core indicated by shaded background. Green nodes are removable: nodes $1$ and $2$ are $\alpha$ removable; nodes $3$ and $5$ are $\beta$ removable; node $4$ is $\gamma$ removable.}  \label{fig3-glr}
\end{figure}

Considering the branching process, the infinite cluster (core) of a tree-like network can also be obtained exactly \cite{Liu2012a}. For that, nodes are divided into four categories, see Fig.\ref{fig3-glr}: (i) $\alpha$ removable: nodes that can become isolated without directly removing themselves, \textit{i.e.}, all neighbors are $\beta$ removable; (ii) $\beta$ removable: nodes that can become a neighbor of a leaf, \textit{i.e.}, at least one neighbor is $\alpha$ removable; (iii) $\gamma$ removable: nodes that can become leaves but are neither $\alpha$ nor $\beta$ removable; (iv) non-removable: nodes that cannot be removed and hence belong to the core. The rule for $\gamma$ removable nodes can be ignored because it is not useful to determine the size of the core. Denoting $\alpha$ (or $\beta$) as the probability that a random neighbor of a random node in a network is $\alpha$ removable (or $\beta$ removable), two self-consistent equations can be derived,
\begin{align}
\alpha & =G_1(\beta),  \\
\beta & =1-G_1(1-\alpha),
\end{align}
where $G_1(x)=\sum_kp_kkx^{k-1}/\langle k\rangle$ is the generating function of the excess-degree distribution. From the two equations, it can be found that $\alpha$ satisfies $\alpha=G_1[1-G_1(\alpha)]$. With this, we can find the solutions of $\alpha$ and $\beta$. For tree-like networks, if the core exists, it must have infinite size (a giant cluster). Hence, the size of the giant cluster, \textit{i.e.}, the normalized core size, can be expressed as
\begin{align}
S_{core} &=\sum^{\infty}_{k=0}p_k\sum^{k}_{s=2} \binom{k}{s} \beta^{k-s} (1-\beta-\alpha)^{s} \nonumber \\
&= G_0(1-\alpha)-G_0(\beta)-c(1-\beta-\alpha)\alpha,
\end{align}
where $G_0(x)=\sum_k p_kx^{k}$ is the generating function of degree distribution. This approach can also be generalized for directed networks with given in- and out-degree distributions \cite{Liu2012a}.

The theoretical results show that, for undirected networks, if the core percolation occurs, then it is always continuous, while for directed networks, it becomes discontinuous with a hyperscaling $\beta=1/2$ when the in- and out-degree distributions are different \cite{Liu2012a}. For ER networks the percolation threshold can be solved theoretically $\langle k\rangle_c=2.7182818\ldots$ \cite{Bauer2001}, and the core of purely SF networks never percolates for any degree exponents larger than $2$. While for the SF networks generated by the static model with $p_k\propto k^{-\gamma}$ only for large $k$ \cite{Catanzaro2005}, the core develops when the average degree is larger than a threshold value, which is actually similar to ER networks.

The leaf removal procedure of the core percolation can also be generalized as that nodes of degree smaller than $k$, together with their nearest neighbors and all incident links are progressively removed from a random network \cite{Azimi-Tafreshi2019}. With this pruning process, the percolation transition can also be well defined, which can be seen as either a generalized core percolation or a generalized $k$-core percolation. Similar to the ordinary $k$-core percolation, this model also shows a discontinuous phase transition for $k\geq3$.

\paragraph{Greedy articulation points removal}

The articulation points of a network are the nodes whose removal disconnects the network, such as nodes $3$, $4$, $5$ and $6$ in Fig.\ref{fig3-glr}. Those nodes play an important role in keeping the connectivity of real-world networks, such as infrastructure networks, protein interaction networks and terrorist communication networks. Tian \textit{et al.} proposed a greedy articulation point removal (GAPR) process to study the organizational principles of complex networks \cite{Tian2017}.

In each iterative step of GAPR, all the articulation points are removed from the network and their sub-clusters disconnected from the giant cluster are also removed. After that new articulation points emerge. This removal process is repeated until there is no articulation point left in the network. Note that at each step all the articulation points are removed simultaneously and the size of the giant cluster in the final network is studied as a key quantity.

\begin{figure}
\centering
\includegraphics[width=0.9\columnwidth]{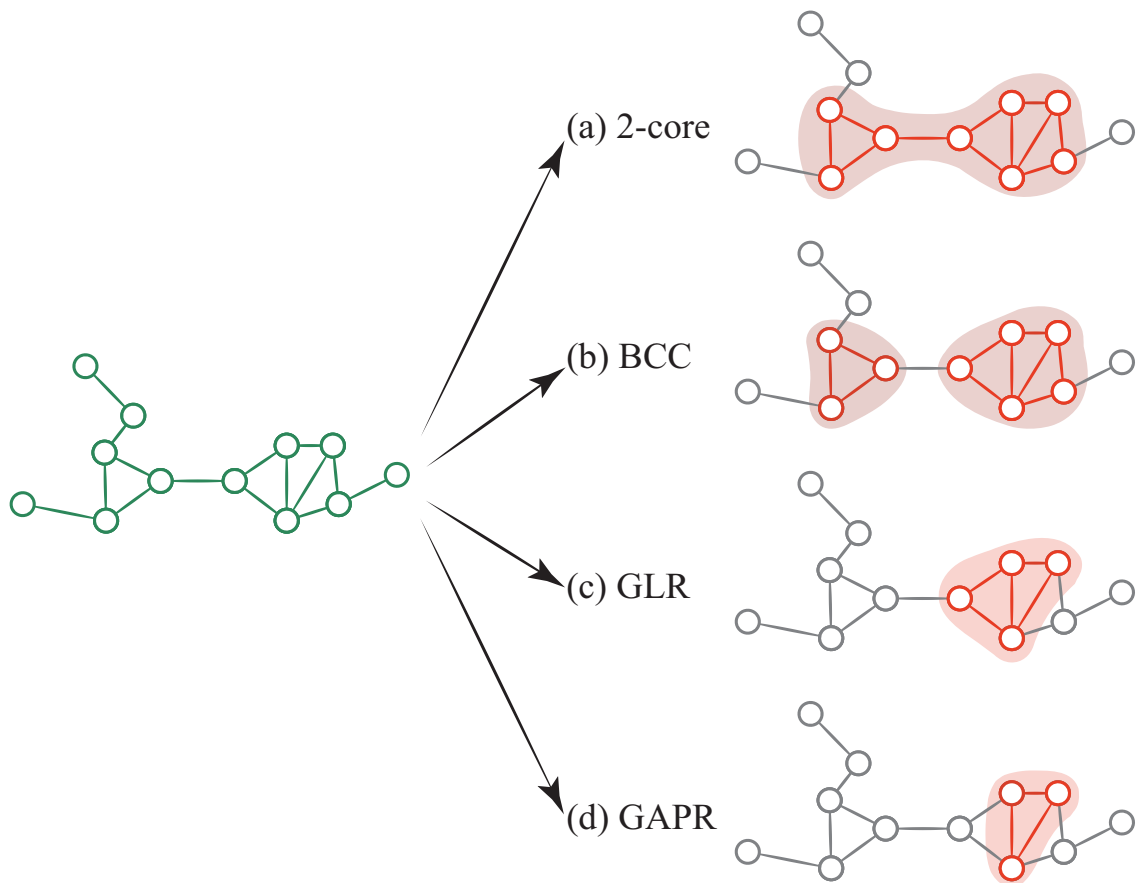}
\caption{(Color online) A comparative sketch of the $2$-core, the biconnected cluster (BCC), the cluster obtained by the greedy leaf removal (GLR), and the one obtained by the greedy articulation points removal (GAPR). For a given network (shown left), the four figures on the right show the corresponding subnetworks (indicated by red) under the four pruning rules, respectively. (a) $2$-core. The cluster is obtained by iteratively removing nodes with degree $1$. (b) BCC. The nodes in a BCC must connect each other with more than one independent paths, which can be found by removing all the nodes and links that do not belong to any loop structure. Here, two separated BBCs are obtained. (c) GLR. The cluster is obtained by iteratively removing nodes of degree $1$ along with their direct neighbors. (d) GAPR. The cluster is obtained by iteratively removing the articulation points, which are the nodes that can bridge two clusters. After these pruning/connection processes, there generally exist more than one clusters. The corresponding percolation theory considers the emergence of such a giant cluster.}  \label{fig3-kbla}
\end{figure}

It is obvious that if the network is sparse, the giant cluster will finally be destroyed with the proceeding of GAPR. For a dense network, a giant cluster, in which all the nodes connect to each other with at least two paths, can be obtained. Note that this cluster is not equivalent to the $2$-core, or the biconnected cluster (BCC) \cite{West1996,Holm2001,Newman2008,Kim2013,Schawe2019}, or the one obtained by GLR. A comparative sketch of the $2$-core, the BBC, the cluster preserved after the GLR, and the one after the GAPR is shown in Fig.\ref{fig3-kbla}.

Next, we briefly cover the relations between these clusters. For convenience, let $S_{2-core}$, $S_{BBC}$, $S_{GLR}$, and $S_{GAPR}$ be the subnetworks after applying the pruning rules of $2$-core, BBC, GLR, and GAPR, respectively. For $2$-core, one only needs to recursively remove the nodes with degree $1$, thus some tree-like structures might exist in $2$-core. The nodes in a BCC must connect each other with more than one independent paths. This further requires removing the nodes and links that do not belong to any loops from the $2$-core, therefore, $S_{BBC}\subseteq S_{2-core}$. For GLR, when we remove the nodes with degree $1$, their root nodes will also be removed, thus we also have $S_{GLR}\subseteq S_{2-core}$.

However, the relation of $S_{BBC}$ and $S_{GLR}$ is network dependent. The case shown in Fig.\ref{fig3-kbla} gives an example that $S_{BBC}\supseteq S_{GLR}$. Considering a network configuration that two triangles are connected by a degree-$2$ node, we can also have $S_{BBC}\subseteq S_{GLR}$. But both $S_{BBC}$ and $S_{GLR}$ are larger than $S_{GAPR}$. This is because all the nodes and links that do not belong to $S_{BBC}$ and $S_{GLR}$ are certainly not in $S_{GAPR}$, whereas the articulation points excluded from $S_{GAPR}$ might contribute to $S_{BBC}$ or $S_{GLR}$, see Fig.\ref{fig3-kbla}. In summary, for a given network, we have the relation $S_{GAPR}\subseteq S_{BBC} (S_{GLR}) \subseteq S_{2-core}$.

An additional consequence of the above relation is that the emergence of the giant cluster after GAPR needs a denser initial configuration than that after GLR. For ER networks, it has been found that the critical point of the percolation induced by GAPR is $\langle k\rangle_c=3.39807\ldots$ \cite{Tian2017}, which is obviously larger than the one induced by GLR (core percolation) \cite{Liu2012a}, see Tab.\ref{tb2-cpp} for the percolation thresholds of some models on ER networks. This percolation is also proved to be a hybrid percolation transition with critical exponent $\beta=1/2$. Note that the core percolation is always continuous for undirected networks. Besides, if one only does a finite iteration of the GAPR, the percolation transition can also be observed, and will have the same nature of the ordinary percolation. This finding has the same manner of the so-called history-dependent percolation on multiplex networks \cite{Li2020}.

There is also a variation of GAPR: iteratively remove the most destructive articulation point that will cause the most nodes disconnected from the giant cluster of the current network. This is called articulation point targeted attack strategy \cite{Tian2017}. Given a limited ``budget'' (that is, the number of nodes to be removed), this strategy is very efficient in reducing the giant cluster, compared with strategies based on other node centrality measures, such as degree \cite{Cohen2000,Albert2000} and collective influence \cite{Morone2015}.

The statistical properties of articulation points, \textit{i.e.}, the probability that a random node in a network is an articulation point, has been also derived in Ref.\cite{Tishby2018a}. It has been found that high-degree nodes are more likely to be articulation points than low-degree nodes. Articulation point is related to another concept of ``bridge'', which is a link whose removal disconnects the network and increases the number of connected clusters \cite{Wu2018b}. Both articulation point and bridge can be used to qualify the importance of a bridge structure in damaging a network and provide useful ideas to destroy a network efficiently.

\paragraph{Color-avoiding percolation}

In the color-avoiding percolation, the nodes in the percolating cluster also need more than one path that connect with each other. The details are as follows. First, each link is assigned a color chosen from $\mathbf{C} =\{c_1, c_2, \ldots, c_n\}$. If there exists a path between two nodes after the removal of all the links of color $c \in \mathbf{C}$, the two nodes are called $c$-avoiding connected. Then, if two nodes are $c$-avoiding connected for all the color $c$, we say that the two nodes are color-avoiding connected. Thus, the color-avoiding percolation studies the behaviors of the giant cluster, in which all the nodes are color-avoiding connected. The color avoiding can also be generalized as the case where arbitrary sets of colors are avoided. Furthermore, colors can also be assigned on nodes, that is the setting when the model was first proposed \cite{Krause2016}.

\begin{figure}
\centering
\includegraphics[width=0.8\columnwidth]{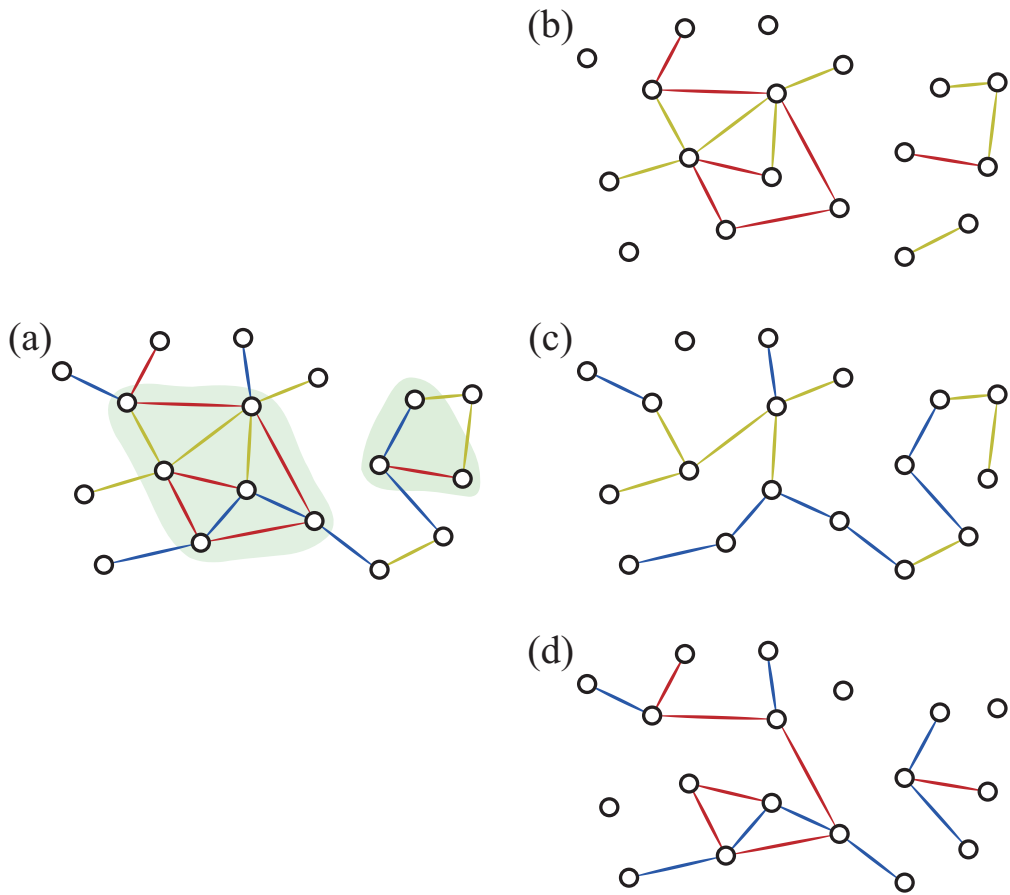}
\caption{(Color online) A schematic for color-avoiding percolation. Here, three colors, namely, red, yellow, and blue, are uniformly distributed among all the links, see (a). If there exists a path between two nodes after the removal of all the blue/red/yellow links, the two nodes are called blue/red/yellow-avoiding connected. Then, if two nodes meet this connection criterion for all the three colors, we say that the two nodes are color-avoiding connected. A cluster, whose nodes are all color-avoiding connected, is thus the color-avoiding connected cluster. (b)-(d) show the clusters after the removal of blue, red, and yellow links, respectively. The intersections of these clusters correspond to the color-avoiding connected clusters indicated by green shades in (a). Note that the color-avoiding connections of the nodes in a color-avoiding cluster could access through some links outside the color-avoiding cluster.}  \label{fig3-cap}
\end{figure}

Different from the previous percolation models, the color-avoiding connected clusters cannot be simply acquired from a single pruning process. Instead, one can find $c$-avoiding connected clusters for all the color $c \in \mathbf{C}$, then the intersections of these clusters are just the color-avoiding connected clusters, see Fig.\ref{fig3-cap} for a schematic of a three-color case. Due to this intersection process, the connectivity of a color-avoiding cluster could access through some nodes/links outside the color-avoiding cluster.

In Ref.\cite{Molontay2019}, it is suggested that the complexity of finding the color-avoiding cluster highly depends on the exact definition, using a strong version the problem is NP-hard, while using a weaker notion makes it possible to find the components in polynomial time. In spite of this, there are some special structures that surely do not contribute to the color-avoiding connected clusters. The first is the bridge link, whose removal disconnects the network. The second is the nodes that all their links have identical colors. A typical example for these two structures is the nodes of degree $1$ and their links, \textit{i.e.}, the leaves of the network.

Besides, the size of $\mathbf{C}$, \textit{i.e.}, the number of colors $n$, also has significant restrictions on the structural features of the color-avoiding connected clusters. First, a small $n$ generally results in a large frequency for each color, thus the emergence of color-avoiding clusters for small $n$ requires a dense connection. Second, the size of the color-avoiding connected cluster has a lower bound, which increases with the color number $n$. All these suggest that the color-avoiding connected cluster must have a relatively dense and complex structure, containing many loops.

If there is only one color $n=1$, the model reduces to a classical percolation with respect to the fraction of colored links/nodes. For $n\geq2$, the critical behaviors have a rich, multifaceted nature, depending on the network topology, the number of avoided colors and the color frequencies \cite{Krause2016,Krause2017,Kadovic2018,Molontay2019}. If all the colors have identical probabilities, the results on ER networks show that the critical exponent $\beta$ is dependent on the number of colors, \textit{i.e.}, $\beta=n$. This is identical for both node and link coloring. Varying the color frequencies, fractal exponents can also be observed. However, topological constraints of SF networks are so strong that the corresponding critical behaviors are independent of the number of the colors (only dependent on the exponent $\lambda$), but still dependent on the existence of the colors and therefore different from standard percolation. Due to the vanished percolation threshold, the breaking of the universality class of the site percolation and the bond percolation can also be observed for $2<\lambda<3$. In Ref.\cite{Kryven2019} a generic analytic theory that describes how structure and sizes of all connected components in the network are affected by simple and color-dependent bond percolation was also established.

\subsection{Percolation on interdependent/multiplex networks}  \label{sct3-pin}

\begin{figure}
\centering
\includegraphics[width=0.8\columnwidth]{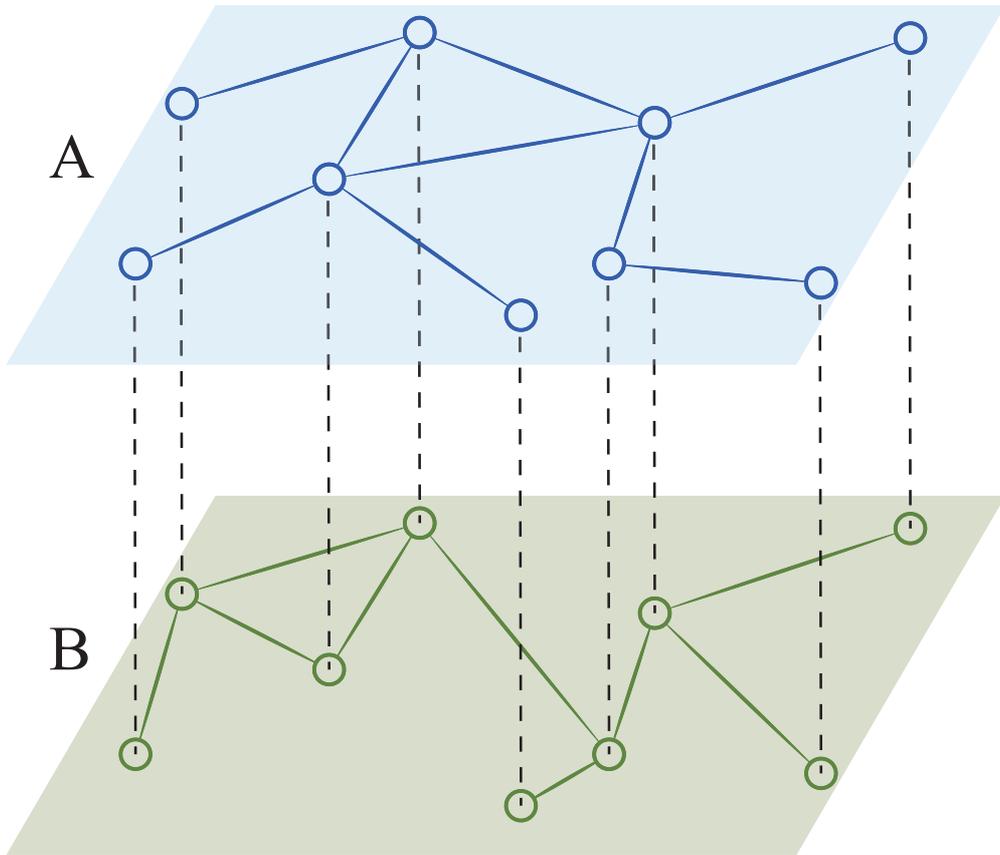}
\caption{(Color online) Schematic of the interdependent/multiplex network. Here, the system has two network layers $A$ and $B$ indicated by different colors, both of which can have their own topology, and dependence links indicated by dashed lines are used to represent the interdependence between nodes from network layers $A$ and $B$. More generally, there can be more than two layers, and the interdependence relation needs not to be restricted to one-to-one.}  \label{fig3-mln}
\end{figure}

The hybrid percolation transition is not unique to the $k$-core percolation \cite{Araujo2014,Saberi2015,Cho2016,Lee2016b,Lee2017,Lee2018,Park2019,Park2020}. In the last decade a major concern in network science, the interdependent networks or multilayer networks \cite{Boccaletti2014,Bianconi2018}, also involves a dependent percolation, for which both abrupt change and critical exponents can be found at the critical point \cite{Buldyrev2010,Parshani2010,Baxter2012}. For convenience, the following discussions are given by means of interdependent networks \cite{Buldyrev2010}, see Fig.\ref{fig3-mln} for an example of the interdependent networks.

This model considers the iterative percolation process between different network layers. If a node does not belong to the giant cluster of one layer, its dependent nodes in other layers are also no longer eligible to be considered in the percolation of their layers, and vice versa. Obviously, one must check the percolation process of each layer interactively to obtain a steady giant cluster, referred to as the mutual giant cluster. Thus, this model checks the emergence of the mutual giant cluster after an initial node removal.

\subsubsection{Model and phase transition characteristics}

\begin{figure}
\centering
\includegraphics[width=0.8\columnwidth]{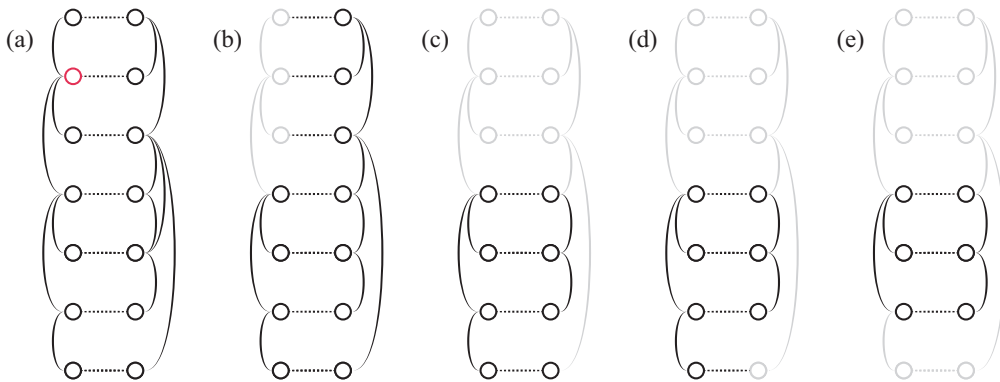}
\caption{(Color online) Schematic of the cascading process on interdependent networks. Here, we consider an interdependent network with two layers, represented by the two columns of circles with the corresponding solid lines, respectively. The dashed lines are the dependence links between the two network layers. With an initial node removal (indicated by red in (a)), the percolation process ((b) and (d)) and the dependence process ((c) and (e)) alternately pruning nodes from the system (indicated by gray) until no more nodes can be removed.}  \label{fig3-idn}
\end{figure}

To provide a direct and precise description for this percolation model, the dependence link is used to connect dependent nodes in different network layers, see Fig.\ref{fig3-mln}. Specifically, a dependence link between node $i$ and node $j$ means that if node $i$ is removed, node $j$ will also be removed, and vice versa. The node removal process caused by dependence links is often called dependence process. Correspondingly, all the finite percolation clusters in each layer will also be removed from the system, which is called percolation process. Thus, after an initial node removal (fraction $1-p$), the removal of nodes caused by the percolation process and the dependence process will occur alternately until no more nodes can be removed, see Fig.\ref{fig3-idn}. The remained nodes (if any) form the mutual giant cluster, in which nodes are reachable through each layer. Note that if we do not remove the finite clusters from the system, the finite mutual clusters can also be well defined in the steady state \cite{Grassberger2015,Li2020}.

The above pruning process to find the mutual giant cluster was first proposed in Ref.\cite{Buldyrev2010}, in which the dependence links are assigned randomly between two network layers, labeled $A$ and $B$ (see Fig.\ref{fig3-mln}). Assuming both the two network layers have tree-like structures, we can express the size of the mutual giant cluster as
\begin{equation}
S = p\left[1-G_0^A(1-R^A)\right]\left[1-G_0^B(1-R^B)\right].  \label{eq3-smn}
\end{equation}
Here, $R^A$($R^B$) represents the probability that a randomly chosen link in network layer $A$($B$) leads to the mutual giant cluster. By analogy with that of classical percolation equations (\ref{eq2-g0p}) and (\ref{eq2-g1p}), we can know that the right hand side of Eq.(\ref{eq3-smn}) just describes the final state of the pruning process, \textit{i.e.}, all the preserved nodes must be reachable in both layers. Similarly, $R^A$ and $R^B$ must satisfy
\begin{align}
R^A &= p\left[1-G_1^A(1-R^A)\right]\left[1-G_0^B(1-R^B)\right], \label{eq3-ramn}  \\
R^B &= p\left[1-G_0^A(1-R^A)\right]\left[1-G_1^B(1-R^B)\right]. \label{eq3-rbmn}
\end{align}
The above three equations (\ref{eq3-smn})-(\ref{eq3-rbmn}) were first used to analyze the percolation on the networks with dependence links by Son \textit{et al.} \cite{Son2011,Son2012}. Considering the iterative process of the pruning process, this model can also be solved. That is the analytical method when this model was proposed \cite{Buldyrev2010,Parshani2010}, which is equivalent to the method listed above \cite{Li2014,Feng2015}.

From Eqs.(\ref{eq3-ramn}) and (\ref{eq3-rbmn}), we can also find the critical point of the system as we do for the classical percolation. The solution of $R$ shows a similar behavior as $k$-core percolation illustrated in Fig.\ref{fig3-kcs}, indicating a discontinuous transition. The value of $p_c$ is significantly larger than that of classical percolation. For example, two interdependent ER networks give $p_c\approx 2.4554 /\langle k\rangle$, while $p_c=1/\langle k\rangle$ for single ER networks. This also means that $\langle k\rangle\approx2.4554$ is the minimum degree to observe a mutual giant cluster in two interdependent ER networks. Even for SF network layers with $2<\lambda<3$, the mutual giant cluster can emerge at a non-vanished $p_c$. Further study have showed that this discontinuous transition is also a hybrid transition with exponent $\beta=1/2$ \cite{Parshani2010,Baxter2012}, which is the same as the $k$-core percolation with $k\geq3$.

Parshani \textit{et al.} also pointed out that the number of iterations in the cascading process of interdependent networks diverges when $p\to p_c$ \cite{Parshani2010}, and thus can be used to identify the transition point in numerical simulations. This is mainly because there is a plateau in the collapse of the system, and the plateau regime increases with the system size. Zhou \textit{et al.} found that during the collapse there is also a random branching process at criticality, \textit{i.e.}, a continuous percolation transition. This simultaneously continuous phase transition is just the origin of the observed long plateau regime in the cascading failures and its dependence on system size \cite{Zhou2014}. This result coincides with the one found in Ref.\cite{Li2020}, in which the percolation transition is defined in each iteration, called generation. Theoretical analysis indicates that for any finite generation, the system demonstrates a continuous percolation transition. Monte Carlo simulations on ER networks further suggest that all these continuous transitions belong to the same universality class. Specifically, SF networks with exponent $2<\gamma<3$ have a vanished critical point for any finite generations. For infinite generation, it just recovers the percolation transition on interdependent networks, \textit{i.e.}, it has a nature of hyper transition. Hu \textit{et al.} further pointed that in two-dimensional systems the infinite generation still presents a continuous transition, but with a different universality class \cite{Hu2020}.

By decreasing the coupled strength $q$ (the fraction of nodes that have dependence links), this discontinuous percolation transition can also reduce to the continuous transition  \cite{Parshani2010}. An exception is the coupled lattices, for which the networks abruptly collapse for any finite $q>0$ \cite{Bashan2013}. But this does not always mean the continuous percolation cannot exist in interdependent spatially embedded networks. If the dependence links establishing the interdependence between two lattices are not random but have a typical length $r$, the percolation transition will be continuous for $r<r_{max}\approx8$ (lattice units), and discontinuous for larger $r$ \cite{Li2012}. For $r<r_{max}$, the percolation threshold increases linearly with $r$ from $0.593$ at $r=0$ and reaches a maximum, $0.738$ for $r=r_{max}$, and then gradually decreases to $0.683$ for $r=\infty$. This is mainly because the spatial embedding induces a similar structure in different network layers, and thus breaks the cascading picture of interdependence \cite{Son2011}. This model has also been generalized to high dimensions \cite{Lowinger2016}. The simulation results show that the value $r_{max}$ decreases with dimension, and for lattices with dimension larger than or equal to $6$, the continuous percolation transition disappears. This suggests that a high degree of freedom (high dimension) decreases the structural similarity of the spatially embedded networks under random node or link dilution, thus reconstructs the cascading picture.

It should be noted that the model, based on which the authors claimed that percolation transitions are not always sharpened by making networks interdependent \cite{Son2011}, is a bond percolation model on the interdependent square lattice with typical length $r=0$, and not the one discussed above. Thus their findings do not contradict that of Ref.\cite{Li2012}, which considers the site percolation with different typical lengths. Grassberger also studied the bond percolation model with typical length $r=0$ in high dimension lattices \cite{Grassberger2015}. The simulation results have showed that there is an upper critical dimension $d=4$, below which the percolation transition is always continuous, and at least for $d\neq3$ -- not in the universality class of ordinary percolation. This suggests that although the similar structure induced by spatial embedding breaks the cascading picture of interdependence, it indeed changes the nature of the branching process in the continuous percolation transition. The bond percolation has also been studied on multiplex networks, and reported a multiple percolation phase transition \cite{Hackett2016}. Here, the multiple percolation transition refers to more than one transition in a percolation process, also is reported in other related percolation models \cite{Nogawa2009,Bianconi2014,Wu2014,Liu2018,Kryven2019a,Oh2019}.

In Ref.\cite{Grassberger2015}, an efficient and implementable algorithm has also been proposed, based on which some other critical behaviors of the percolation on interdependent networks have been rechecked, numerically. The result suggests that for the model of Ref.\cite{Li2012} there exists such a $r_{max}$ above which the transition seems to be discontinuous, but a strong evidence that this is related to very large finite-size corrections was also found. Moreover, this work also provided some evidences that the cluster statistics of independent ER networks can be exactly described by mean-field theory, while the cascade process cannot. For interdependent lattices with dimension $d=5$ a discontinuous transition was observed as ER networks, but a nontrivial distribution of finite clusters also exists at the transition point.

Besides, some other percolation model can also be generalized to multiplex networks, such as $k$-core percolation \cite{Azimi-Tafreshi2014}, bootstrap percolation \cite{Baxter2014}, and group percolation \cite{Wang2018a}.

\subsubsection{Algorithms for reducing time complexity}

For facilitating the studies on non-tree-like networks, some efficient algorithms have been developed to find the mutual giant cluster. Tracing the largest cluster during the whole pruning process, the algorithm proposed by Schneider \textit{et al.} can achieve time complexity $O(N\log N)$ \cite{Schneider2013}. Nevertheless, this processing is too loose to study the critical phenomena, since an initially large cluster could be taken over by a cluster that had started smaller after the pruning process, especially near the critical point. Apart from the critical phenomena or limited to tree-like networks, this algorithm might be a good choice to avoid brute-force searching.

Although the finite mutual clusters are ignored in the original definition of this percolation model \cite{Buldyrev2010,Parshani2010}, all mutual clusters must be taken into count for providing a more reliable simulation. A viable way to define such clusters can be found in Refs.\cite{Grassberger2015,Li2020}, in which the percolation transition can be defined in each generation. Based on a fully dynamic graph algorithm \cite{Holm2001}, Hwang \textit{et al.} proposed an efficient algorithm designed to proceed as links are deleted, whose time complexity is approximately $O(N^{1.2})$ for random networks \cite{Hwang2015}. With this algorithm, large scale simulations for ER and $2$-dimensional interdependent networks have been carried out, which verifies the exponents for the order parameter and its fluctuations numerically, as well as the finite avalanches \cite{Lee2016a}. Stippinger \textit{et al.} later generalized this algorithm to efficiently simulate interdependent networks with healing \cite{Stippinger2018}.

Besides, Grassberger proposed an easily implemented algorithm to find all the mutual clusters by mapping the problem onto a solid-on-solid model \cite{Grassberger2015}. This method is simply done by alternately performing Leath-like processes on different network layers \cite{Leath1976}, each is referred to as a ``wave" from the starting node. Waves are confined to the nodes that have the same ``height" as the starting node. All the nodes' heights are initially set as $0$, and after the spreading of a wave, the heights of the involved nodes increase by $1$. Do the searching process for all the possible starting nodes, a landscape of node heights can be obtained, from which the mutual giant cluster and the distribution of mutual cluster sizes can be found. With some techniques this algorithm can also simulate the percolation process on interdependent networks, effectively. In Ref.\cite{Stippinger2018}, it is also pointed out that compared to the algorithm of Ref.\cite{Grassberger2015} their algorithm is generally intended for fast updates at the cost of higher memory use.

\subsubsection{Variants and related models}

With the growing trend of multiplex networks, this percolation model has also been employed frequently to figure out the robustness of multiplex networks \cite{Boccaletti2014,Kivelae2014,Li2014,Lee2015,Bianconi2018}, which can be generally classified into three categories, single networks with dependence links, networks of networks, and multiplex networks with general and special dependence. Most of these studies demonstrate a hybrid transition as shown above, and a crossover between the continuous and discontinuous percolation transition can also be found in some of them. Besides, the correlation between different network layers can also be represented by interconnections, in which the percolation transition can be also observed \cite{Leicht2009}.

More importantly, these works show the robustness of multiplex networks in various environments, and are meaningful for understanding the structures and the robustness of multiplex networks. More results will be reviewed later when we turn our thoughts to the applications of percolation model. Of course, one can find more information in the special papers and books on multiplex networks \cite{Boccaletti2014,Kivelae2014,Li2014,Lee2015,Bianconi2018}.

\subsection{Explosive percolation}

The explosive percolation was proposed by Achlioptas \emph{et al.} in 2009 \cite{Achlioptas2009}, hence this type of percolation process is named as Achlioptas process. In general, Achlioptas process is a competitive process that can be realized by many rules, and the key is suppressing the emergence of a giant cluster. There is a specialized review article for this issue \cite{Boccaletti2016}, here we only give a brief introduction of the main findings.

\subsubsection{Achlioptas process}

\begin{figure}
\centering
\includegraphics[width=0.9\columnwidth]{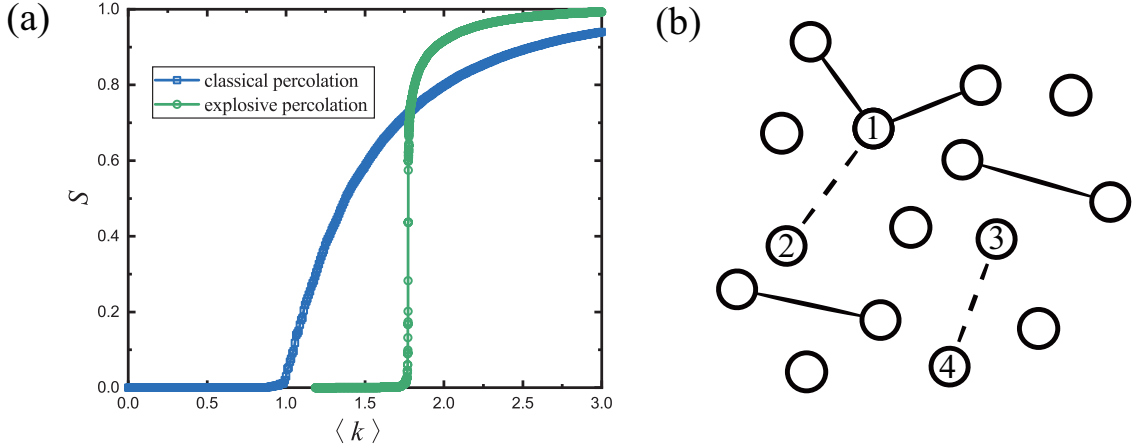}
\caption{(Color online) (a) Illustrations for the classical percolation and the explosive percolation on ER networks. Here the explosive percolation is manufactured by the best-of-$2$ rule. (b) Schematic of the best-of-$2$ rule. At each time step, $2$ potential links indicated by dashed lines are chosen randomly from all the possible ones. According to the sum/product rule, here the potential link $l_{34}$ will be inserted, and link $l_{12}$ is discarded.}  \label{fig3-ep}
\end{figure}

As we know, ER networks can be generated by randomly inserting links into a set of nodes one by one. The Achlioptas process is just a variation of this process \cite{Achlioptas2009}. Assuming there are initially $N$ nodes and no links, at each time step, $m$ potential links are arbitrarily chosen, each of which is supposed to bridge two distinct clusters with size $s_1$ and $s_2$, respectively. (The two nodes can belong to the same cluster with no impact on the following process). Then, according to a function $f(s_1,s_2)$, the potential link with the smallest $f(s_1,s_2)$ is inserted eventually, and other potential links are discarded, see Fig.\ref{fig3-ep} (b) as an example. Repeating this process until the expected number of links is met. Such a selection mechanism is often called the best-of-$m$ rule or the min-cluster-$m$ rule \cite{Friedman2009}, and the function $f(s_1,s_2)$ represents the competitive rule.

In order to suppress the emergence of the giant cluster, the competitive rule $f(s_1,s_2)$ must be of positive correlation with $s_1$ and $s_2$. Two commonly used ones are the product rule $f(s_1,s_2)=s_1s_2$ and the sum rule $f(s_1,s_2)=s_1+s_2$. For $m=1$, this model reduces to the ER network model, which percolates when $N/2$ links have been inserted, \textit{i.e.}, $\langle k\rangle=1$. For $m\geq2$, the emergence of a giant cluster will be suppressed, resulting in an explosive percolation. A comparison of the two cases is shown in Fig.\ref{fig3-ep} (a). One can find that the explosive percolation shows a sharper transition with a larger threshold.

Since Ref.\cite{Achlioptas2009}, many other generalized Achlioptas processes have been proposed and investigated. A typical example is the $l$-node ($l$-vertex) model. In this model $l$ nodes are chosen randomly at each step, and two of them are connected according to a selection criteria, for example, the size difference \cite{Nagler2012}. The original Achlioptas process just corresponds to $l=4$, and the case $l=3$ with different selection criteria is used in many studies \cite{DSouza2010,Nagler2011,Riordan2012}.

Besides, there are some other models can realize the Achlioptas process, such as BFW model \cite{Bohman2004,Riordan2011,Chen2011,Chen2012,Schrenk2012,Zhang2012,Chen2013,Chen2014}, dCDGM model \cite{Costa2010,Cho2011,Li2012a,Yi2013,Costa2014,Costa2014a,Costa2015,Costa2015a}, Hamiltonian model \cite{Moreira2010}, diffusion-limited cluster aggregation model \cite{Cho2011a,Cho2012}, spanning cluster-avoiding model \cite{Ziff2013,Cho2013}, and hybrid model \cite{Araujo2011,Cao2012,Fan2012}. One can find a contrast analysis of these models in Ref.\cite{DSouza2015}. Most of these models can be generalized to any networks by constraining the potential links in a given configuration. One can refer to Ref.\cite{Boccaletti2016} for more information.

\subsubsection{Phase transition characteristics}

For a deep understanding of the explosive percolation, one can consider a complete network with size $N$ \cite{Friedman2009,Nagler2011}. In each step $t$, all the unconnected node pairs provide potential links, \textit{i.e.}, $m=N(N-1)/2-t+1$. In this case, up to $N/2$ new links are added, only connect isolated nodes to result in clusters with size $2$ such that the maximum cluster size remains $2$ for the first $N/2$ steps. Each of the subsequent $N/4$ links connects clusters of size $2$ to clusters of size $4$ keeping the maximum cluster size $4$ until step $3N/4$. As an analogy, in step $N−1$, the remaining two clusters of size $N/2$ join. The jump of the largest cluster is thus clearer, indicating a discontinuous transition \cite{Riordan2011,Waagen2014}. In general, one can connect a link chosen from all the possible positions at each step, thus the jump of the largest cluster is excepted in any network configurations. By the way, such successive jumps of the largest cluster are named microtransition cascades \cite{Chen2014}, which is probably used to warn signals anticipating phase transitions in complex systems.

A small $m$ will relax the suppression for the emergence of the giant cluster, thus the jump of the largest cluster is blurred. Although a sharp transition might also be observed as shown in Fig.\ref{fig3-ep}, Riordan \emph{et al.} pointed out that any rule based on picking a fixed number of random nodes gives a continuous transition \cite{Riordan2011}. They also suggested a lower end of the range: Whenever $m\to\infty$ as $N\to\infty$, the Achlioptas process exhibits explosive percolation. The findings of da Costa \emph{et al.} show that the Achlioptas process is irreversible, which also indirectly suggests a continuous transition \cite{Costa2010}. However, these findings do not mean that the Achlioptas process behaves exactly like a continuous percolation. The discontinuities can also be observed in the Achlioptas process \cite{Nagler2012}. The double humped distribution of the sizes of the largest cluster and the hysteresis, which are both recognized as the sign of discontinuous transition, can be also found in finite systems \cite{Grassberger2011,Bastas2011}. These works also reported a finite-size behavior with non-analytic scaling functions, and demonstrated that the explosive percolation transitions are indeed continuous but with some unusual scaling properties \cite{Costa2010,Cho2010,Radicchi2010,Chen2011,Grassberger2011,Lee2011,Tian2012,DSouza2015}. Riordan \emph{et al.} also pointed out that the fluctuation of the order parameter in explosive percolation does not disappear in the thermodynamic limit \cite{Riordan2012}.

\subsection{Percolation transition during the growth of networks}      \label{sct3-ptng}

In contrast to the network models with a fixed size, such as the configuration model and ER network model, there is a family of network models with growing size featuring the growing property of real networks \cite{Albert2002,Dorogovtsev2002a,Newman2003b}. A typical example is the famous BA networks \cite{Barabasi1999}. However, there is always only one single cluster in the growing process, thus the system always percolates. To observe a percolation transition in the growth of a network, it must be allowed that new nodes enter the system without connecting to the existing nodes. In this way, the network could contain isolated nodes and finite clusters.

Instead of looking at the network configuration, one can solve the growing network model by checking the evolution process via the master equation. The key point is featuring the increment and decrement of the parameters in one time step, which might vary from model to model \cite{Barabasi1999a,Callaway2001,Dorogovtsev2001a}.

\subsubsection{Growing random network}

A simple rule to achieve a percolation transition in network growing is the one proposed in Ref.\cite{Callaway2001}, called growing random network. This model begins with no node, and at each time step a new node is inserted into the system, then two nodes are chosen randomly from all the existing nodes and joined by a link with probability $p$.

At time step $t$, there are $N(t)=t$ nodes in the system. Let $p_k(t)$ be the degree distribution at time step $t$, then a rate equation for the evolution of the degree distribution $p_k(t)$ can be established,
\begin{equation}
N(t+1) p_k(t+1) -N(t) p_k(t) =2pp_{k-1}(t) -2pp_{k}(t).  \label{eq3-pkt}
\end{equation}
The two terms on the right hand side of this equation correspond to the expected increment and decrement of nodes with degree $k$ in one time step, respectively. For $t\to\infty$, $p_k(t+1)=p_k(t)\equiv p_k$, and Eq.(\ref{eq3-pkt}) reduces to a recursion formula $p_k/p_{k-1}=2p/(1+2p)$. Note that Eq.(\ref{eq3-pkt}) holds only for $k>0$, and for $k=0$ it is specialized as
\begin{equation}
N(t+1) p_0(t+1) -N(t) p_0(t) =1 -2pp_{0}(t),  \label{eq3-p0t}
\end{equation}
which gives $p_0=1/(1+2p)$ for $t\to\infty$. Consequently, the degree distribution has a closed form for $t\to\infty$,
\begin{equation}
p_k=\frac{1}{1+2p}\left(\frac{2p}{1+2p}\right)^{k}.   \label{eq3-pkrgn}
\end{equation}
Obviously, the degree distribution $p_k$ decreases with the increase of $k$. In fact, the power-law degree distribution can also be produced in this model by introducing preferential attachment  \cite{Dorogovtsev2001a,Weaver2015}.

From Eq.(\ref{eq3-pkrgn}), the generating functions are also available, those are $G_0(x)=1/(1+2p-2px)$ and $G_1(x)=1/(1+2p-2px)^2$. Substituting $G_1(x)$ into the Molloy-Reed criterion Eq.(\ref{eq2-mrc}), we can find the percolation threshold $p_c=1/4$. This result is for an uncorrelated tree-like network with degree distribution Eq.(\ref{eq3-pkrgn}), whereas the growing random network has degree correlation. Specifically, earlier nodes are sure to form a core, in which there is a higher average degree. This indicates that the giant cluster forms more readily than in a network whose links are uniformly distributed.

To find the percolation threshold of the growing random network, we need to examine the cluster size distribution \cite{Callaway2001}. Let $n_s(t)$ be the normalized cluster number at time step $t$, \textit{i.e.}, the ratio of the number of clusters with size $s$ to the total number of nodes, then a rate equation for the evolution of $n_s(t)$ can be expressed as
\begin{align}
N(t+1) n_1(t+1)-N(t) n_1(t) &= 1 -2pn_1(t),   \label{eq3-n1t}  \\
N(t+1) n_s(t+1)-N(t) n_s(t) &= p\sum_{i=1}^{s-1}in_i(t)(s-i)n_{s-i}(t) -2psn_s(t), ~~~s\geq2.  \label{eq3-nst}
\end{align}
Note that $sn_s$ can be interpreted as the size of the cluster that a randomly chosen node belongs to. Therefore, $p\sum_{i=1}^{s-1}in_i(t)(s-i)n_{s-i}(t)$ is the increment of clusters with size $s\geq2$ by combining two small clusters in one time step, while for $s=1$ the increment just is the newly added node; $2psn_s(t)$ is the decrement of clusters with size $s$ in one time step by combining two clusters that at least one of them has size $s$.

For $t\to\infty$, $n_s(t+1)=n_s(t)\equiv n_s$, then Eqs.(\ref{eq3-n1t}) and (\ref{eq3-nst}) can be simplified to
\begin{align}
n_1 &=1- 2pn_1,    \label{eq3-ns1} \\
n_s &=p\sum_{i=1}^{s-1}i(s-i)n_in_{s-i} -2psn_s, ~~~ s\geq2.   \label{eq3-ns2}
\end{align}
Although the two equations cannot give a closed form of $n_s$, we can find a recursive function for the generating function $H_0(x)=\sum_s\pi_sx^s=\sum_ssn_sx^s$ by multiplying both sides of Eqs.(\ref{eq3-ns1}) and (\ref{eq3-ns2}) by $sx^s$ and then summing over $s$. The result reads
\begin{align}
H_0(x) &=\sum_{s=1}sn_sx^s   \nonumber \\
& = x -2pn_1x +p\sum_{s=1}\sum_{i=2}^{s-1}i(s-i)n_in_{s-i} sx^s   -2p\sum_{s=2}s^2n_sx^s  \nonumber    \\
& = x +px\frac{d}{dx}\left( \sum_{s=1} sn_s x^s \right)^2 -2px\frac{d}{dx}\sum_{s=1}sn_sx^s  \nonumber \\
& = x +2pxH_0(x)H_0'(x)-2pxH_0'(x),
\end{align}
where $\sum_na_nx^n\sum_nb_nx^n=\sum_n\sum_ma_mb_{n-m}x^n$ is used. Now, we have
\begin{equation}
H_0'(x) =\frac{H_0(x)-x}{2px\left[H_0(x)-1\right]}. \label{eq3-h0x}
\end{equation}

Note that the mean cluster size $\chi=H_0'(1)/H_0(1)\propto H_0'(1)$, then the critical behavior of $\chi$ can be derived from Eq.(\ref{eq3-h0x}). In the supercritical phase, the giant cluster exists, thus $H_0(1)<1$, Eq.(\ref{eq3-h0x}) gives the mean cluster size $\chi=H_0'(1)/H_0(1)=1/2pH_0(1)$. For the subcritical phase, all the clusters have finite sizes, \textit{i.e.}, $H_0(1)=1$, then $H_0'(1)$ can be solved by letting $x\to1$ in Eq.(\ref{eq3-h0x}), that is
\begin{equation}
H_0'(1) =\frac{1-\sqrt{1-8p}}{4p}.
\end{equation}
For a physical meaning, $p\leq 1/8$ is required here, which gives the percolation threshold $p_c=1/8$. This threshold is smaller than that for the configuration network with the same degree distribution, confirming that the growing mechanism facilitates the forming of the giant cluster, theoretically \cite{Callaway2001}.

As just described, the critical behavior of $\chi$ can be summarized as
\begin{equation}
\chi \propto H_0'(1) =\left\{
                          \begin{array}{lr}
                            \frac{1-\sqrt{1-8p}}{4p}, &  p\leq\frac{1}{8},\\
                            \frac{1}{2p},             & p>\frac{1}{8}.
                          \end{array}
                       \right.
\end{equation}
This means that the mean cluster size jumps discontinuously at the percolation threshold, whereas diverges for the standard percolation.

Moreover, compared with the static network, the existence of the core makes the size of the giant cluster increases more slowly in the supercritical phase. The simulation results imply that the size of the giant cluster obeys $S\propto e^{\alpha (p-p_c)^{-\beta}}$ with $\beta=1/2$, and suggest an infinite order phase transition \cite{Callaway2001,Kim2002,Bollobas2004,Krapivsky2004}, like the Berezinskii-Kosterlitz-Thouless phase transition in the condensed matter \cite{Berezinskif1971,Kosterlitz1972}. In addition, such a phase transition was also found on the substrate formed by XY spin configurations \cite{Hu2011}. Some simulation verification for these findings and the fractal geometry can be found in Refs.\cite{Hasegawa2013c,Wu2017}. In a growing network with a renormalization group treatment, Dorogovtsev also found that the critical behavior of percolation on growing networks differs from that in uncorrelated networks \cite{Dorogovtsev2003a}.

\subsubsection{Variants and related models}

Similar results can also be found in a general model with preferential attachment \cite{Dorogovtsev2001a}, \textit{i.e.}, the link between nodes $i$ and $j$ is inserted with a probability proportional to $(k_i+a)(k_j+a)$, where $k_i$ ($k_j$) is the degree of node $i$ ($j$), and $a$ is a positive constant which plays the role of additional attractiveness of node for new links \cite{Dorogovtsev2000}. The power-law form of the cluster size distribution in standard percolation has a decay for large sizes both above and below the percolation threshold. However, in the growing network the power-law form can be observed in the entire phase without the giant cluster. This indicates that the system is in the critical state for the entire phase.

Instead of connecting nodes randomly, other typical mechanisms for generating networks can also be incorporated into the growing network model, such as BA network model \cite{Zen2007} and the explosive percolation model \cite{Vijayaraghavan2013,Yi2013,Oh2016,Oh2019}. In these models, when a new node enters the system, with probability $p$ links are inserted following the given rule, otherwise, do nothing. When the rule of BA network model is applied, the percolation transition is still of infinite order \cite{Zen2007}. This mainly because the rule of BA network model does not break the picture of the forming of the giant cluster, which is the crucial factor for the nature of infinite order transition \cite{Dorogovtsev2001a}.

Conversely, the Achlioptas process induces a suppression effect against the growth of the large clusters. Thus, when the Achlioptas process is used, the percolation of growing network can revert to a nature of the second order transition with scaling behaviors depending on the specific rules \cite{Yi2013,Oh2016}. It should be pointed out that if there are not enough potential nodes/links in the Achlioptas process, the growth mechanism will still dominate the nature of the percolation transition, hence an infinite order transition \cite{Vijayaraghavan2013}; rather, if the Achlioptas process is done with global information, the growth of large clusters can be completely suppressed, so that the continuous percolation transition will change to a discontinuous one \cite{Oh2018,Oh2019}.

Besides, the infinite order transition can occur even in the hierarchical networks with small-world effects \cite{Boettcher2009,Berker2009,Hasegawa2010,Boettcher2012}. For the growing network without percolating state \cite{Lancaster2002}, the preferentially growing network \cite{Kullmann2001}, and the growing directed network \cite{Krapivsky2001,Coulomb2003}, the power-law distribution of cluster sizes can be also observed. The percolation transition of growing networks has also been studied by diluting links or nodes in a generated network \cite{Hasegawa2010a}. The dense structures were also found to undergo phase transitions in the growth of networks\cite{Bhat2016,Lambiotte2016}.

%*****************************************************************************
% Section IV
% Applications on network structures
%*****************************************************************************

\section{Applications to network structural analysis}  \label{sct4}

The previous section mainly reviews the typical percolation models on complex networks, focusing on the phase transition and the critical phenomena. In fact, the percolation theory, including conceptions, theoretical methods, and algorithms, has a broader range of applications in the study of network structures and dynamics. This is mainly because the percolation model is by nature for featuring the cluster forming process in random media, while network science is exploring the properties of various topological structures and the correlation with dynamics. In this section we will review the applications to network structural analysis. We should point out that some of these problems are not firstly motivated by percolation, however, a good interpretation can be given from the perspective of percolation.

\subsection{Hierarchical structure of networks}

Hierarchical structure is one of the important characteristics of real networks \cite{Albert2002,Boccaletti2006,Dorogovtsev2008,Newman2010,Barabasi2016}, which can be generally identified by either a branching process or a pruning process. For tree-like networks, the analytical method for network percolation is often employed to find theoretical results, while for non-tree-like networks percolation process can also be used as an algorithm to identify hierarchical structures.

\subsubsection{Tree-like networks}

\begin{figure}
\centering
\includegraphics[width=0.7\columnwidth]{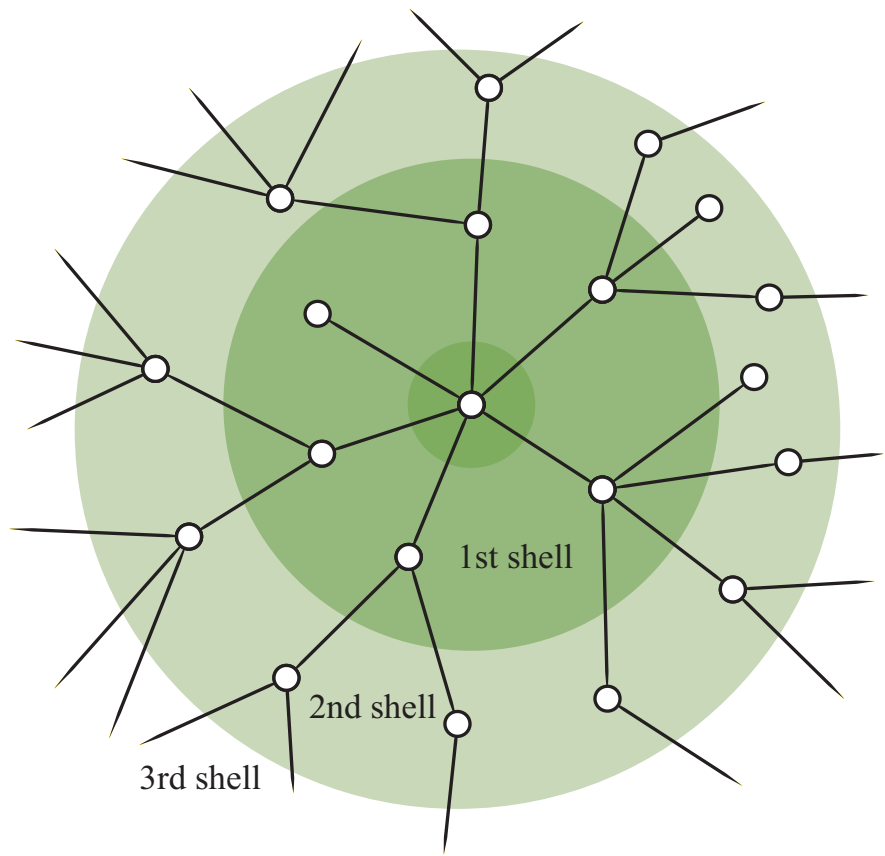}
\caption{(color online) Schematic for the hierarchical structure of a tree-like network. The number of nodes in the $(m+1)$-th shell is just the total excess degrees of the nodes in the $m$-th shell.}  \label{fig4-hs}
\end{figure}

As shown in Fig.\ref{fig4-hs}, the hierarchical structure of a tree-like network can be figured out by a spanning tree. For a large enough network with tree-like structures, any node can be seen as the root of the spanning tree. Then the direct neighbors of this node are the first shell of the network, and the neighbors' neighbors belong to the second shell, and so on. This is just the branching process we discussed in Sec.\ref{sct2-ampb}. Thus, the generating function can be also employed to study this shell structure \cite{Newman2001}.

Since the root node is arbitrarily chosen, the sub-branchings from it can be described by the generating function $G_0(x)$. Here, an $x$ refers to an outgoing link from the root node. Following these outgoing links, we can find the nodes in the first shell of the network. As the explanation for Fig.\ref{fig2-csd}, the generating function $G_1(x)$ can be used to represent the state of the node at the end of a randomly chosen link, where $x$ also refers to the outgoing link from this node. Therefore, replacing $x$ in $G_0(x)$ with $G_1(x)$, we will obtain the sub-branchings from the nodes in the first shell $G^{(1)}(x)=G_0(G_1(x))$. Similarly, the sub-branchings from the second shell can be represented by $G^{(2)}(x)=G_0(G_1(G_1(x)))$. In general, the sub-branchings coming from the $m$-th shell can be written as
\begin{align}
&G^{(m)}(x)  \nonumber \\
=&G_0(\underbrace{G_1(G_1(...G_1(x)...))}_{m~\text{generations}})   \nonumber \\
=&G^{(m-1)}(G_1(x)),~~~m>1.  \label{eq4-gm}
\end{align}
With this iteration relation, Shao \textit{et al.} found that the distribution of node numbers in these shells follows a power law for a broad class of complex networks \cite{Shao2008}. This branching process can also be used to realize the configuration model \cite{Kalisky2006a}, \textit{i.e.}, connecting nodes with given degrees shell by shell.

Note that loops are forbidden here, so there must be a one-to-one correspondence between the parameter $x$ and the sub-branching. Consequently, the exponent for the power of $x$ in Eq.(\ref{eq4-gm}) is just the total number of sub-branchings from the $m$-th shell. In other words, the derivative of Eq.(\ref{eq4-gm}) with respect to $x$ at $x=1$ must equal to the average number of nodes in the $(m+1)$-th shell,
\begin{align}
\langle k_{m+1}\rangle &= \frac{dG^{(m)}(x)}{dx}\bigg|_{x=1} \nonumber \\
&= \frac{dG^{(m-1)}(x)}{dx}\bigg|_{x=1}G_1'(1)   \nonumber \\
&= \langle k_{m}\rangle G_1'(1),~~~m\geq1, \label{eq4-km}
\end{align}
where $\langle k_{m}\rangle$ is the average number of nodes in the $m$-th shell, and $\langle k_1\rangle = \langle k\rangle$ is just the average degree of the network. From Eq.(\ref{eq4-km}), we can also find
\begin{equation}
\frac{\langle k_{m+1}\rangle}{\langle k_{m}\rangle}=G_1'(1),~~~m\geq1. \label{eq4-fkm}
\end{equation}
This indicates that the growth rate of the node number with shells is a constant $G_1^\prime(1)$. For an infinite network, the shell must be endless, so the node number cannot decrease with the increasing of shells, \textit{i.e.},
\begin{equation}
G_1'(1)=\frac{\langle k^2\rangle-\langle k\rangle}{\langle k\rangle}\geq 1.
\end{equation}
This is the existence condition of the giant cluster in a tree-like network, namely, percolation threshold, see Sec.\ref{sct2-ampb}.

For a finite network, the number of shells $l$ must be finite. Next, we will show how to estimate $l$. Using Eq.(\ref{eq4-fkm}), the number of nodes in the $m$-th shell can be written as
\begin{align}
\langle k_m\rangle=&\langle k_1\rangle\left(\frac{\langle k_2\rangle}{\langle k_1\rangle}\right)^{m-1}  \nonumber \\
=&\langle k\rangle\left(\frac{\langle k^2\rangle-\langle k\rangle}{\langle k\rangle}\right)^{m-1},~~m\geq1.  \label{eq4-kma}
\end{align}
Summing over all the shells, we obtain the network size
\begin{equation}
N=1+\sum_{m=1}^{l}\langle k_m\rangle.
\end{equation}
This yields
\begin{equation}
l=\frac{\ln\left[(N-1)\left(\langle k^2\rangle-2\langle k\rangle\right)+\langle k\rangle^2\right]-2\ln\langle k\rangle}{\ln\left(\langle k^2\rangle-\langle k\rangle\right)-\ln\langle k\rangle}.
\end{equation}
Real networks usually have $N\gg\langle k\rangle$ and $\langle k^2\rangle\gg\langle k\rangle$, thus $l$ can be approximately represented as
\begin{equation}
l=\frac{\ln N-\ln\langle k\rangle}{\ln\langle k^2\rangle-\ln\langle k\rangle}+1.   \label{eq4-al}
\end{equation}
This result can be seen as an approximation of the average distance or the diameter of real networks \cite{Newman2001b,Newman2001c}, which indicates that $l\propto\ln N$. This is just the small-world effect of random networks.

The above derivations depend on a finite $\langle k^2\rangle$ (when $N\to\infty$). However, SF networks with $2<\lambda<3$ give a divergent moment $\langle k^2\rangle \propto \int_{k_0}^{N^{1/(\lambda-1)}} k^{2-\lambda}dk \propto N^{(3-\lambda)/(\lambda-1)}$, where $k_0$ is the lower boundary of degrees and $N^{1/(\lambda-1)}$ is the natural cutoff of degrees. Substituting $\langle k^2\rangle\propto N^{(3-\lambda)/(\lambda-1)}$ into Eq.(\ref{eq4-al}), we can find a size-independent $l$, indicating that SF networks are ultrasmall.

Actually, the diameter $l$ of SF networks increases with the system size in the form of $\ln\ln N$ \cite{Chung2002,Cohen2003,Dorogovtsev2003}. To find this, the degree distribution of the nodes in each shell must be checked, separately. In general, an outgoing link is more likely to lead to a node with large degree, which can be seen from the expression of the excess-degree distribution $q_k\propto p_kk$. For finite networks, this means that earlier shells will bag almost all the nodes with large degrees, so that the small degree nodes are located in the periphery of the hierarchical structure. This effect is more notable when networks have strong heterogeneity like SF networks. This mainly indicates that the degree cutoff of nodes in each shell decreases with the growth of shells. With this, the form $l\propto\ln\ln N$ can be found in \cite{Chung2002,Cohen2003,Dorogovtsev2003,Kalisky2006a,Shao2009}.

\subsubsection{k-shell structure}

\begin{figure}
\centering
\includegraphics[width=0.7\columnwidth]{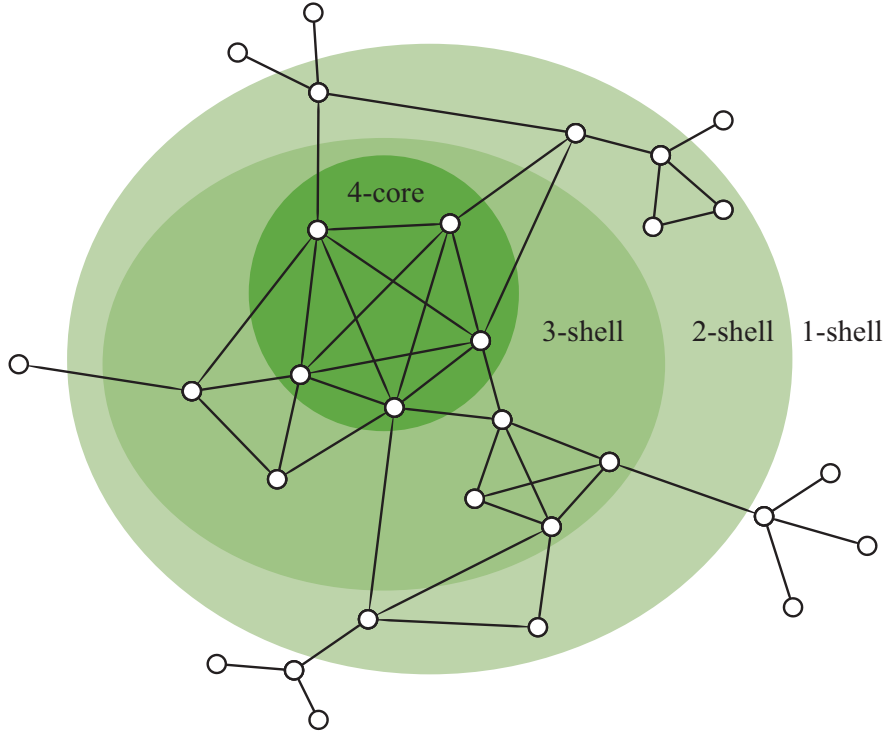}
\caption{(color online) Schematic for the $k$-shell structure of networks. In a $k$-core, nodes are connected to one another by at least $k$ paths. The nodes that belong to the $k$-core but not belong to the $(k+1)$-core are called as $k$-shell of the network.}  \label{fig4-ks}
\end{figure}

As pointed above, the hierarchical structure shown in Fig.\ref{fig4-hs} has an arbitrary root. In network science, there is another hierarchical structure with deterministic shells, \textit{i.e.}, $k$-shell structure \cite{Dorogovtsev2006}. For a given network, the $k$-shell refers to the nodes that belong to the $k$-core but not belong to the $(k+1)$-core, see Fig.\ref{fig4-ks}.

For tree-like networks, one can use Eqs.(\ref{eq3-sk}) and (\ref{eq3-rk}) to obtain the $k$-core size $S_k$ for any $k$, the size of $k$-shell is thus $S_k-S_{k+1}$. In general, the pruning process of $k$-core percolation is just used as an algorithm to identify the $k$-shell structure, called $k$-core/$k$-shell decomposition. The related discussions are far away from the percolation transition. So we will not go into details on this topic here, and one can refer to the specialized review on this topic for details \cite{Kong2019}.

\subsection{Network robustness}   \label{sct4-nr}

In network science, percolation process has already become a paradigm for studying the network robustness. Specifically, one often removes a fraction $1-p$ of nodes from a given network, then checks the existence of the giant cluster. If the giant cluster remains, the resulted network is still considered functional. Thus the percolation threshold $p_c$ can be used to evaluate the robustness of the network. A large threshold $p_c$ indicates that only a small amount of failed nodes can disconnect the network, so its robustness is poor, and vice versa. With different node removal methods, the percolation model can further explain the robustness of networks under various failure mechanisms, some of which are also used to model network dynamics. Here, we review the main findings of the network robustness based on the percolation model from two aspects, single networks and multiplex networks.

\subsubsection{Single networks}

As mentioned above, the percolation threshold of SF networks is much smaller than that of ER networks with the same link density, especially for SF networks with $2<\lambda<3$ which have a vanished percolation threshold \cite{Callaway2000,Cohen2000,Albert2000,Cohen2001,Cohen2010}. This suggests that SF networks are more robust than ER networks. More generally, for networks with the same average degree, the broader the degree distribution, the more robust the network is \cite{Yuan2015}.

For a given link density, the clustering structure compacts node clusters with the price of diluting global connections. Therefore, the corresponding percolation threshold is smaller than that of tree-like networks, indicating that the networks with clustering are fragile \cite{Kiss2008,Newman2009,Miller2009,Gleeson2010}. However, highly clustered networks are often characterized as a resilient system \cite{Newman2003,Serrano2006,Gleeson2009}. In fact, the two are reconcilable, since strong clustering could induce the so-called core-periphery structure \cite{Rombach2014,Rombach2017}, for which the percolation model shows a double transition \cite{Colomer-de-Simon2014}.

More generally, the loop structure, including clustering, is a meaningful concept for network robustness, captured by the connection between a pair of nodes through at least two disjointed paths, termed biconnectivity. Since a pair of biconnected nodes in networks can communicate under the removal of one route, biconnectivity can play a significant role in network robustness \cite{Lee2006,Newman2008,Kim2013,Schawe2019}, as well as other high order connectivity. Some of studies on this topic in terms of the percolation transition can be found in Sec.\ref{sct3-kcp}. Moreover, the resilience of networks with community structure has also been studied by the percolation model, which shows a behavior similar to the spin system under an external field \cite{Dong2018a}.

In addition, the degree correlation can also influence the percolation process on networks, see Sec.\ref{sct2-ccn}. Generally speaking, increasing the assortativity of a network makes the network more robust against nodes removal \cite{Newman2003a,Vazquez2003}. In fact, the clustering structure and the degree correlation are inseparable for real networks. Even for network models, it is impossible to adjust the clustering structure arbitrarily without inducing any degree correlation, and vice versa. Therefore, the robustness of real networks, even of some network models, cannot be evaluated by their clustering coefficients or assortative coefficients, individually.

If the hub nodes are preferred to be removed initially, \emph{i.e.}, intentional attack, the percolation model can also be used to study the robustness of networks under targeted attack. A simple implementation approach is to remove a fraction $1-p$ of the nodes with the highest degrees \cite{Cohen2001}, which reports that SF networks, known to be resilient to random removal, are sensitive to intentional attack. More generally, the intentional attacks are commonly realized by removing nodes with probability proportional to $k^\alpha$ \cite{Gallos2005}, where $k$ refers to degree. When $\alpha>0$, nodes with larger degrees are more vulnerable under the intentional attacks, while for $\alpha<0$, nodes with larger degrees are defended and so have a lower probability to fail. The case $\alpha=0$ just represents the random removal of nodes. This mechanism can also be developed to study the robustness of networks under intentional link attack \cite{Hooyberghs2010}.

For $\alpha>0$, SF networks become extremely fragile \cite{Albert2000}, and the percolation threshold will be larger than that of ER networks with the same link density \cite{Albert2000} or even equal to $1$ \cite{Hasegawa2013b}. At criticality, the topology of the network depends on the details of the removal strategy, implying that different strategies may lead to different kinds of percolation transitions \cite{Gallos2005,Hooyberghs2010}. An optimal attack is also directly related to optimal graph partitioning and immunization of epidemics \cite{Paul2007}. Different structures can also emerge from these pruning processes \cite{Valente2004,Srivastava2012,Peixoto2012}. Moreover, by mapping the intentional attack to a normal percolation problem, this robustness model can also be solved \cite{Huang2011}. Further research also shows that an onion-like structure could be a nearly optimal structure against simultaneous random and targeted high degree node attacks \cite{Schneider2011,Herrmann2011,Tanizawa2012}.

There is also a special percolation model for the case $\alpha\to-\infty$, called degree-ordered percolation \cite{Lee2013}, where nodes are occupied in degree-descending order (or alternatively, removed in degree-ascending order). The mean-field results on SF networks suggest that the critical exponents depend on the heterogeneity of the network, and do not belong to the universality class of the standard percolation \cite{Caligiuri2020}.

Besides, the localized attack, meaning that an attack on a node can destroy the node together with its neighbors within a range, has also been introduced into the percolation model to show the robustness of networks under attacks \cite{Shao2015,Yuan2015,Berezin2015,Yuan2016,Dong2016,Vaknin2017,Dong2019}. When the intentional attacks are dependent on spatial distances, the spatial networks are more fragile than expected \cite{McAndrew2015}. Moreover, the finite-size scaling analysis suggests that for some of the intentional attack strategies, the critical exponents seem to deviate from the mean field, but the evidence is not conclusive \cite{Almeira2020}.

The percolation model is a random process, and the corresponding theory only characterizes the results on average. Nevertheless, two different realizations with the same magnitude of removal could result in very different giant clusters. For the study of the robustness of real networks, the fluctuations of different percolation realizations become extremely important. Bianconi introduced a message-passing algorithm able to predict the fluctuations in a single network, and an analytic prediction of the expected fluctuations in ensembles
of sparse networks \cite{Bianconi2017a,Coghi2018}. The following works also contributed to the stability of the giant cluster \cite{Kitsak2018}, the large deviation of percolation \cite{Bianconi2018a,Bianconi2019}, and the largest biconnected cluster for random graphs \cite{Schawe2019}. The results mainly showed that the large-deviation approach to percolation can provide a more accurate characterization of system robustness.

As shown in Sec.\ref{sct2-dn}, the percolation transition can also be well defined on directed networks, so that with similar consideration the robustness of directed networks can also be studied in the framework of percolation transition. It mainly demonstrates that the GWCC is more vulnerable, and the directed network may have the GWCC and, simultaneously, may not have the GSCC \cite{Dorogovtsev2001,Schwartz2002,Boguna2005,Meyers2006,AngelesSerrano2007,Gleeson2008a,Restrepo2008,Hoorn2015,Liu2017}.

In addition, if the communication of nodes in the network is effective only if the shortest path between nodes $i$ and $j$ after removal is shorter than $\alpha l_{ij}$, where $l_{ij}$ is the shortest path before removal, a much smaller failure of the network can lead to an effective network breakdown. This problem was also studied in a percolation form called limited path percolation \cite{Lopez2007,Lopez2012}. Note that this mechanism was also used to feature the entanglement percolation in quantum complex networks \cite{Cuquet2011}.

\subsubsection{Multiplex networks}

In recent years, multiplex network has become a research hotspot. The consideration of multiplex network naturally introduces the interaction between different network layers, thus constructs a cascading picture, which changes the nature of percolation transition, see Sec.\ref{sct3-pin}. Consequently, the robustness of multiplex networks under the framework of percolation transition has its own characteristics, or even in a completely opposite manner as the single network.

\paragraph{Topological structure}

As shown in Sec.\ref{sct3-pin}, multiplex networks are more fragile than single networks due to the cascading failures induced by the interdependence between nodes from different network layers. More interestingly, the effects of the topology structure on the robustness of the network are also very different from those of the single network.

In Refs.\cite{Parshani2011,Buldyrev2010}, researchers found that a network with a broader degree distribution results in a larger $p_c$, which is opposite to ordinary networks. As we know, the hub nodes have a dominant role in the robustness of a network. However, when the node dependence is involved, a hub node could depend on a small degree node, which is quite vulnerable during the iterated removal process triggered by the initial node removal. Moreover, a broader degree distribution with the same average degree implies that there are more small degree nodes. Therefore, the advantage of a broad degree distribution for the ordinary network becomes a disadvantage for the network with dependence links. For this reason, the assortativity and high clustering structures will also make such networks fragile \cite{Zhou2012,Huang2012}. These features are in consonance with the robustness of single networks under intentional attack \cite{Valente2004,Gallos2005,Hooyberghs2010,Srivastava2012,Peixoto2012,Shao2015}. For directed networks, the in-degree and out-degree correlations could increase the robustness of interdependent networks with heterogeneous degree distributions, but decrease the robustness of interdependent networks with homogeneous degree distributions and with strong coupling strengths \cite{Liu2016c}.

\paragraph{Inter similarity}

In contrast to single networks, the multiplex networks are difficult to defend by strategies such as protecting the large degree nodes (intentional attack with $\alpha<0$) that have been found useful to significantly improve the robustness of single networks \cite{Huang2011,Dong2012,Zhou2013}. This is mainly because the dependence partners could have very different degrees, and defend strategies based on degrees cannot protect the two nodes at the same time. By this consideration, the system will obviously become robust, when the dependence partners have identical degrees \cite{Buldyrev2011,Min2014}. Such correlations between dependent partners are called inter similarity of different network layers.

Parshani \textit{et al.} introduced two parameters to evaluate the level of inter similarity between networks: the inter degree-degree correlation and the inter clustering coefficient. The simulation results demonstrated that the inter-similar multiplex networks are significantly more robust to random failures than a randomly interdependent system \cite{Parshani2010a}. Theoretically, Hu \textit{et al.} developed an approach to analyze the robustness of multiplex networks with inter similarity \cite{Hu2013}. In their model, there are some probabilities that the dependence nodes of two adjacent nodes in one network are also connected. The studies showed that this inter similarity can improve the robustness of the interdependent networks and change the critical behaviors of the percolation transition. In addition, a similar work has also been done on this problem \cite{Cellai2013b}. The inter-similarity of dependence partners is also studied in single networks. The results show that the local dependence, \textit{i.e.}, dependence partners are adjacent, can result in a robust network \cite{Li2013}.

\paragraph{Dependence relations}

The dependence links were proposed to reflect the dependence relationships in reality, but not all the nodes in a networked system have such dependence. In this way, the percolation on networks with weak dependence has been studied \cite{Parshani2010,Zhou2013,Valdez2013}. In this model, instead of each node having a dependence partner, only a fraction $q$ of nodes have dependence links. When $q=0$, the multiplex networks reduce to some single networks, while $q=1$, the fully dependence model discussed above is covered. Thus the system becomes robust with the decreasing of $q$.

Besides, without reducing the number of dependence links, the dependence can also be adjusted by the so-called dependence/coupling strength $\alpha\in [0,1]$ \cite{Liu2016a,Kong2017,Liu2019}. When a node loses its dependence partner in the pruning process, each of its links is removed from the network with a probability $\alpha$. With this mechanism, heterogeneous dependence can be realized by assigning different dependence links with different dependence strength, and the results demonstrate that the heterogeneous dependence strength makes the system more robust \cite{Kong2017}. In addition, Liu \textit{et al.} proposed a model that integrates interdependent networks and single networks with dependence links \cite{Liu2016b}. They found that when most of the dependence links are inner- or inter-ones, the coupled network system is fragile and shows a discontinuous percolation transition. However, when the two types of dependence links have nearly the same number, the system is robust and shows a continuous percolation transition.

The one-to-one dependence in the original model is also generalized to multiplex dependence or group dependence. Obviously, the multiplex/group dependence makes the spreading of failures easier, so that the vulnerability of interdependent network under various dependence mechanisms is demonstrated \cite{Parshani2011,Shao2011,Bashan2011,Bashan2011a}. Wang \textit{et al.} further showed that a larger dependence group did not always make the network fragile, which is also dependent on the grouping mode and the cascading rules in the group \cite{Wang2015,Wang2018}. Li \textit{et al.} also pointed out that the asymmetric dependence can break the cascading picture between dependence partners, thus results in a robust network \cite{Li2014a}.

\paragraph{Network of networks}

Considering the real situation that more than two networks are connected by the dependence links, Gao \emph{et al.} extended the model of the interdependent networks into the model of a network of networks \cite{Gao2011,Gao2012,Gao2013,Gao2014,DAgostino2014,Havlin2014,Kenett2015}. In this model, each network is composed of a set of nodes and connectivity links, then the dependence links connect them to form a larger network. The common factors that influence the network robustness, such as clustering \cite{Shao2014}, and targeted attack \cite{Dong2013}, have also been explored in the network of networks. Liu \textit{et al.} proposed a ``weak” interdependence mechanism capturing the differences of networks, where network layers at different positions within the multiplex system experience distinct percolation transitions \cite{Liu2018}.

In general, the robustness of network of networks decreases with the increasing of network layers. Due to the cascading failures caused by the dependence of networks, the percolation transition of a network of networks could be discontinuous and even a single node failure may lead to an abrupt collapse of the system \cite{Gao2011,Gao2012,Gao2013}. Based on a network model composed of spatially embedded networks, Shekhtman \textit{et al.} showed the extreme sensitivity of coupled spatial networks and emphasized the susceptibility of these networks to sudden collapse \cite{Shekhtman2014}. In real life, infrastructural networks are governed and operated separately, and interactions are only allowed within well-defined boundaries. Therefore, for different situations, the percolation on network of networks may be very different. For example, Radicchi \textit{et al.} introduced a model of percolation where the condition that makes a node functional is that the node is functioning in at least two network layers of the system, and found that the addition of a new layer of interdependent nodes to a preexisting multiplex network will improve its robustness \cite{Radicchi2017}.

In addition, the network of networks is also studied as interacting networks \cite{Leicht2009,Dickison2012,Rapisardi2019}, in which the connections between the network layers are ordinary links. A system of this kind is therefore equivalent to a single modular network, for which the system can be easily separated into isolated modules, and different modules might percolate at different thresholds \cite{Melnik2014,Shai2015,Faqeeh2016,Dong2018a}. It should mention that there are some special reviews on the resilience of network of networks \cite{Gao2015,Shekhtman2016}. One can refer to those for more details.

\subsection{Community detection}  \label{sct4-cd}

Community structures are the subsets of nodes within a network that have a high density of within-group connections but a lower density of between-group connections. Traditional community detection methods are usually deterministic and used for finding separated communities, whereas most of the actual networks are made of highly overlapping cohesive groups of nodes. To identify the overlapping communities in large real networks, Der\'{e}nyi and co-workers have proposed a dependent percolation model, called clique percolation \cite{Derenyi2005,Palla2005}.

\begin{figure}
\centering
\includegraphics[width=0.7\columnwidth]{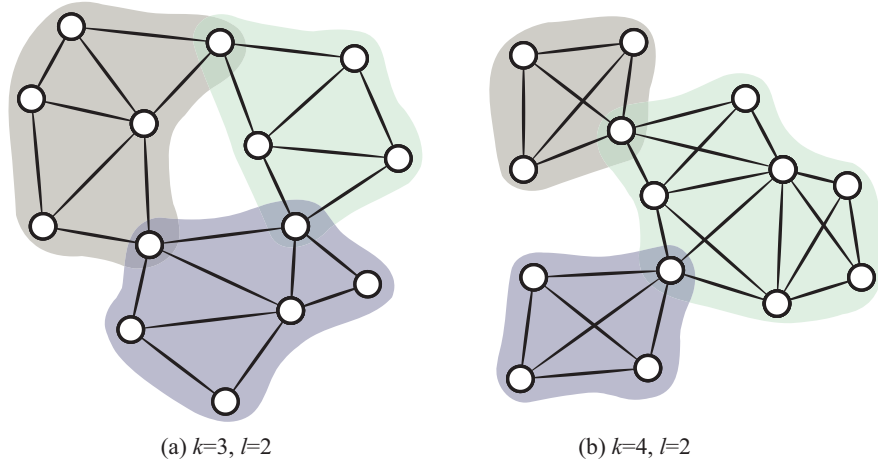}
\caption{(color online) Schematics of the clique percolation. Two $k$-cliques are regarded as adjacent, if they share $l$ ($\leq k-1$) nodes. Different clusters are distinguished by different colors. (a) $k=3$ and $l=2$. (b) $k=4$ and $l=2$.}  \label{fig4-cp}
\end{figure}

A clique is a fully connected subset of nodes of a network; such a subset with $k$ nodes is often called a $k$-clique. In their model, two $k$-cliques are regarded as adjacent, if they share $l$ ($=k-1$) nodes. With this connection rule, a node can belong to more than one clique cluster, see Fig.\ref{fig4-cp}. Then, by identifying these clique clusters as communities, the overlapping communities are thus detected.

This method also has a clear limitation. That is the network must have a large number of cliques, so it may fail to give meaningful community structures for networks with just a few cliques. For more discussions on this problem, one can refer to Ref.\cite{Fortunato2010}.

Aside from the analysis of network structure, clique percolation transition itself provides a set of very interesting problems. As classical percolation in ER networks, $k$-clique percolation considers the behaviors of clusters formed by connected $k$-cliques for a different connection probability $p$ of links. The simulation results in Ref.\cite{Derenyi2005} indicate that the fraction of nodes in the largest $k$-clique cluster makes a discontinuous percolation transition with increasing connection probability $p$; however, the fraction of $k$-cliques in the largest $k$-clique cluster demonstrates a continuous percolation transition. Bollob\'{a}s and Riordan gave a rigorous mathematical result of the critical point for a more general percolation model, in which two $k$-cliques are regarded as adjacent if they share $l~(=1,2,...,k-1)$ or more nodes \cite{Bollobas2009}. The Monte Carlo simulation indicates that for different $k$ and $l$, this general clique percolation model could demonstrate rich critical phenomena \cite{Fan2014}.

In Ref.\cite{Li2015}, Li \emph{et al.} proved theoretically that the fraction of nodes in the giant clique cluster $\phi$ for $l>1$ makes a step-function-like phase transition in the thermodynamic limit and a continuous phase transition for $l=1$. More interestingly, they showed that the order parameter at the critical point $\phi_c$ is neither $0$ nor $1$, but a constant depending on $k$ and $l$. In addition, the fraction of cliques in the giant clique cluster $\psi$ always makes a continuous phase transition as the classical percolation. Through analysis of the clique cluster number distribution, they found that there is no essential distinction between the two processes measured by $\psi$ and $\phi$, and the different behaviors are mainly caused by the quantitative relation between the total numbers of cliques and nodes. Besides, the clique percolation has also been studied in two-dimensional systems \cite{Dong2018}. The finite-size scaling analysis shows that, in contrast to the clique percolation on an ER network where the critical exponents are parameter dependent, the two-dimensional clique percolation simply shares the same critical exponents with traditional site or bond percolation, independent of the clique percolation parameters \cite{Dong2018}.

\subsection{Vital nodes identification}

The heterogeneity of complex networks shows up in the different roles that nodes play in the structure resilience and the dynamical behaviors. The vital nodes identification classifies nodes based on their roles played in network structure or dynamics \cite{Lue2016}. As a model of cluster forming, percolation model has also been employed to identify vital nodes.

In Ref.\cite{Piraveenan2013}, Piraveenan \emph{et al.} proposed a centrality measure based on the percolation state of networks, called percolation centrality. In this method, percolation state of node $i$ is denoted as $x_i$. Specifically, $x_i=1$ indicates that node $i$ is in the fully percolated state, and $x_i=0$ for a non-percolated state, while a partially percolated state corresponds to $0<x_i<1$. Accordingly, the percolation centrality of a node $i$ in a network with size $N$ is defined as
\begin{equation}
PC(i)=\frac{1}{N-2}\sum_{s\neq i\neq r}\frac{\sigma_{s,r}(i)}{\sigma_{s,r}}\frac{x_s}{\sum_j x_j-x_i},
\end{equation}
where $\sigma_{s,r}$ is the number of shortest paths between nodes $s$ and $r$, and $\sigma_{s,r}(i)$ is that pass through node $i$. With this, nodes in large clusters are considered to be more important. If all nodes are in the same cluster, $x_s/(\sum_j x_j-x_i)=1/(N-1)$, so that the percolation centrality reduces to the well-known betweenness centrality. From this perspective, percolation centrality is just a betweenness centrality with weighting based on percolated state.

In percolation process, if a node's removal breaks down the network into many small clusters, the nodes must be very important for the network structure. From this basis, Morone and Makse proposed a method to find the superspreaders in network spreading, in which the influence of a node is the size of the largest cluster after removing this node \cite{Morone2015}. Accordingly, the influence maximization problem is to find the minimal set of nodes which, if removed, would turn the network into the non-percolating state. Such an operation is often called optimal percolation. The method is also used to find the influential nodes in brain networks \cite{Ferraro2018} and multiplex networks \cite{Osat2017}.

Due to the inherent relation with spreading dynamics, bond percolation is also used to identify the optimal combination of a given number of influential spreaders \cite{Ji2017}. This method needs first to perform a bond percolation on the network, \textit{i.e.}, randomly removing some links. In the resulting network, the largest degree nodes in the top-$n$ largest clusters are selected and one score is assigned. With different realizations, the average score of each node can be obtained, and the ones with large scores are suggested to be the influential spreaders. Furthermore, Hu \textit{et al.} offered an analogy of the correlation length in percolation transition to explain the local length scale in spreading dynamics, which could also be used to measure a node’s or a group of nodes’ global influence \cite{Hu2018}.

In addition, percolation process is also employed as a criterion of the node ranking algorithm \cite{Lue2016}. In this criterion, the size of the giant cluster after a node's removal is assumed to be the structural importance of the node, such that the smaller the obtained giant cluster is the more important the node is. For a node ranking, one can remove nodes from the network one by one in the order of the ranking, then the average size of the giant clusters for each stage can be expressed as
\begin{equation}
R=\frac{\sum_{i=1}^N S_i}{N}. \label{eq4-rsn}
\end{equation}
Here, $N$ is the number of nodes in the network, and $S_i$ is the size of the giant cluster when the $i$-th node has been removed. Obviously, a good node ranking will give a small $R$. From Eq.(\ref{eq4-rsn}), if the network is large enough, $R$ is approximately the area surrounded by function $S_i$ and the horizontal axis.

\subsection{Network observability}

In Ref.\cite{Yang2012}, a fundamental problem in network science, namely, the determination of the state of the system from limited measurements, was abstracted as a percolation model, called network observability transition. In this model the initial node occupation is understood as placing a sensor device on the corresponding node, making both the node and all of its neighbors observable. Then, network observability transition focuses on the emergence of the giant cluster formed by observable nodes.

Obviously, the threshold of such percolation must be smaller than that of the classical percolation. By generating function technique, this model can also be solved theoretically. The percolation threshold depends dominantly on the average degree and the variance of degree distribution. The transition occurs earlier in denser and more heterogeneous networks, meaning that only a small amount of sensor devices can observe the key structure of such networks. However, Allard \textit{et al.} further found a nontrivial coexistence of an observable and of a nonobservable extensive cluster in this percolation model \cite{Allard2014}. This suggests that monitoring a macroscopic portion of a network does not prevent a macroscopic event to occur or be unknown to the observer. Further studies also showed that both the negative degree correlation and the betweenness-based sensor placement can make networks more observable \cite{Hasegawa2013,Shunkun2016}, and were applied to many real networks \cite{Yang2016,Osat2018}. In Ref.\cite{Osat2018}, a discontinuous observability transition was found in synthetic multiplex networks.

There is another percolation model called $l$-hop percolation \cite{Shang2011}, following a very similar percolation rule. In this model when a node is removed all the nodes in the $l$-hop distance, \textit{i.e.}, the nodes no farther than $l$ away from this node, will also be removed, therefore it can cause more damage than the normal robustness model based on the classical percolation. For the study of networks robustness, such a removal rule is often named localized attack \cite{Shao2015,Yuan2015,Berezin2015,Yuan2016,Dong2016,Vaknin2017,Dong2019}.

%*****************************************************************************
% Section V
% Applications on network dynamics
%*****************************************************************************

\section{Applications to network dynamics}  \label{sct5}

The origin and development of percolation are inextricably bound up with the spreading process, therefore it is completely natural that in network science the percolation mechanism has been widely used to embody the spreading dynamics. In this section we will first review some typical models of spreading dynamics and their relations with the percolation transition. Besides, the percolation theory has also been borrowed to analyzing the cluster forming of individuals in the dynamics that the node or individual group plays a key role in the evolution of systems, such as the cascading process, traffic jam, and cooperation evolution. These works will also be reviewed in this section.

\subsection{Epidemic spreading on networks}

The modeling of epidemics is a very active field of research across different disciplines \cite{Anderson1992}, and complex networks provide a powerful framework to describe the contact structure among individuals \cite{Vespignani2012}. As mentioned in Sec.\ref{sct2-drnp}, the epidemic spreading on networks is closely related to percolation process. When the spanning cluster does not exist or the susceptible individuals are isolated into small clusters, infectious diseases will be constrained to some small areas and a global outbreak is impossible. Similarly, when some individuals are immunized to the disease, \textit{i.e.}, recovered from the disease, being quarantined or get vaccinated, the susceptible individuals can be also separated as small isolated clusters and thus the disease cannot spread across the network. From this perspective, regardless of the specifics, the outbreak of the epidemics corresponds to a percolation transition among connected individuals, and there is a direct correlation between the percolation threshold and the epidemic threshold.

In this section, we introduce the framework of epidemic modeling in complex networks from a percolation perspective, although most of them are not inspired by percolation theory initially. Due to the focus of this article, we will not deeply explore various models and the corresponding analytic methods. For details, one can refer to the special reviews \cite{Hethcote2000,Wang2017,Pastor-Satorras2015}.

It should be noted that the spreading of epidemics is a dynamic process, while the percolation process results in a static configuration. Because of this, one often employs the master equation to feature the system state during the spreading process, instead of considering the static configuration of the system. Nevertheless, we will find that the results obtained by the master equation directly reflect the nature of percolation transition. With some mathematical techniques, some spreading models can also be exactly mapped to a percolation process \cite{Newman2002,Pastor-Satorras2015}.

Generally, the epidemic model assumes that the individuals can be categorized as different classes depending on the stage of the disease in the population, such as susceptible ($S$, those who have not got infected), infectious ($I$, those who have been infected and are contagious), and recovered ($R$, those who have recovered from the disease or death) \cite{Pastor-Satorras2015}. Based on this classification, three epidemiological models, susceptible–infected (SI), susceptible–infected–susceptible (SIS) and susceptible–infected–recovered (SIR) are usually used to study the dynamics of disease evolution.

The simplest one is the SI model in which individuals can only exist in two discrete states, namely, $S$ and $I$. The probability that an $S$ node gets infected by any given neighbor in an infinitesimal time interval $dt$ is $\beta dt$, where $\beta$ is the infection rate of the epidemic. The SIS model considers that individuals change stochastically through a reversible process: $S\to I \to S$, where the infection rate is described as in the $SI$ model, but infected individuals may recover and become susceptible again with a recovery rate $\mu$. In the SIR model, infected individuals are removed permanently from the network with a rate $\mu$ and labeled as the state $R$ corresponding to immune or death. There is a second version of SIS/SIR model, where infected individuals remain in state $I$ for $\tau$ time steps, and then become either $R$ in SIR model or $S$ again in SIS model \cite{Parshani2010b}.

\subsubsection{SI model}

As the setting of SI model, even for a very small infection rate $\beta$, all the individuals will be finally infected. Therefore, the epidemic threshold is meaningless for SI model, and the interest of research is the effect of the heterogeneity of degree distributions on the spreading velocity \cite{Barthelemy2004}.

Let $i_{k}(t)$ be the fraction of infections in nodes with degree $k$ at time step $t$, then the growth of $i_{k}(t)$ per time step can be expressed as $\beta k s_{k}(t) \Theta_{k}(t)=\beta k[1-i_{k}(t)]\Theta_{k}(t)$, where $s_{k}(t)$ is the fraction of susceptible in nodes with degree $k$, and $\Theta_{k}(t)$ is the probability that a neighbor of a node of degree $k$ is infected. Note that $i_{k}(t)$ cannot decrease with time for this model, so the rate equation reads
\begin{align}
\frac{di_{k}(t)}{dt} &=\beta k[1-i_{k}(t)]\Theta_{k}(t)  \nonumber \\
&=\beta k\Theta_{k}(t)-\beta k i_{k}(t)\Theta_{k}(t).  \label{eq5-si1}
\end{align}
For uncorrelated networks, $\Theta_{k}(t)\equiv\Theta(t)$ is independent of $k$. Considering that an infected node at least has an infected neighbor, from which the disease has been transmitted, we have
\begin{equation}
\Theta(t)=\frac{\sum_{k'}(k'-1)p_{k'}i_{k'}(t)}{\langle k\rangle}. \label{eq5-si2}
\end{equation}
In the early epidemic stages, $i_k(t)$ is small, then one can neglect the second term of the right hand side of Eq.(\ref{eq5-si1}), which is in an order of $O([i(t)]^{2})$. Together with Eq.(\ref{eq5-si2}), we have
\begin{align}
\frac{di_{k}(t)}{dt} &= \beta k\Theta(t),  \\
\frac{d\Theta(t)}{dt} &= \beta (\kappa-1)\Theta(t),
\end{align}
where $\kappa=\langle k^{2}\rangle/\langle k\rangle$. For a uniform initial condition $i_{k}(t=0)\equiv i(0)$, the two differential equations have solution
\begin{equation}
i_{k}(t)=i(0)\left[1+\frac{k}{\langle k\rangle}\frac{\langle k\rangle-1}{\kappa-1}\left(e^{t/\tau}-1\right)\right],
\end{equation}
with the outbreak's time scale
\begin{equation}
\tau=\frac{1}{\beta(\kappa-1)}. \label{eq5-sitau}
\end{equation}
The average prevalence can also be obtained,
\begin{align}
i(t) &=\sum_{k}p_k i_{k}(t)   \nonumber  \\
&=i(0)\left[1+\frac{\langle k\rangle-1}{\kappa-1}\left(e^{t/\tau}-1\right)\right].
\end{align}
These results indicate that the prevalence shows an exponential growth in the early stages, and larger degree nodes display larger prevalence levels. By the way, the spreading dynamics on the networks with small-world effects also reported a power-law growth with a tunable exponent \cite{Grassberger2013,Juhasz2015}, which has already been observed in the epidemic data of COVID-19 \cite{Ziff2020,Li2020a,Singer2020,Maier2020}. Such growth patterns are often referred to the sub-exponential growth dynamics in the phenomenological models \cite{Chowell2016}.

Equation (\ref{eq5-sitau}) implies that the outbreak's time scale of an epidemic is related to the heterogeneity of the degree distribution as measured by $\kappa$. A strong heterogeneity corresponds to a large $\kappa$, and thus the time scale of outbreak $\tau$ is very small, which signals a very fast spreading of the infection. The physical reason is that once the disease has occupied the hubs, it can spread very fast among the network following a ``cascade" of decreasing degree classes \cite{Barthelemy2004,Barthelemy2005}.

Note that the percolation threshold corresponds to $p_c=1/(\kappa-1)$ \cite{Cohen2001}. When the infection rate $\beta>p_c$, the outbreak's time scale becomes $\tau<1$. This means that an epidemic with an infection rate larger than the percolation threshold can quickly spread far and wide. Moreover, a special case is SF networks with an exponent $2<\lambda<3$, for which any infected rates $\beta>0$ will lead to a rapid spreading due to the vanished percolation threshold.

\subsubsection{SIR model}

The above discussion can be easily extended to the SIR model. To describe the transition from $I$ state to $R$ state, an extra term $− \mu i_{k}(t)$ defining the rate at which infected individuals of degree $k$ recover and permanently removed must be added into Eq.(\ref{eq5-si1}) \cite{Moreno2002},
\begin{align}
\frac{di_{k}(t)}{dt} &=\beta k s_{k}(t)\Theta_{k}(t) − \mu i_{k}(t),  \label{eq5-siri}  \\
\frac{ds_{k}(t)}{dt} &=-\beta k s_{k}(t)\Theta_{k}(t), \label{eq5-sirs}    \\
\frac{dr_{k}(t)}{dt} &=\mu i_{k}(t),  \label{eq5-sirr}
\end{align}
where $r_{k}(t)$ is defined as the density of removed individuals of degree $k$ and $s_{k}(t)+i_{k}(t)+r_{k}(t)=1$. With the initial conditions $s_k(0)\simeq 1$, $i_k(0)\simeq 0$, and $r_k(0)=0$, the solution of Eq.(\ref{eq5-sirs}) can be written as
\begin{equation}
s_k(t)=e^{-\beta k \phi(t)}.        \label{eq5-skt}
\end{equation}
Here, considering the uncorrelated networks, \textit{i.e.}, $\Theta_k(t)\equiv\Theta(t)$, $\phi(t)$ reads
\begin{align}
\phi(t) &=\int_0^t\Theta(t') dt'   \nonumber \\
&=\frac{1}{\langle k\rangle}\sum_{k}p_{k}(k-1)\int_0^ti_{k}(t')dt'   \nonumber \\
&=\frac{1}{\mu\langle k\rangle}\sum_{k}p_{k}(k-1)r_k(t).
\end{align}
In the last equality we have utilized the integral of Eq.(\ref{eq5-sirr}).

Different from SI model, for $t\to\infty$, the infections vanish and the individuals are either in state $S$ or in state $R$, depending on the infection rate $\beta$ and the recovery rate $\mu$. We can use $d\phi(t)/dt=0$ to find the steady state, that is
\begin{align}
0=\left.\frac{d\phi(t)}{dt}\right|_{t\to\infty} &=\frac{1}{\langle k\rangle}\sum_{k}p_{k}(k-1)i_k(t)  \nonumber  \\
 &=\frac{1}{\langle k\rangle}\sum_{k}p_{k}(k-1)\left[1-r_k(t)-s_k(t)\right]  \nonumber \\
 &=\frac{\langle k\rangle-1}{\langle k\rangle} -\mu\phi(t) -\frac{1}{\langle k\rangle}\sum_{k}p_{k}(k-1)s_k(t).
\end{align}
Together with Eq.(\ref{eq5-skt}), we find a self-consistent equation for $\phi(\infty)$,
\begin{equation}
\phi(\infty) = \frac{\langle k\rangle-1}{\mu\langle k\rangle} -\frac{1}{\mu\langle k\rangle}\sum_{k}p_{k}(k-1)e^{-\beta k \phi(\infty)}.         \label{eq5-phi}
\end{equation}
Solving this equation, we can find $\phi(\infty)$, and thus all other parameters for the steady state can be found.

It is easy to find that if
\begin{equation}
\frac{d}{d\phi(\infty)}\left.\left[\frac{\langle k\rangle-1}{\mu\langle k\rangle} -\frac{1}{\mu\langle k\rangle}\sum_{k}p_{k}(k-1)e^{-\beta k \phi(\infty)}\right]\right|_{\phi(\infty)=0}<1,
\end{equation}
Eq.(\ref{eq5-phi}) only has a solution $\phi(\infty)=0$, indicating the epidemic cannot break out. This yields the epidemic threshold
\begin{equation}
\delta_c=\frac{\beta_c}{\mu_c}=\frac{1}{\kappa-1},  \label{eq5-dlt}
\end{equation}
above which the epidemic incidence attains a finite value. Note that $1/(\kappa-1)$ is just the percolation threshold of the network. This suggests that there is a tight correspondence between the emergence of the giant cluster in the bond percolation and the outbreak of epidemic in the system.

Similar to SI model, Eq.(\ref{eq5-dlt}) also suggests that a high level of connectivity heterogeneity (large $\kappa$) facilitates the spreading of epidemics. When $\kappa$ diverges (SF networks with $2<\lambda<3$), an epidemic with any non-zero infection rate $\beta$ can break out in the system \cite{Barthelemy2004,Barthelemy2005,Volz2009}. This is analogous to those concerning the network resilience under random damages, which can be studied by the standard percolation on networks.

The above discussion focuses the steady state of spreading process. The time evolution of SIR model can be solved by other approaches, such as the effective degree-based method \cite{Eames2002} and the pair-based mean-field method \cite{Sharkey2008,Sharkey2011}. The extended degree-based method is used to derive an expression for the basic reproduction ratio $R_{0}$ \cite{Eames2002}, and also provides an excellent agreement with numerical simulations for both the temporal evolution and the final outbreak size \cite{Lindquist2011}. The threshold condition derived from the effective degree-based method turns out to be equal to the one obtained by percolation theory. The pair-based method was proven to be an exact deterministic description of the infection probability time course for each individual in the tree-like network \cite{Sharkey2015}. For networks with loops, a precise closures can be also derived according to the detailed loop structure \cite{Kiss2015}. At the same time, the pair-based approach can be easily extended to models with heterogeneous infection and recovery rates \cite{Sharkey2008}.

The correspondence between the steady properties of SIR model and the bond percolation is also studied in Refs.\cite{Newman2010,Newman2002,Grassberger1983,Andersson2000}. Considering the second version of SIR model with a uniform infection time $\tau$, \textit{i.e.}, infected nodes will recover after being infected for $\tau$ time steps. For continuous-time dynamics, the transmissibility $T$, defining the probability that the disease will be transmitted from an infected node to a susceptible neighbor before recovery takes place, can be computed as
\begin{equation}
T=1-\lim_{\Delta t \rightarrow 0}(1-\beta\Delta t)^{\tau/\Delta t}=1-e^{-\tau\beta}.
\end{equation}
With this relation, we can exactly map the steady state of an SIR model to a bond percolation with occupied probability $T$. Then, from the critical transmissibility
\begin{equation}
T_{c}=\frac{1}{G_1'(1)}=\frac{\langle k\rangle}{\langle k^{2}\rangle-\langle k\rangle}=\frac{1}{\kappa-1},
\end{equation}
one can find the critical infection rate for a given $\tau$, that is
\begin{equation}
\beta_{c}=\frac{1}{\tau} \ln\frac{\kappa-1}{\kappa-2}.
\end{equation}
For discrete time, particularly computer simulations, the transmissibility $T=1-(1-\beta)^\tau$, thus
\begin{equation}
\beta_c=1-\left(\frac{\kappa-2}{\kappa-1}\right)^{1/\tau}.
\end{equation}
These results also show that different combinations of $\beta$ and $\tau$ could result in the same spreading result, related to the percolation cluster under occupied probability $T$ \cite{Newman2002}.

\subsubsection{SIS model}

Based on the degree-based mean-field theory \cite{Pastor-Satorras2001}, the complete evolution equation for the SIS model on a network with arbitrary degree distribution can be written as
\begin{equation}
\frac{di_{k}(t)}{dt}=\beta [1-i_{k}(t)]k\Theta_{k}(t) − \mu i_{k}(t).
\end{equation}
The creation term considers the density $1−i_{k}(t)$ of susceptible nodes with degree $k$ that can get the infection from a neighboring individual \cite{Lloyd2001}. For uncorrected networks, $\Theta(t)=\sum_{k'}k'p_{k'}i_{k'}(t)/\langle k\rangle$ is used in Ref.\cite{Pastor-Satorras2001}, which is different from Eq.(\ref{eq5-si2}). This is because the infected nodes can recover to be a susceptible, and all the excess-degree of an infected node at the end of a link could lead to a node in state $S$. However, we think that around or above the epidemic threshold, on average an infected individual is very unlikely to recover before the individuals infected by it transmit the disease to others. Thus, we still use Eq.(\ref{eq5-si2}) to represent $\Theta_{k}(t)$ here. It must be pointed out that both the two choices are an approximation for $\Theta_{k}(t)$, and do not affect the basic conclusion.

With the $t\rightarrow\infty$ limit, the system can reach the stationarity state, \textit{i.e.}, $di_{k}(t)/dt=0$. This yields
\begin{equation}
i_k(\infty)=\frac{k\beta\Theta(\infty)}{\mu+k\beta\Theta(\infty)}.  \label{eq5-sisi}
\end{equation}
This equation shows that the larger the degree of a node, the higher its probability to be infected. Substituting Eq.(\ref{eq5-si2}) into Eq.(\ref{eq5-sisi}), a self-consistent equation can be found
\begin{equation}
\Theta(\infty)=\frac{1}{\langle k\rangle}\sum_{k}p_{k}(k-1)\frac{k\beta\Theta(\infty)}{\mu+k\beta\Theta(\infty)}.
\end{equation}
This equation always has a trivial solution $\Theta(\infty)=0$, and the nontrivial solution exists only for
\begin{equation}
\frac{d}{d\Theta(\infty)}\left.\left[\frac{1}{\langle k\rangle} \sum_{k}p_{k}(k-1) \frac{k\beta\Theta(\infty)}{\mu+k\beta\Theta(\infty)}\right] \right|_{\Theta(\infty)=0} \geq 1.
\end{equation}
Thus, the epidemic threshold reads
\begin{equation}
\delta_c=\frac{\beta_c}{\mu_c}=\frac{1}{\kappa-1}.
\end{equation}
This result is the same as that found in SIR model, the qualitative analysis is not repeated here. Note that for SF networks with $2<\lambda<3$, neither assortativity nor disassortativity can reconstruct a non-vanished epidemic threshold for SIS model \cite{Boguna2003a}.

For the version that infected nodes remain in state $I$ for $\tau$ time steps, and then become $S$ again, the epidemic threshold of SIS model can be also derived by the branching process in percolation theory \cite{Parshani2010b}. In SIS model, it has been found that there exist dynamic correlations coming from that if one node is infected, its neighbors are likely to be infected, which has been neglected by the degree-based mean-field theory. To account for this, effective degree approach \cite{Lindquist2011,Cai2014} and the combination of degree-based mean-field theory and effective degree approach \cite{Cai2016} have been proposed to obtain accurate expressions for the average prevalence and the epidemic threshold.

\subsubsection{Immunization}

Not only the epidemic model itself but also the immunization strategy for suppressing the spreading is related to percolation process. The simplest immunization strategy is the random one, which chooses immune individuals randomly from a population. After immunization, the infected individuals cannot transmit disease to the immune individuals. Thus it is equivalent to a spreading process on a diluted network. If there is no giant cluster in the diluted network, the epidemic will not break out, regardless of infection rate. The corresponding immunization ratio is called immunization threshold. A detailed discussion for this problem is the same as that for the network robustness, and thus immunization can be treated as a percolation process.

The introduction of a fraction $g$ of immune individuals chosen randomly is equivalent to dilute nodes with probability $1-g$. According to Eqs.(\ref{eq2-g0p}) and (\ref{eq2-g1p}), the immunization process results in $\kappa-1\to(\kappa-1)(1-g)$ \cite{Pastor-Satorras2002}. Thus, the epidemic threshold is rescaled as
\begin{equation}
\delta_c=\frac{\beta_c}{\mu_c}=\frac{1}{1-g}\frac{1}{\kappa-1}.   \label{eq5-bgk}
\end{equation}
This generally covers the common sense that the more immune individuals, the larger the epidemic threshold. However, for SF networks with $2<\lambda<3$, $\kappa$ is divergent, implying that random immunization cannot bring the epidemic into healthy region except for the case $g=1$. From Eq.(\ref{eq5-bgk}), we can also find that if $1-g<1/(\kappa-1)\equiv p_c$, even if $\beta=1$ Eq.(\ref{eq5-bgk}) does not make sense. This indicates that when the fraction of unimmunized individuals is below the percolation threshold $1-g_c=p_c$, the epidemic cannot break out \cite{Pastor-Satorras2002}. This is mainly because below this threshold the susceptible individuals are segregated in smaller regions by immune individuals.

Obviously, whoever the immune individuals are, they can suppress the infection locally. However, an apparent inadequacy of random immunization strategies is that it gives the same importance to both high-connected nodes (with the largest infection potential) and low-connected nodes, and thus is unable to find the critical individuals to make them immune for the eradication of infection. Targeted immunization is referred as the immunization of most highly connected individuals (potentially the largest spreaders), \textit{i.e.}, a fraction $g$ of the individuals with the highest degrees is immunized \cite{Anderson1992}. The equivalent diluted result depends on the details of the targeted immunization, which is very similar to the intentional attack in the study of network robustness (see Sec.\ref{sct4-nr}).

In Ref.\cite{Pastor-Satorras2002}, the authors showed that the immunization threshold takes the form $g_{c}\sim e^{-2\mu/m\beta}$ for SF networks with $\lambda=3$, where $m$ is the minimum degree. This clearly highlights that the targeted immunization program is extremely convenient, with an immunization threshold that is exponentially small in a wide range of infection rates. Based on the global knowledge of network, the nodes with the highest recalculated degree or betweenness centrality are selected for immunization during the selection procedure \cite{Holme2002}. Schneider \textit{et al.} sought to minimize a properly defined size for the connected cluster of susceptible individuals, which is found to be more effective than the methods based on immunizing the highest-betweenness links or nodes \cite{Schneider2011a}. Similarly, seeking to fragment the network into connected clusters of approximately the same size, ``equal graph partitioning'' strategy can achieve the same degree of immunization as targeted immunization by fewer immunization doses \cite{Chen2008}.

An efficient targeted immunization requires global information about the network, which makes it to be impractical in many real situations. Acquaintance immunization can make immune a small fraction of random acquaintances of randomly selected nodes \cite{Cohen2003a}. In this way, the highly connected nodes are more likely to be chosen and immunized, and the process prevents epidemics with a small finite immunization threshold and without requiring global knowledge of the network. In acquaintance immunization, a random fraction $q$ of nodes is chosen and then selects their random neighbors, \textit{i.e.}, acquaintances. The acquaintances, rather than the originally chosen nodes, are the ones immunized. The fraction $q$ may be larger than $1$, for a node might be queried more than once, on average, while the fraction of nodes immunized $g$ is always less than or equal to $1$. The critical threshold $q_{c}$ for a complete immunization can be solved analytically based on a branching process on a network as that in percolation model.

If more local information is available, the acquaintance immunization strategy can be further improved in efficiency. For example, allowing initial selected node to have knowledge of the degrees of its nearest neighbors, immunizing the neighboring nodes with the largest degree is smart for efficiency \cite{Holme2004}. Similarly, if the information within a distance $d$ of initial selected node is available, one can consider the immunization of the nodes with the highest degrees within that scope.

\subsection{Cascading process on networks}

It has been widely recognized that there are interdependencies or hidden interactions among the nodes in a networked system, which allows the viruses, information, opinion or failures to propagate from node to node via a certain topology. Although the specific scenarios might be different, most models present an iterative process: each individual in a population must decide a state between two or more alternatives and their decisions are made based on the options of their neighbors rather than relying on their own information about the problem, thus after an initial state setting the state flipping of individuals will occur recursively until no node meets the flipping criterion. This type of processes is often called cascading process.

With a given amount of excitation sources, the emergence of a giant cluster formed by individuals with identical states is just a dependent percolation process. Most of the models reviewed in Sec.\ref{sct4} belong to this category, \textit{i.e.}, the system evolves in generations. Here we mainly concentrate on another model widely used for the cascading process, called threshold model.

%The epidemic model is just a simple cascading process, as well as the $k$-core percolation and its variants (see Sec.\ref{sct3-kcp}). Here, we concentrate on another model widely used in the modeling of cascading process, called threshold model.

Threshold model was first proposed by Watts for the study of cascading failures in networks \cite{Watts2002}. In this model, there are two states for nodes, namely, functional and failed, to be more generic, we use state $0$ and state $1$ instead. At each time step, the binary decision rule with externalities is outlined as follows: each node in state $0$ observes the current states (either $0$ or $1$) of all its $k$ neighbors, and adopts state $1$ if at least a threshold fraction $\beta$ of its $k$ neighbors are in state $1$, else remains unchanged. No transition from state $1$ back to state $0$ is possible. With a small fraction of seed nodes (state $1$) in a population which is initially set as state $0$, the system evolves at successive time steps with all nodes updating their states according to the threshold rule above.

In a sufficiently large random network with sparse connections and only a few seeds, it can be assumed that the network is locally tree-like, hence, no node has more than one neighbor in state $1$. Under this condition, only the nodes with degrees $k\leq1/\beta$ have the potential to flip their states to be $1$, called vulnerable nodes. If the vulnerable nodes percolate in the system, the global cascades can be triggered by the initial seeds; otherwise, the propagation of cascades will be limited by the stable nodes and global cascade is impossible.

That is to say, this dynamic problem can be mapped to a static percolation problem, which can be solved by the generating function approach, see Sec.\ref{sct2-ampb}. According to percolation threshold of random networks $G_1'(1)=1$ and denoting the probability that a node with degree $k$ is vulnerable as $\theta_{k}$, the largest vulnerable cluster percolates when
\begin{equation}
\frac{1}{\langle k\rangle}\sum_{k}\theta_{k}p_{k}k(k-1)=1.          \label{eq5-tmc}
\end{equation}

With the above analysis, we know that $\theta_{k}=1$ for $k\leq1/\beta$, otherwise $\theta_{k}=0$. For large thresholds $\beta$, Eq.(\ref{eq5-tmc}) has no solution, meaning that the vulnerable nodes cannot percolate and there is no global cascade in the system. For small thresholds $\beta$, Eq.(\ref{eq5-tmc}) has two solutions resulting in two phase transitions. Below the smaller critical point, the network is composed of some small clusters and the propagation of cascades is thus constrained by these segregated clusters. Above the larger critical point, the network becomes dense enough, so that the propagation of cascades is blocked by the local stability of nodes. In summary, the global cascades can only occur in the network with a structure lying between the two phase transition points. Moreover, the transition of the cascade upon the average degree is continuous at the smaller critical point and discontinuous at the larger critical point.

The above discussion is based on that the seeds are scarce. Gleeson and Cahalane further studied the effects of seed size on the cascading process \cite{Gleeson2007}. Their method is based on a level-to-level update process, the final cascade size $\rho$ (rescaled by the system size) of active nodes (state $1$) is given by
\begin{equation}\label{chap-5-2}
\rho=\rho_{0}+(1-\rho_{0})\sum_{k=1}^{\infty}p_{k}\sum_{m=0}^{k} \binom{k}{m} q^{m}(1-q)^{k-m}F\left(\frac{m}{k}\right),
\end{equation}
where $\rho_{0}$ is seed size, and $F(m/k)$ gives the probability that the node with degree $k$ and $m$ active neighbors meets the excitation condition. In addition, $q$ is the probability that a link leads to an active node, which satisfies the recursion relation
\begin{equation} \label{chap-5-3}
q=\rho_{0}+(1-\rho_{0})\sum_{k=1}^{\infty}\frac{p_kk}{\langle k\rangle}\sum_{m=0}^{k-1} \binom{k-1}{l}\binom{l}{n} q^{m}(1-q)^{k-m-1}F\left(\frac{m}{k}\right).
\end{equation}
With these equations, the cascade size can be solved. The results show that the cascade transition at low $\langle k\rangle$ may in fact be discontinuous in certain parameter regimes, and the seed size $\rho_0$ as low as $0.1\%$ has dramatic effects on the location of cascade transition points, that is
\begin{equation}
\frac{1}{\langle k\rangle}\sum_{k=1}p_{k}k(k-1)\left[F\left(\frac{1}{k}\right)-F(0)\right]>\frac{1}{1-\rho_{0}}.
\end{equation}
Obviously, this result reduces to Eq.(\ref{eq5-tmc}) found in Ref.\cite{Watts2002} by letting $\rho_0\to0$, for which $F(0)=0$ and $\theta_{k}=F(1/k)$.

This analysis method is also valid for large seed $\rho_0$, which has been also extended to analyze networks with community structures and degree-degree correlations \cite{Gleeson2008,Dodds2009}. For networks with clusterings \cite{Hackett2011}, it turns out that for large and small values of $\langle k\rangle$ clustering reduces the size of cascades, while the converse occurs for intermediate values of the average degree. In Ref.\cite{Liu2012}, Liu \textit{et al.} showed that using the seed size $\rho_0$ as the control parameter, the crossover of the continuous and discontinuous phase transitions can be found.

Watts's threshold model can be seen as a particular instance of a more generalized model of contagion \cite{Dodds2004}, and has been used to study the effects of network properties that may influence the spreading process, such as boolean networks \cite{Samuelsson2006}, SW networks \cite{Centola2007}, multiple layers \cite{Brummitt2012}, temporal networks \cite{Backlund2014} where the associated bursty activity of individuals may either facilitate \cite{Takaguchi2013} or hinder \cite{Karimi2013} the spreading process.

\subsection{Traffic and transportation}

As travel demand increases in modern cities, traffic congestion becomes more frequent. As a result, the study on the dynamical transition from free flow to congestion in traffic flow has attracted much attention. To understand the traffic transition in a network scale (representing a city or an urban region), the percolation model has also been applied to the urban traffic network.

With percolation theory, the typical question of how the traffic in the network collapses from a global efficient state to isolated local flows in small clusters is clearly demonstrated, which might be useful for improving transportation efficiency, reducing traffic delay, and facilitating emergency evacuation. Percolation theory also provides useful tools for urban planning. Traffic network connectivity under disasters, such as earthquake or rain, has also been investigated by percolation theory.

\subsubsection{Percolation in urban traffic networks}

Li \textit{et al.} considered the percolation process in Beijing road traffic network, and identified bottleneck roads that connect different functional clusters \cite{Li2014b}. The velocity data of $5$-min segment records measured in a central region of Beijing is used to construct the road network, where nodes represent the intersections and links represent the road segments between two intersections. For each road, the velocity varies with time during a day. A dynamic functional traffic network can be constructed by the links with velocity higher than a given threshold $q$. With a small threshold $q$, a giant cluster of functional network can extend to almost the full scale of the road network. As $q$ increases, the size of the giant component decreases, and the second-largest cluster becomes maximal at a threshold $q_c$, which signifies the percolation transition from the connected phase to the fragmented phase of the functional traffic network. The critical value $q_c$ can be seen as an indicator of the efficiency of global traffic. Obviously, an efficient traffic system will have a higher $q_c$.

In a static network, bottleneck links are identified usually based on structural connectivity information. Traffic, as a dynamical non-equilibrium system, evolves with time and shows varying $q_c$ during the day. The bottleneck links of global traffic will also evolve accordingly and appear in different locations in different hours. By investigating the functional cluster at percolation criticality, Li \textit{et al.} identified some bottleneck links that play a critical role in bridging different functional clusters of traffic \cite{Li2014b}. With the congestion of these links, the giant cluster will disintegrate and result in some small clusters. Conversely, improvement of these critical bottleneck roads can significantly increase the threshold $q_c$ of the traffic network, thus benefit the global traffic. This kind of improvement of $q_c$ is higher than that by increasing the velocity of a random link or the link with the highest path-based betweenness. This proves the uniqueness of the bottleneck links found by percolation theory.

Zeng \textit{et al.} further identified two modes of percolation behaviors in Beijing and Shenzhen traffic network \cite{Zeng2018}. To investigate the differences of the dynamic traffic network in rush hours and nonrush hours, they performed a percolation analysis by calculating the size of the functional clusters. At criticality, a well-defined Fisher exponent $\tau$ as that in percolation theory can be observed. With this, Zeng \textit{et al.} found that the disintegration transition of urban traffic can be characterized by two sets of exponents. During rush hours on workdays, the critical exponent $\tau$ is in general smaller than that during nonrush hours or days off. During workday rush hours, $\tau$ is close to the theoretical result of the two dimensional lattice ($\tau=187/91\approx2.055$), while during other periods, it is close to the mean-filed value for high dimensional systems ($\tau=5/2$).

This phenomenon can be understood as follows. The urban traffic network can be generally seen as a system embedded in two dimensions with some long-range connections corresponding to urban highways. During nonrush hours, all the roads are in the free flow state, the system is thus a two dimensional lattice with long-range connections. In contrast, during rush hours, the highways become congested and the urban traffic network reduces to a two dimensional system. With the aid of traffic management methods, the traffic system can be tuned to the desired class by adjusting the amount of effective long-range connections. Using the corresponding signal control or road pricing on urban highways, this mechanism can help to create a significantly better traffic system.

The above investigations are based on the analysis of the empirical traffic data. Obviously, the percolation properties vary with traffic parameters, such as the level of traffic load and the detailed behaviors of drivers. Such analysis can also be performed by modeling simulations. Wang \textit{et al.} studied the traffic percolation properties on a traffic system based on an agent-based model \cite{Wang2015a}. Using the traffic model in Ref.\cite{Echenique2005}, the level of traffic demand and routing preference can be tuned to verify their impacts on the percolation property. It has been widely proved that the traffic will undergo a phase transition from the free state to the congested state with the increase of traffic load in the network. In Ref.\cite{Wang2015a}, the functional network is constructed by the free state nodes with load lower than a threshold. The emergence of the giant cluster indicates the percolation transition and the formation of global traffic flow. It is also found that the threshold of the traffic percolation $q_c$ shows a sharp minimum value at the traffic transition threshold, indicating that the global flow can be maintained with minimal number of local flows. The value of $q_c$ is also influenced by the routing choice. In the giant percolation cluster, optimal paths will accumulate to a high level. This can provide a possible routing choice to mitigate traffic congestion.

In the modern urban traffic system, different transportation modes have been widely constructed, including bus, metro, and road networks, which can be represented as a multiplex network. In such a system, a fraction of node failures in one layer can trigger a cascade of failures that propagate in the multiplex network, see Sec.\ref{sct3-pin}. Baggag \textit{et al.} studied the resilience and robustness of multi-modal transportation networks using percolation theory \cite{Baggag2018}. In this study, percolation is used to investigate the impact of link failures in the multiplex transportation network. By calculating the size of the mutually connected giant cluster under random or targeted failures, the robustness of multiplex transportation networks can be evaluated. The four cities studied in Ref.\cite{Baggag2018} behave similarly in terms of coverage degradation under random failure, with Paris network being the most robust.

\subsubsection{Percolation in connected vehicle networks}

Connected Vehicles (CV) technology is the leading edge of intelligent traffic systems \cite{Mahmassani2016}, and has great potential to mitigate traffic congestion through the creation of a CV network \cite{Ubiergo2016}. Once the vehicle cloud network is formed, travel information such as speed, density, and signal timing can be retrieved from the network. This information would help the driver's decision-making and eventually enhance the mobility of CV in terms of travel time. In an urban traffic system, the benefits of the corresponding applications will depend on the CVs' communication range and the market penetration, which can be featured by the percolation model.

The study of CV network was pioneered by the analysis of network connectivity for wireless sensor networks and ad-hoc networks. Ammari and Das studied the sensing-converge and network-connectivity in wireless sensor networks using percolation theory \cite{Ammari2008}. A probabilistic approach is created to compute the covered area fraction at percolation threshold. Under various scenarios, the critical percolation density and radius of the covered components are identified. Similarly, Khanjary \textit{et al.} conducted percolation studies in the two-dimensional fixed-orientation directional sensor network \cite{Khanjary2015}. The critical density of the nodes can be analytically computed from the percolation theory framework.

Jin \textit{et al.} first assessed the connectivity of Vehicular Ad-hoc NETwork (VANET) through a percolation framework \cite{Jin2011}. The relationship among network connectivity, vehicle density and communication range was analyzed. When vehicle density or communication range is big enough, VANET will show a jump of network connectivity, corresponding to the percolation transition. Given vehicle density, the minimum communication range can be calculated to achieve good network connectivity. As a large transmission range can cause serious collisions in wireless links, the percolation analysis can help to determine the tradeoff and guide the deployment of VANETs in real world.

Talebpour \textit{et al.} investigated the effect of CV information availability on the stability of traffic flow through continuum percolation theory \cite{Talebpour2017}. The impact of CV density and communication range on the connectivity of the network is determined by percolation properties. The results show that as the communication range increases, the traffic flow system becomes more stable.

Mostafizi \textit{et al.} investigated the impacts of CV network on the mobility of traffic systems at varying levels of market penetration and communication range \cite{Mostafizi2017}. When more CVs are present, the network tends to be more connected. When the percentage of CV decreases, there exists a percolation threshold where the giant cluster breaks apart. It has been found that the percolation transition is a function of both market penetration and communication range. The emergence of mobility benefits is highly related to the percolation transition in CV network. Only when the system reaches the percolation threshold, the benefits of reducing mean travel time begin to appear. Below the threshold, although small clusters can form in the network, they will not create a significant impact on the mean travel time due to the scarce information available from the CV network.

\subsubsection{Percolation in urban traffic planning}

The analysis of the inter-relations between traffic flows and urban street networks can provide a meaningful tool for urban planning processes. From this perspective, Serok \textit{et al.} identified functional spatio-temporal street clusters of London and Tel Aviv using network percolation analysis \cite{Serok2019}. The dynamics of these clusters and their spatial stability over time are analyzed, which can help to develop new, adaptive, decision-making tools for urban and transportation planners. Traditionally, urban planning is a long-term and static process. The application of traffic percolation will enable planners to keep their role in the flexible and dynamic urban system.

Behnisch \textit{et al.} studied the settlement connectivity in Germany using the percolation concept \cite{Behnisch2019}. The percolation is introduced to quantify the connectivity of buildings. They found that at a critical distance of $830\pm10$ m the buildings settlement network in Germany will undergo a transition from isolated clusters to a country-spanning building cluster. This rather short critical distance indicates that the landscape is already overbuilt in Germany. For urban planning, limiting urban sprawl and further land consumption is crucial. The application of percolation can provide a useful monitoring tool for settlement and landscape degradation.

Transport accessibility is an important issue for industrial relocation and the related sustainable development of economics. Jiang \textit{et al.} studied the impact of transportation accessibility on industry relocation pattern of China's Yangtze River Economic Belt using percolation theory \cite{Jiang2018}. In the network, the nodes are cities, and the links are defined according to the Yangtze River's waterways and the highways in the Yangtze River Economic Belt. Their results proved the existence of percolation transitions during the process of industry relocation, and identified the bottlenecks for industry relocation as cities located in border regions.

Piovani \textit{et al.} studied the relation of urban retail location distribution and road network properties in London from percolation theory. Road network and retail location are two interwoven factors in an urban ecosystem. The emergence of clustering in retail centers depends greatly on road network \cite{Piovani2017}. Piovani \textit{et al.} compared the road network's hierarchical percolation structure with the retail location distribution. To evaluate the road network, the subgraph is constructed by the road segments with length below a threshold, and the transition is defined at the threshold that generates the maximal entropy configuration. The results show that clusters of the road network and retail location are very similar in spatial characteristics.

\subsubsection{Percolation in post-disaster traffic networks}

Post-disaster traffic networks are very important since they offer access to affected areas, sustain evacuation and medical operations after a disaster. If a disaster is very severe and many roads are malfunctioned, the traffic network could collapse into isolated subnetworks, and there will be no route connecting some locations. Therefore, the connectivity of post-disaster traffic networks also provide an important index for the assessment of vulnerability, robustness, and resilience of the road system.

Zhou \textit{et al.} studied the connectivity of post-earthquake road networks using percolation theory \cite{Zhou2019}. To assess the road network, they proposed the concepts of ``global connectivity" and ``local connectivity". Global connectivity measures the extent to which the whole network is connected, and local connectivity measures the distances between each node to its neighbors. Specifically, the global connectivity is quantified by the percolation threshold, the size of the giant subnetwork, and average sizes of small subnetworks, while local connectivity is quantified by the number of neighbors. In particular, Zhou \textit{et al.} proposed an integrated percolation model combining localized attacks and random failures to capture the feature of earthquake-impacted \cite{Zhou2019}. Namely, the impact of the earthquake on the inner network is modeled as a localized attack (where node disruption is dominant), while it is modeled as random link disruption in the outer network. The model can be used to assess the connectivity of road networks impacted by different magnitudes of earthquakes.

Extreme weather, such as torrential rain and hurricane, could greatly influence urban traffic system. Although the traffic flows under extreme weather have been widely studied, little work has been done to understand whether and how the destruction of local roads can degrade the global traffic. Guo \textit{et al.} studied the impact of torrential rain on traffic percolation properties in Beijing and Wuhan \cite{Guo2018}. In Beijing, whose street layout is more symmetrical, the dysfunctional roads are scattered uniformly. However, for Wuhan with a river valley, the dysfunctional roads appear in a cluster way and partially along the riverbank. Interestingly, the percolation threshold of traffic network is stable against weather perturbation, even if some roads are significantly influenced. This is mainly because extreme weather (torrential rain) will not only damage road conditions in the supply end, but also reduce the traffic demand correspondingly.

\subsection{Evolutionary game}

Evolutionary game theory offers a powerful mathematical framework to study the emergence of cooperation among selfish individuals. When the contact network of individuals is taken into account, \textit{i.e.}, the spacial evolutionary game, the focus of the research just translates to the emergence of the clusters formed by adjacent cooperators. Therefore, the percolation theory is also employed to explore the spacial evolutionary game.

In Ref.\cite{Yang2014}, Yang \textit{et al.} studied the cooperation percolation in spatial prisoner’s dilemma game. Their model is defined on a two-dimensional square lattice. Initially, with probability $f$ a node is occupied by a cooperator and with probability $1−f$ a node is occupied by a defector. At each time step, each individual plays the prisoner’s dilemma game with its four nearest neighbors. Specifically, if both players cooperate, then both get the payoff $1$. If one cooperates while the other defects, the defector gets $b$ while the cooperator gets nothing. If both defect, no gains for both of them. After the games, individuals update their strategies asynchronously in a random sequential order. A randomly selected player compares its payoff with its nearest neighbors and changes strategy by following the one (including itself) with the highest payoff.

They found that the percolation transition can be well-defined by the control parameter $f$, the phase diagram can be divided into several regions based on whether the cooperator cluster can percolate. With the analysis of finite size scaling, they claimed that the region $1<b<4/3$ belongs to the universality class of the classical percolation in two dimensions, whereas region $3/2<b<2$ is subject to that of invasion percolation with trapping. However, an exhaustive simulation revealed that all these regions belong to the universality class of the classical percolation in two dimensions \cite{Choi2015}. In Ref.\cite{Choi2015}, the authors also studied the model on the triangular lattice and the hexagonal lattice, and concluded that the observed percolation transitions on various two-dimensional lattices belonged to the universality class of the standard percolations.

In Ref.\cite{Yang2019} the percolation transition was also observed in the Stag Hunt game and the Snowdrift game. More interestingly, they found a double transition in ER networks. With the increase of $f$, the giant cooperator cluster increases continuously from zero above a critical point, then at a second critical point the giant cooperator cluster jumps from a low value to a very high value. These rich phenomena highlight the role of the social contact in the emergence of the cooperation.

In Ref.\cite{Lin2010} Lin \textit{et al.} also found a scaling feature in the evolutionary game on the interaction graph extracted from percolation process. The power-law distribution of cooperator clusters like that in percolation model was also observed in fractal hierarchical networks \cite{Yun2011}. Besides, an interaction graph at the percolation threshold has also been proved to be an optimal population density for public cooperation \cite{Wang2012,Wang2012a}. Peng and Li suggested that this is due to the fractal structure formed at the percolation threshold \cite{Peng2020}, which constructs an asymmetric barrier that the defection strategy is almost impossible to cross, but the cooperation strategy has a not too small chance. In the system with mobile individuals, the geometric percolation of the contact network (\textit{i.e.}, irrespective of the strategy) also enhances cooperation \cite{Vainstein2014}.

%*****************************************************************************
% Section VI
% Discussion
%*****************************************************************************

\section{Discussion and outlook}  \label{sct6}

Percolation describes the patterns of connections under a random or semi-random connection mechanism, while network science focuses on the non-trivial topological features inspired largely by the empirical study of real-world networks. In this sense, although the proposal of percolation model predates the rise of the research of complex networks by about fifty years, it inherently provides a framework for the studies of complex networks. So it is no surprise that the application of percolation theory in network science has achieved fruitful results.

First of all, the study of percolation process on a non-trivial topology, including strong heterogeneity, high clustering, assortativity or disassortativity, modular, hierarchical structures and so on, has brought new power to percolation theory, especially for the fractal and non-physical dimension systems. For examples, the heavy-tailed degree distribution of the SF network induces a special mean-field nature of percolation transition (see Sec.\ref{sct2-sb}), and the percolation in the growing network demonstrates an infinite order transition like that found in the Berezinskii-Kosterlitz-Thouless phase transition in condensed matter (see Sec.\ref{sct3-ptng}). In recent years, we have also witnessed a wide range of discussion on the explosive percolation and the hybrid percolation transition on network systems (see Sec.\ref{sct3}). Besides, quite a number of pruning rules used in the identification of special network structures are also defined in terms of percolation process, such as $k$-core percolation (see Sec.\ref{sct3-kcp}), clique percolation (see Sec.\ref{sct4-cd}), and core percolation (see Sec.\ref{sct3-kcp}). These percolation models also attracted a lot of studies on the system with physical dimensions, and enriched the research of the percolation theory \cite{Araujo2014,Saberi2015}.

In addition to providing theoretical supports for the findings from empirical researches, the ideas and conceptions of percolation theory furnish a quantitative analysis approach to the methodology of network science. A typical representative is the giant cluster, which has already been widely used to evaluate the state of a network. For the study of network resilience, if a giant cluster exists, the network is considered functional, otherwise the network is paralyzed. In this way, the percolation threshold can be used to quantify the robustness of networks (see Sec.\ref{sct4-nr}). Similarly, if the infections form a giant cluster, it can be concluded that the epidemic breaks out, so that an epidemic with a smaller percolation threshold is more infectious (see Sec.\ref{sct5}). Even for the problems that seemed to be irrelative to percolation process, the percolation analysis can offer a good interpretation for their findings, such as the evolution of traffic state of an urban network, and the emergence of the cooperation clusters (see Sec.\ref{sct5}). Although almost all of these studies are not originally motivated by the percolation process and the evolution rules might be far away from the standard percolation, the emergence of the giant cluster is employed as a picture of the inherent problem. Because of this, some of the modeling of network dynamics just take the form of percolation process, such as the cascading failures, and the spreading of information/opinion/epidemics (see Sec.\ref{sct5}).

The classical percolation in physical dimensions often refers to regular systems, such that the percolation results are totally random, and the focus is more on the statistic characteristic of the clusters. However, the important characteristic of empirical networks, \textit{i.e.}, strong heterogeneity, can bring determinacy into the percolation configuration. For example, the hubs of a SF network almost always belong to the giant cluster of different realizations. Once the hubs are destroyed, the network will be disintegrated. From this perspective, the percolation results in part reflect the connective features of networks, and thus can be used as an algorithm for identifying the special subsets of networks. Some typical examples have been reviewed in Sec.\ref{sct4}, including $k$-core decomposition, community detection, vital node identification, and so on.

In summary, the percolation theory has already percolated into many research aspects of network science, ranging from structure analysis to dynamics modeling. Loosely speaking, one would hardly evade conceptions, analytic methods, and algorithms of percolation in the studies of complex networks. For this reason, almost all the review articles on the network science spill some ink on the corresponding percolation problems, such as network structures and dynamics \cite{Boccaletti2006,Albert2002,Dorogovtsev2008,Yeung2013}, multiplex networks \cite{Boccaletti2014,Havlin2014,Kivelae2014,Shekhtman2016,Kenett2015}, spatial networks \cite{Barthelemy2011}, epidemic process \cite{Pastor-Satorras2015,Wang2017}, explosive transition \cite{Boccaletti2016}, community detection \cite{Fortunato2010,Malliaros2013}, $k$-core \cite{Kong2019}, and vital nodes identification \cite{Lue2016}. Besides, the significant findings of network percolation are also included in the articles on recent advances of percolation transition \cite{Saberi2015,Araujo2014}.

By tracing the progress of percolation on networks and its applications, in this paper we briefly review the percolation on complex networks from models, theoretical methods, to applications for network problems. Overall, the studies of percolation on networks yield a rich harvest from theory to applications. Meanwhile, many challenges still remain. First, real networks have a rich mixture of properties, such as degree correlation, clustering, modularity, heterogeneity, small world, and spatial constraint. For these complex structures, only some approximate theory can be applied to simplify the structural models, while whether these structures can separately or collectively change the nature of the percolation transition or not remains an open question. Moreover, the mixture pattern often refers to the high-order structure of networks. This is a new research hotspot of network science in recent years, for which the percolation theory also plays an important role \cite{Bianconi2018b,Bianconi2019a,Kryven2019b,Bianconi2020,Fountoulakis2020,Sun2020,Battiston2020}.

%To answer this question, an exhaustive examination on the phase transition via Monte Carlo simulation is needed. Before this, the model-dependent network ensemble may still have to be overcome. Taking SF networks as an example, the hidden parameter model can only generate the power-law degree distribution for large degrees, and show a negative deviation for small degrees. Conversely, the configuration model shows a good power-law for small degrees and large fluctuations for large degrees.

Furthermore, as a random process, the percolation result of a given network can differ from realization to realization, while the network identification often requires a deterministic result. How to solve this conflict when employing percolation process to analyze network structure? Are there some ways to take advantage of the fluctuation of percolation results to give more information on network structures? Moreover, it is known that the percolation clusters in the subcritical phase are dominated by local connections, and the global connections for the supercritical phase. In this way, is there a unified algorithm based on percolation process for identifying network structures in different scales? To answer these questions, more researches should also be done to explore the percolation process on networks.

\section*{Acknowledgments}

We would like to thank Pan Zhang, Manuel S. Mariani, and Xu Na for useful discussions and comments. The research was supported by the National Natural Science Foundation of China under Grant Nos.11622538, 61673150, 61773148, and 12072340. L.L. also acknowledges Zhejiang Provincial Natural Science Foundation of China under Grant No.LR16A050001, and the Science Strength Promotion Programme of UESTC.

\bibliography{ref}

\end{document}